\documentclass[a4paper,10pt]{article}
\usepackage{amsmath}
\usepackage{amssymb}
\usepackage{lmodern}

\usepackage{sectsty}
\sectionfont{\fontsize{12}{15}\selectfont}

\usepackage{extsizes}
\usepackage{textgreek}
\usepackage{multicol}
\usepackage{fullpage}
\usepackage{setspace}
\usepackage{pdflscape}
\usepackage[export]{adjustbox}
\usepackage{caption}
\usepackage{chngcntr}

\usepackage[usenames,dvipsnames]{color}
\usepackage{sansmath}
\usepackage{txfonts}
\usepackage[a4paper,margin=1in,headsep=7.5mm,footnotesep=2.2\baselineskip]{geometry}
\usepackage{times}
\usepackage{framed}
\usepackage{relsize}
\usepackage{mathtools}
\usepackage{xfrac}
\usepackage{graphicx}
\usepackage{tikz}
\usepackage[Symbol]{upgreek}
\usepackage{array}
\usepackage{accents}
\usepackage[T1]{fontenc}
\usepackage{esint} 
\usepackage{lipsum}
\usepackage{float}
\usepackage{epstopdf}
\usepackage{color}
\usepackage[font=small,labelfont=bf]{caption}
\usepackage{fancyhdr}

\allowdisplaybreaks

\usepackage[hang,flushmargin,bottom]{footmisc} 

\makeatletter
\DeclareSymbolFont{largesymbolsB}{U}{esint}{m}{n}
\re@DeclareMathSymbol{\intop}{\mathop}{largesymbolsB}{'001}
        \def\int{\intop\nolimits}
\makeatother

\newcommand*{\dt}[1]{
  \accentset{\mbox{\large\bfseries .}}{#1}}
\newcommand*{\ddt}[1]{
  \accentset{\mbox{\large\bfseries .\hspace{0.00ex}.}}{#1}}

\newcommand{\tit}[1]{{\fontfamily{ppl}\selectfont \textit{#1}}}
\def\ccos{\text{\,\tit{cos}\,}}
\def\csin{\text{\,\tit{sin}\,}}

\usepackage{rotating}
\def\infinity{\rotatebox{90}{8}}

\usepackage[pscoord]{eso-pic}
\definecolor{armygreen}{rgb}{0.01, 0.55, 0.01}
\definecolor{darkgreen}{rgb}{0.08, 0.48, 0.18}
\definecolor{darkred}{rgb}{0.86, 0.153, 0.153}
\definecolor{azure}{rgb}{0.03, 0.35, 0.84}
\definecolor{bole}{rgb}{0.9, 0.32, 0.05}

\usepackage[numbers]{natbib}
\usepackage[colorlinks=true,
            linkcolor=blue,
            urlcolor=blue,
            citecolor=red]{hyperref}

\makeatletter
\newcommand*{\textlabel}[2]{%
  \edef\@currentlabel{#1}
  \phantomsection
  #1\label{#2}
}
\makeatother

\newcommand{\qag}[1]{{\fontfamily{qag}\selectfont #1}}
\begin{document}
\newpage
\topskip15pt
\fancyhead[L]{\footnotesize\tit{Avneet Singh}}
\fancyhead[R]{{\footnotesize \tit{published as} \href{https://journals.aps.org/prd/abstract/10.1103/PhysRevD.95.024022}{\tit{\textbf{Physical Review D}} 95(2):024022}}}
\begin{flushleft}
\textbf{\large Gravitational wave transient signal emission via Ekman pumping in neutron stars during post-glitch relaxation phase\vspace{0.1in}}\linebreak
{{Avneet Singh$^\mathrm{1,\,2,\,3,\,\dagger,\,\ddagger}$\let\thefootnote\relax\footnote{$^\dagger$ avneet.singh@aei.mpg.de}\let\thefootnote\relax\footnote{$^\ddagger$ avneet.singh@ligo.org}}}\linebreak\linebreak
{{\footnotesize $^1$ Max-Planck-Institut f{\"u}r Gravitationphysik, am M{\"u}hlenberg 1, 14476, Potsdam-Golm\\
$^2$ Max-Planck-Institut f{\"u}r Gravitationphysik, Callinstra{$\upbeta$}e 38, 30167, Hannover\\$^3$ Leibniz Universit{\"a}t Hannover, Welfengarten 1, 30167, Hannover}}\linebreak\setcounter{footnote}{0}
\end{flushleft}
\begin{center}
\begin{abstract}
Glitches in the rotational frequency of a spinning neutron star could be promising sources of gravitational wave signals lasting between a few $\mu$s to a few weeks. The emitted signals and their properties depend upon the internal properties of the neutron star. In neutron stars, the most important physical properties of the fluid core are the viscosity of the fluid, the stratification of flow in the equilibrium state and the adiabatic sound speed. Such models were previously studied by \citet{melatos2008} and \citet{melatos2010} following simple assumptions on all contributing factors, in which the post-glitch relaxation phase could be driven by the well-known process of {\fontfamily{ppl}\selectfont \textit{Ekman pumping}} \citep{walin,epstein}. We explore the hydrodynamic properties of the flow of fluid during this phase following more relaxed assumptions on the stratification of flow and the pressure-density gradients within the neutron star than previously studied. We calculate the time-scales of duration as well as the amplitudes of the resulting gravitational wave signals, and we detail their dependence on the physical properties of the fluid core. We find that it is possible for the neutron star to emit gravitational wave signals in a wide range of decay time-scales and within the detection sensitivity of aLIGO for selected domains of physical parameters.
\end{abstract}
\end{center}
\begin{multicols}{2}
\section{Introduction}
\label{section:intro}
Pulsar glitches are sudden fractional increases in the rotational velocity of a neutron star. Several pulsars, observed in radio, X-ray and $\gamma$-ray bands of the electromagnetic spectrum, have been repeatedly observed to glitch \citep{magnetars,npulsars1,glitchstats}. The fractional spin-up $\delta\Omega$ of the rotational velocity $\Omega$ of the neutron star lies in the range of $\displaystyle \frac{\delta\Omega}{\Omega}\in[O(10^{-11}), O(10^{-4})]$\citep{melatos2008,glitchstats,glitchstats2}.

Gravitational wave emission is typically associated with a non-zero derivative of the quadrupole moment stemming from accelerated flow of non-axisymmetrically distributed bulk of matter. It is possible that such non-axisymmetric motions are excited following a glitch; possible mechanisms for producing such non-axisymmetric motions, besides {Ekman pumping}, include bulk two-stream instabilities \citep{tsi1}, surface two-stream instabilities \citep{tsi2}, crust deformation and precession \citep{crustd}, meridional circulation and super-fluid turbulence driven by crust-core differential rotation \citep{mcst}, crust-core coupling via magnetic field \citep{crustcore2}, excitation of pulsation modes \citep{mcc,qro1,qro2}, and mutual friction in two-fluid model for superfluid core \citep{superfluid}. These mechanisms have been briefly reviewed by \citet{melatos2008} and \citet{melatos2010}. In this paper, we solely consider the hydrodynamic properties of the fluid core following a glitch and concentrate on the mechanism of {Ekman pumping}. 

In this work, we consider the hydrodynamic evolution the post-glitch relaxation phase via the mechanism of {Ekman pumping}, pioneered by \citet{walin} and \citet{epstein}. We extend the previous works on this by \citet{melatos2008} and \citet{melatos2010}, where an initial non-axisymmetric perturbation introduced by the glitch induces {Ekman pumping} in the core of the star. {Ekman pumping} is briefly described as the induced flow of the bulk matter in the core when it is acted upon by a tangential force (in this case, Coriolis force) at its boundary i.e. the crust-core interface. In our case, the Coriolis force results from the differential rotation of the crust with respect to the bulk fluid, resulting from the glitch in the star's rotational velocity. This induced flow of the bulk matter could then have a time-varying quadrupole moment and lead to emission of gravitational waves. In this context, a glitch can lead to gravitational wave emission in two phases. Initially, a burst-type emission occurs during the fast spin-up of crust at time-scales of at most a few seconds \citep{Burst}. Secondly, a decaying continuous-wave signal during the post-glitch relaxation phase is emitted on much longer time-scales. The initial non-axisymmetric motion of the bulk with respect to the crust in the second case is excited by the glitch. The resulting damped continuous-wave-like signal arises as the internal fluid dynamics evolve to set the bulk in co-rotation or a steady differential rotation with the crust, erasing the non-axisymmetric motions in the bulk \citep{melatos2008,melatos2010}.

In this paper, we relax certain assumptions in more recent works \citep{melatos2008,melatos2010} on the stratification length and the adiabatic sound speed; we explore a regime of {Ekman pumping} where these quantities are allowed to vary across the star and study their effect on the emitted gravitational wave signal. This extends the parameter space and introduces more generality to the analysis.

We will keep other simplifying assumptions made in \citep{epstein,melatos2008}, and analyze the system in a pure hydrodynamical sense, ignoring the two-stream dynamics, sidestepping the crust-core interface, neglecting the effects of pinning and unpinning of quantum vortices, and disallowing the crust to precess, and ignoring the affects of magnetic field. In particular, the inclusion of magnetic field in the current model would make it analytically intractable. In the spirit of first tackling these two aspects separately i.e. magnetic field and {Ekman pumping}, this paper concentrates on the latter process only; such an approach has been extensively considered in the past \citep{epstein, melatos2008, melatos2010}. Moreover, for analytic simplicity, we approximate our spherical neutron star to a fluid-filled rigid cylinder \citep{melatos2010}, as opposed to the choice of semi-rigidity \citep{melatos2008}. The correctness of this choice will be explained in the next section. In nutshell, as a result of all these simplifications, a toy-model for a neutron star is studied for the possibility of emission of gravitational waves.\vspace{-0.13in}
\section{Hydrodynamics of the system}
\label{section:equationmain}
\begin{figure}[H]
\centering\includegraphics[width=65mm]{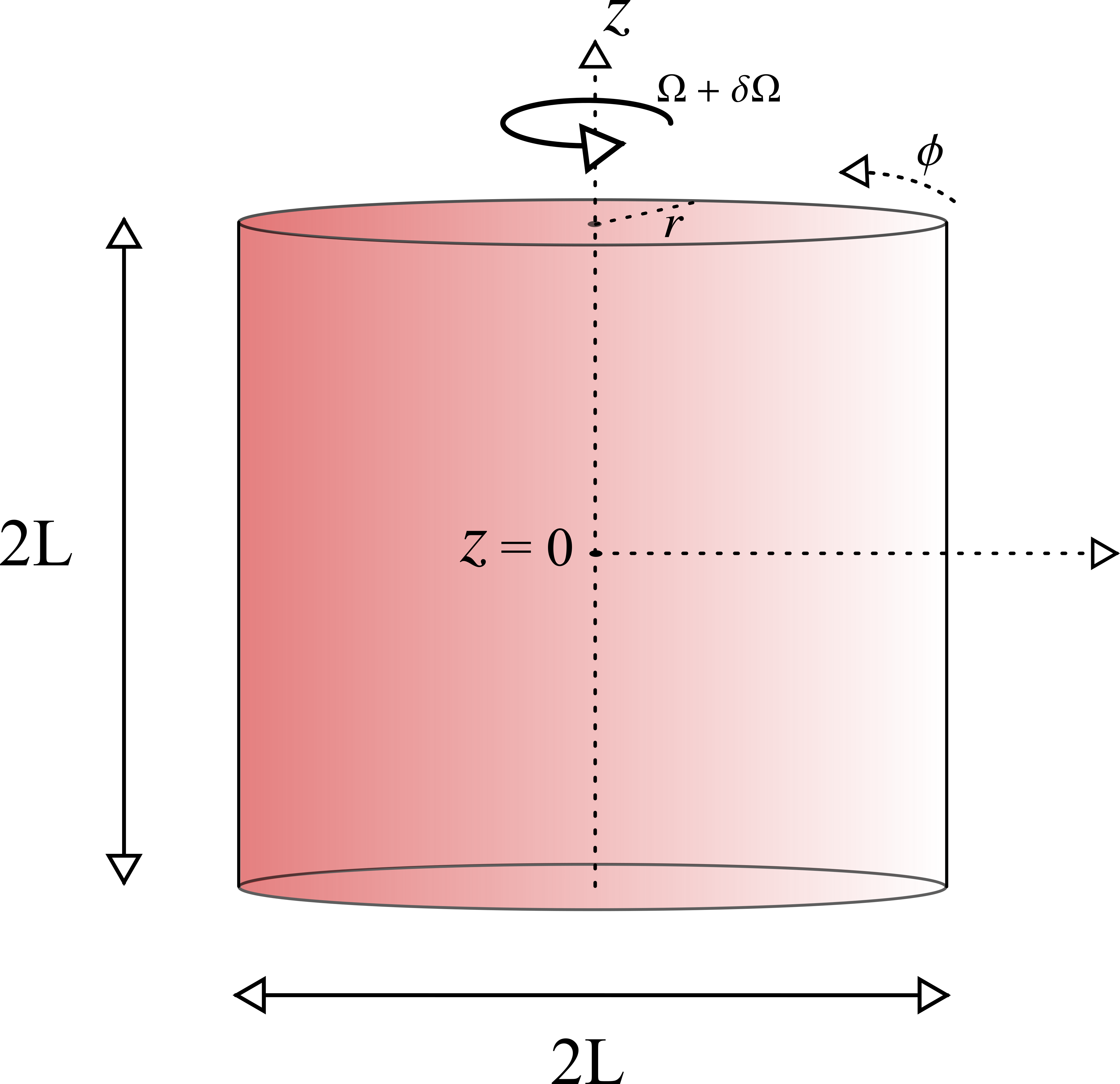} 
\caption{{\small Idealized system}}
\label{fig:cylinder}
\end{figure}

Let us consider a rotating neutron star of radius of order $O(\mathrm{L})$ with a solid crust around a compressible and viscous fluid with viscosity $\upnu$, pressure $p$ and density $\uprho$. We now approximate this spherical system with a rigid cylindrical container of height $2\mathrm{L}$ and radius $\mathrm{L}$ rotating at an angular frequency of $\Omega$ along the \textit{z}-axis (figure \ref{fig:cylinder}). We represent the glitch as a sudden perturbation in the angular velocity of magnitude $\delta\Omega$ along the \textit{z}-axis. This geometric simplification doesn't lead to an order-of-magnitude change in the amplitude or duration of the emitted gravitational wave signal from {Ekman pumping} \citep{cylinder}, and has extensively been used in majority of literature on neutron star modeling. Moreover, this reformulation to the cylindrical system leads to simpler analytic solutions.
\subsection{Governing equations}
\label{section:equations}
Our physical system is described by the velocity field $\vec{v}$, the pressure $p$ and the density $\uprho$ of the fluid. The forces acting on fluid elements of the bulk volume are the viscous force, the Coriolis force, the centrifugal force, the compressible strain, pressure gradients and gravitational force. The Navier-Stokes equation, the conservation of mass equation, and the `energy equation' (i.e. equation of state) govern our physical system. The Navier-Stokes equation for a fluid element in the rotating Lagrangian frame of the cylinder for a compressible fluid is given by
\begin{align*}
\frac{\partial\vec{v}}{\partial t}+\vec{v}\cdot\mathbf{\boldsymbol{\nabla}}\vec{v}+2\vec{\Omega}\times\vec{v}=-\frac{1}{\uprho}\boldsymbol{\nabla} p+\upnu\,\boldsymbol{\nabla}^2\vec{v}+\\\frac{\upnu}{3}\boldsymbol{\nabla}(\vec{\boldsymbol{\nabla}}\cdot\vec{v})+\boldsymbol{\nabla}[\Omega\times(\Omega\times r)]+\vec{g},\tag{1}\label{eq:navierstokes}
\end{align*}
where, $\vec{v}$ is the fluid velocity and $\vec{g}$ is the gravitational acceleration. The Navier-Stokes equation relates the restoring forces on a fluid element (written on the right-hand side: pressure gradients, viscous force, gravitational force, compressible strain, centrifugal force) to the impulsive change in momentum of the fluid element (written on the left-hand side: Coriolis force etc). We have ignored terms from the magnetic field of the neutron star as previously stated, restricting ourselves to a purely hydrodynamic analysis. The gravitational acceleration is taken to be of the following form:
\setcounter{equation}{1}
\begin{equation}
\vec{g}=-\frac{z}{|z|}g\hat{z}.\label{eq:gravity}
\end{equation}
This form for $\vec{g}$ is unphysical since it is generated by a singular and planer mass distribution located at $z=0$. However, such a choice is standard in neutron star literature \citep{epstein,melatos2008}; it is equivalent to a radial field for a sphere which compares well to numerical simulations comprising of more realistic mass distributions \citep{gravity}. This assumption leads to symmetric flow across the mid-plane of the cylinder and we can restrict ourselves to $z\ge 0$. \\

\noindent The second governing equation is the `Continuity equation' i.e. the conservation of mass equation
\begin{equation}
\frac{\partial\uprho}{\partial t}+\boldsymbol{\nabla}\cdot(\uprho\vec{v})=0.\label{eq:continuity}
\end{equation}
Lastly, we write the `energy equation', i.e.  the equation of state in terms of the adiabatic sound speed $v_\mathrm{c}$ (where, the sub-script $\mathsf{S}$ represents derivative taken at constant entropy) of the form
\begin{equation}
\Bigg[\frac{\partial\hspace{-0.01in}p}{\partial\uprho}\Bigg]_\mathsf{S}=v^2_\mathrm{c},\label{eq:equationofstate}
\end{equation}
which, in the adiabatic limit\footnote{The adiabatic limit for an isolated neutron star allows us to drop the  sub-script $\mathsf{S}$ in \eqref{eq:equationofstate}; more discussion in section \ref{eos}.} and expressed in Lagrangian frame, takes the following form:
\begin{equation}
\Bigg[\frac{\partial}{\partial t}+\vec{v}\cdot\boldsymbol{\nabla}\Bigg]\uprho=
\Bigg[\frac{\partial}{\partial t}+\vec{v}\cdot\boldsymbol{\nabla}\Bigg]\frac{p}{v_\mathrm{c}^2}\label{eq:energyequation}
\end{equation}
Note that we do not impose invariance of $v_\mathrm{c}$ in either space or time as previously done in \citep{melatos2008,melatos2010}.\\

\noindent In addition, we scale our variables to dimensionless form by redefining them as, $r\rightarrow \mathrm{L}r,\, z\rightarrow \mathrm{L}z,\, \vec{v}\rightarrow (\delta\Omega)\mathrm{L}\vec{v},\, \uprho\rightarrow \uprho_\mathrm{o}\uprho,\, p\rightarrow \uprho_\mathrm{o}g\mathrm{L}p, \,\boldsymbol{\nabla}\rightarrow {\mathrm{L}}^{-1}\boldsymbol{\nabla},\, t\rightarrow t_et$; where we define $\uprho_\mathrm{o}$ as the equilibrium mass-density at $z=0$, and $t_e$ and the {\fontfamily{ppl}\selectfont \textit{Ekman number}} $\mathsf{E}$ as,
\begin{equation}
t_e=\mathsf{E}^{\frac{1}{2}}\Omega^{-1}, \;\text{and}\;\mathsf{E}=\frac{\upnu}{\mathrm{L}^2\Omega}.\label{eq:scaleing}
\end{equation}
One can then write the governing equations \eqref{eq:navierstokes}, \eqref{eq:continuity} and \eqref{eq:energyequation} in the re-scaled form as
\begin{equation}
\begin{multlined}
\epsilon\mathsf{F}\Bigg[\mathsf{E}^\frac{1}{2}\frac{\partial\vec{v}}{\partial t}+\epsilon\vec{v}\cdot\mathbf{\boldsymbol{\nabla}}\vec{v}+2\hat{e}_z\times\vec{v}\Bigg]=-\frac{1}{\uprho}\boldsymbol{\nabla}\hspace{-0.01in}p-\vec{e}_z+\\\epsilon\mathsf{F}\mathsf{E}\Bigg[\boldsymbol{\nabla}^2\vec{v}+\frac{1}{3}\boldsymbol{\nabla}\{\boldsymbol{\nabla}\cdot\vec{v}\}\Bigg]+\mathsf{F}\boldsymbol{\nabla}\Bigg[\frac{1}{2}r^2\Bigg],\label{eq:snavierstokes}
\end{multlined}
\end{equation}
\begin{equation}
\begin{multlined}
\mathsf{E}^\frac{1}{2}\frac{\partial\uprho}{\partial t}+\epsilon\boldsymbol{\nabla}\cdot(\uprho\vec{v})=0,\label{eq:scontinuity}
\end{multlined}
\end{equation}
\begin{equation}
\begin{multlined}
\mathsf{E}^\frac{1}{2}\frac{\partial[\uprho\eta]}{\partial t}+\epsilon\vec{v}\cdot\boldsymbol{\nabla}[\uprho\eta]=
\mathsf{K}\Bigg[\mathsf{E}^\frac{1}{2}\frac{\partial p}{\partial t}+\epsilon\vec{v}\cdot \boldsymbol{\nabla}p\Bigg],\label{eq:senergy}
\end{multlined}
\end{equation}
where $\eta$, the {\fontfamily{ppl}\selectfont \textit{Froude number}} $\mathsf{F}$, the maximum {\fontfamily{ppl}\selectfont \textit{scaled compressibility}} $\mathsf{K}$, and $\epsilon$ are defined as
\begin{equation}
\begin{multlined}
\eta=\frac{v^2_\mathrm{c}}{c^2},\;\;\;\mathsf{K}=g\frac{\mathrm{L}}{c^2},\;\;\;\mathsf{F}=\Omega^2\frac{\mathrm{L}}{g},\;\;\;\epsilon=\frac{\delta\Omega}{\Omega}.\label{eq:parameters}
\end{multlined}
\end{equation}
\subsection{Orders of magnitude}
\label{section:magnitudes}
Here, we list the physical constants and parameters introduced in the previous section, and their order-of-magnitude values in \textsf{SI} units for the interior of a neutron star from estimates on the expected and observed physical properties.
\begin{equation}
\begin{multlined}
g= O(10^{12}),\;\mathrm{L}=O(10^4),\;\frac{\Omega}{2\uppi}=O(1 \mathrm{Hz}-10^2 \mathrm{Hz}),\;\\\mathsf{E}\in[O(10^{-17}),O(10^{-7})],\;\frac{v_\mathrm{c}}{c}\in [O(10^{-2}),1].\,\label{eq:parametersorders}
\end{multlined}
\end{equation}
The viscosity $\upnu$ of the fluid for a neutron star is currently unknown and widely debated. The associated value of {Ekman number} $\mathsf{E}$ also remains volatile. Yet, there are estimates on the value of $\mathsf{E}$ from results of heavy-ion collision experiments \citep{melatos2008,E1e,E2e,E3e,E4e} and from theoretical calculations of neutron-neutron scattering in the superfluid limit \citep{largeE1,E1t}. The results from such analysis lead to the wide range of possible values for $\mathsf{E}$ quoted above in \eqref{eq:parametersorders}. The parameters listed in \eqref{eq:parameters} then take the following values\footnote{We will use the values quoted in \eqref{eq:parametersorders} and \eqref{eq:sparametersorders} when making order-of-magnitude estimates on the emitted gravitational wave signals.}
\begin{equation}
\begin{multlined}
\mathsf{K}=O(10^{-1}),\;\;\mathsf{F}\in [O(10^{-9}),O(10^{-3})],\\\epsilon\in [O(10^{-11}),O(10^{-4})].\;\label{eq:sparametersorders}
\end{multlined}
\end{equation}  
\section{Solution}
\label{section:solution}
In this section, we solve to the governing equations given in the section \ref{section:equations}.\vspace{-0.15in}
\subsection{Equilibrium solution}
\label{section:equilibrium}
In equilibrium, due to the symmetry of the system across the $z=0$ mid-plane, the flow is steady and axisymmetric, and the density and pressure are functions of $z$ and $r$ only. Since $\epsilon$ and $\mathsf{F}$ are quite small in their absolute magnitude, we can ignore the centrifugal term \citep{epstein} in the re-scaled equations \eqref{eq:snavierstokes}, \eqref{eq:scontinuity} and \eqref{eq:senergy}. With this approximation, \eqref{eq:snavierstokes} reduces to
\begin{equation}
\begin{multlined}
\frac{1}{\uprho}\boldsymbol{\nabla} p+\vec{e}_z=0.
\label{eq:equilibrium}
\end{multlined}
\end{equation}
In order to solve the above equation, we need to make an assumption for either the mass-density or the pressure. We introduce the dimensionless parameter $\mathsf{K_s}$, following \citep{melatos2008}, and assume the following:
\begin{equation}
\begin{multlined}
\uprho^{-1}(\mathrm{d}\uprho/\mathrm{d}z)=-\mathsf{K_s}(z).
\label{eq:sequilibrium}
\end{multlined}
\end{equation}
The stratification length, $z_\mathsf{s}$, is defined in terms of the dimensionless quantity $\mathsf{K_s}$ as $z_\mathsf{s}=\mathrm{L}\,\mathsf{K}^{-1}_\mathsf{s}$. The above expression defines a steady-state density profile of the system. The equilibrium pressure and density profiles are then given by solving \eqref{eq:equilibrium} and \eqref{eq:sequilibrium} respectively, as
\vspace{-0.1in}\begin{equation}
\begin{multlined}
\uprho_e(z)=e^{\displaystyle -\int_0^z\!\mathsf{K_s}(z')\,\mathrm{d}z'},
\label{eq:edensity}
\end{multlined}
\end{equation}
\begin{equation}
\begin{multlined}
p_e(z)=\mathsf{K}^{-1}_\mathsf{s}(z)\,e^{\displaystyle -\int_0^z\!\mathsf{K_s}(z')\,\mathrm{d}z'}.
\label{eq:epressure}
\end{multlined}
\end{equation} 
Here, we have not assumed $\mathsf{K_s}$ to be a constant, as previously done in \citep{melatos2008,melatos2010}. The introduction of the form $\mathsf{K_s}$ is not trivial. In fact, the assumption of a certain form of $\mathsf{K_s}$ incorporates the nature of entropic or compositional gradients, which in turn incorporate the deviation of equilibrium state from an adiabatic state. We introduce {\fontfamily{ppl}\selectfont \textit{equilibrium sound speed}} accordingly, given by $v^2_\mathrm{eq}=\displaystyle\frac{g\mathrm{L}}{\mathsf{K_s}(z)}$. We note that in the equilibrium state, gravity acts to vary density and pressure along the axis of the cylinder.
\subsection{Induced Perturbations}
\label{section:spinup}
Let us assume that a glitch induces perturbations in pressure, density and velocity fields of the internal bulk fluid of the neutron star and that the resultant bulk fluid flow may be non-axisymmetric. When such non-axisymmetric perturbations are induced, the density $\uprho(r,\phi,z,t)$ and pressure $p(r,\phi,z,t)$ are functions of all spatial coordinates and time, as opposed to the case of stable equilibrium. In order to solve for the perturbed fluid motion, we treat the system `perturbatively', given the small magnitude of $\epsilon$. In the perturbative treatment, the density, pressure and the velocity field can be expanded as $\uprho\rightarrow\uprho+\epsilon\delta\uprho$, and $p\rightarrow p+\epsilon\delta\hspace{-0.01in}p$, where we have let the magnitude of $\delta\hspace{-0.01in}p$ and $\delta\uprho$ run free and normalized it by $\epsilon$. The velocity field, however, is written simply as $\vec{v}\rightarrow\delta\vec{v}$. We do not perturb $\eta$ explicitly; the variation in $\eta$ occurs naturally from variation in $v_\mathrm{c}$.  Now, ignoring all terms larger then $O(1)$ in $\epsilon$, the set of three re-scaled governing equations \eqref{eq:snavierstokes}, \eqref{eq:scontinuity} and \eqref{eq:senergy} reduce to
\begin{equation}
\begin{multlined}
\mathsf{F}\Bigg[\mathsf{E}^\frac{1}{2}\frac{\partial[\delta\vec{v}]}{\partial t}+2\hat{e}_z\times[\delta\vec{v}]\Bigg]=-\frac{1}{\uprho}\boldsymbol{\nabla} [\delta\hspace{-0.01in}p]-\frac{\delta\uprho}{\uprho}\vec{e}_z+\\\mathsf{F}\,\mathsf{E}\Big[\boldsymbol{\nabla}^2[\delta\vec{v}]+\frac{1}{3}\boldsymbol{\nabla}\{\boldsymbol{\nabla}\cdot[\delta\vec{v}]\}\Bigg]+\mathsf{F}\frac{\delta\uprho}{\uprho}\boldsymbol{\nabla}\Bigg[\frac{1}{2}r^2\Bigg],\;\;\;\;\label{eq:pnavierstokes}
\end{multlined}
\end{equation}
\begin{equation}
\begin{multlined}
\mathsf{E}^\frac{1}{2}\frac{\partial[\delta\uprho]}{\partial t}+\boldsymbol{\nabla}\cdot(\uprho\hspace{0.01in}\delta\vec{v})=0,\label{eq:pcontinuity}
\end{multlined}
\end{equation}
\begin{equation}
\begin{multlined}
\mathsf{E}^\frac{1}{2}\frac{\partial[\eta\delta\uprho]}{\partial t}+\delta\vec{v}\cdot\boldsymbol{\nabla}[\uprho\eta]=\mathsf{K}\Bigg[\mathsf{E}^\frac{1}{2}\frac{\partial[\delta\hspace{-0.01in}p]}{\partial t}+\delta\vec{v}\cdot\boldsymbol{\nabla} \hspace{-0.01in}p\Bigg].\label{eq:penergy}
\end{multlined}
\end{equation}
\subsection{Method of multiple scales}
\label{section:scales}
In the perturbative treatment, we employ the {\fontfamily{ppl}\selectfont \textit{method of multiple scales}} \citep{melatos2008,melatos2010,walin,epstein}. The perturbations in the density, pressure and velocity field are expanded into scales of order $\mathsf{E}^0$, $\mathsf{E}^\frac{1}{2}$ and $\mathsf{E}^1$, such that for density perturbations,
\begin{equation}
\begin{multlined}
\delta\uprho=\delta\uprho^{(0)}+\mathsf{E}^\frac{1}{2}\delta\uprho^{(1)}+\mathsf{E}^{1}\delta\uprho^{(2)},
\label{eq:pdensity}
\end{multlined}
\end{equation}
and for pressure perturbations,
\begin{equation}
\begin{multlined}
\delta\hspace{-0.01in}p=\delta\hspace{-0.01in}p^{(0)}+\mathsf{E}^\frac{1}{2}\delta\hspace{-0.01in}p^{(1)}+\mathsf{E}^{1}\delta\hspace{-0.01in}p^{(2)},
\label{eq:ppressure}
\end{multlined}
\end{equation}
and for velocity perturbations,
\begin{equation}
\begin{multlined}
\delta\hspace{-0.0in}\vec{v}=\delta\hspace{-0.0in}\vec{v}^{(0)}+\mathsf{E}^\frac{1}{2}\delta\hspace{-0.0in}\vec{v}^{(1)}+\mathsf{E}^{1}\delta\hspace{-0.0in}\vec{v}^{(2)}.
\label{eq:pvelocity}
\end{multlined}
\end{equation}
The idea behind the {method of multiple scales} is to separate sub-process that occur at time-scales in increments of $\mathsf{E}^\frac{1}{2}$, and solve them individually. 

One can now solve \eqref{eq:pnavierstokes} for the the velocity field, its radial, azimuthal and vertical components of $v^{(i)}_r$, $v^{(i)}_\phi$ and $v^{(i)}_z$, up to $i$th order in $\mathsf{E}^\frac{i}{2}$,
\begin{equation}
\begin{multlined}
\mathsf{F}\Bigg[\mathsf{E}^\frac{1}{2}\frac{\partial\delta{v}_r}{\partial t}-2\delta{v}_\phi\Bigg]=-\frac{\partial}{\partial r}\Bigg[\frac{\delta\hspace{-0.01in}p}{\uprho}\Bigg]+\mathsf{F}\hspace{0.01in}\mathsf{E}\;\times \\ \Bigg[\Bigg\{\boldsymbol{\nabla}^2-\frac{1}{r^2}\Bigg\}\delta{v}_r-\frac{2}{r^2}\frac{\partial[\delta{v}_\phi]}{\partial\phi}+\frac{1}{3}\frac{\partial}{\partial r}[\boldsymbol{\nabla}\cdot\delta\vec{v}]\Bigg],
\label{eq:vr}
\end{multlined}
\end{equation}\vspace{-0.075in}	
\begin{equation}
\begin{multlined}
\mathsf{F}\Bigg[\mathsf{E}^\frac{1}{2}\frac{\partial\delta{v}_\phi}{\partial t}-2\delta{v}_r\Bigg]=-\frac{1}{r}\frac{\partial}{\partial \phi}\Bigg[\frac{\delta\hspace{-0.01in}p}{\uprho}\Bigg]+\mathsf{F}\hspace{0.01in}\mathsf{E}\;\times \\ \Bigg[\Bigg\{\boldsymbol{\nabla}^2-\frac{1}{r^2}\Bigg\}\delta{v}_\phi-\frac{2}{r^2}\frac{\partial[\delta{v}_r]}{\partial\phi}+\frac{1}{3r}\frac{\partial}{\partial \phi}[\boldsymbol{\nabla}\cdot\delta\vec{v}]\Bigg],
\label{eq:vphi}
\end{multlined}
\end{equation}\vspace{-0.075in}	
\begin{equation}
\begin{multlined}
\mathsf{F}\hspace{0.01in}\mathsf{E}^\frac{1}{2}\frac{\partial\delta{v}_z}{\partial t}=-\frac{1}{\uprho}\frac{\partial}{\partial z}\Bigg[\frac{\delta\hspace{-0.01in}p}{\uprho}\Bigg]-\frac{\delta\uprho}{\uprho}+\mathsf{F}\hspace{0.01in}\mathsf{E}\;\times\\\Bigg[\boldsymbol{\nabla}^2[\delta{v}_z]-\frac{1}{3}\frac{\partial}{\partial\phi}[\boldsymbol{\nabla}\cdot\delta\vec{v}]\Bigg].\!\!\!\!\!\!\!\!\!\!
\label{eq:vz}
\end{multlined}
\end{equation}
Two additional relations are derived from the energy equation \eqref{eq:penergy} and the continuity equation \eqref{eq:pcontinuity}, and they are given by
\begin{equation}
\begin{multlined}
\mathsf{E}^\frac{1}{2}\frac{\partial}{\partial t}\Bigg[\frac{\delta\uprho}{\uprho}\Bigg]+\boldsymbol{\nabla}\cdot\delta\vec{v}=\mathsf{K_s}(z)\delta v_z,
\label{eq:pe}
\end{multlined}
\end{equation}
and, \vspace{-0.1in}	
\begin{equation}
\begin{multlined}
\eta(z)\mathsf{E}^\frac{1}{2}\frac{\partial}{\partial t}\Bigg[\frac{\delta\uprho}{\uprho}\Bigg]=\mathsf{K}\mathsf{E}^\frac{1}{2}\frac{\partial}{\partial t}\Bigg[\frac{\delta\hspace{-0.01in}p}{\uprho}\Bigg]+\mathsf{F}\mathsf{N}^2(z)\,\delta v_z,
\label{eq:pc}
\end{multlined}
\end{equation}
where, we have introduced $\mathsf{N}(z)$ - the redefined {\fontfamily{ppl}\selectfont \textit{Brunt-V{\"a}is{\"a}l{\"a} frequency}}\footnote{Note that the {{Brunt-V{\"a}is{\"a}l{\"a} frequency}} $\mathsf{N}$ is a well-known quantity in fluid mechanics and atmospheric sciences. It is a measure of the buoyant force experienced by a fluid element when displaced from equilibrium.},
\begin{equation}
\begin{multlined}
\mathsf{N}^2(z)=\frac{[\eta\mathsf{K_s}-\partial_z\eta]-\mathsf{K}}{\mathsf{F}}=\frac{\mathsf{K}}{\mathsf{F}}\Bigg[\frac{v^2_\mathrm{c}}{v^2_{\mathrm{eq}}}-1\Bigg]-\frac{\partial_z\eta}{\mathsf{F}}.\label{eq:bvf}
\end{multlined}
\end{equation}
We can reformulate \eqref{eq:bvf} by introducing $\mathsf{K}'_\mathsf{s}(z)$ as
\begin{equation}
\begin{multlined}
\mathsf{K}'_\mathsf{s}(z)=\eta\mathsf{K_s}-\partial_z\eta=\mathsf{K}\frac{v^2_\mathrm{c}}{v^2_{\mathrm{eq}}}-\partial_z\eta,\label{eq:Kp}
\end{multlined}
\end{equation}
such that \eqref{eq:bvf} takes the form of
\begin{equation}
\begin{multlined}
\mathsf{N}^2(z)=\frac{\mathsf{K}'_\mathsf{s}(z)-\mathsf{K}}{\mathsf{F}}.\label{eq:rbvf}
\end{multlined}
\end{equation}
Here, $\eta$ and $\mathsf{K_s}$ are allowed to vary with $z$ only. In the set of equations \eqref{eq:vr}-\eqref{eq:pc}, the terms on different $O(\mathsf{E}^0)$, $O(\mathsf{E}^\frac{1}{2})$ and $O(\mathsf{E}^1)$ scales are reducible at each order. Moreover, we can distinguish and deduce the time-scales at which several processes contribute to the overall perturbed flow of the bulk matter, such as the formation of viscous {\fontfamily{ppl}\selectfont \textit{Rayleigh shear layer}} layer, followed by partial spin-up of the interior fluid via {Ekman pumping}, followed by complete spin-up of the interior on longer time-scales. These processes have been discussed briefly by \citet{melatos2008} and in much greater detail by \citet{epstein}. We will also discuss them in the later section(s). These times scales are $\mathsf{E}^0\Omega^{-1}$, $\mathsf{E}^{-\frac{1}{2}}\Omega^{-1}$ and $\mathsf{E}^{-1}\Omega^{-1}$. One can now isolate solutions at these different scales since they are effectively independent due to very small magnitude of the {Ekman number} $\mathsf{E}$.
\subsection{$\mathbf{O}({\mathbf{E^\textbf{0}}})$ solutions}
\label{section:sol0}
On the order of $\mathsf{E}^0$, the expressions \eqref{eq:vr}-\eqref{eq:pc} yield
\begin{equation}
\begin{multlined}
\delta\hspace{-0.0in}{v}_r^{(0)}=-\frac{1}{2\mathsf{F}r}\frac{\partial}{\partial\phi}\Bigg[\frac{\delta\hspace{-0.01in}p^{(0)}}{\uprho}\Bigg],\label{eq:0vr}
\end{multlined}
\end{equation}
\begin{equation}
\begin{multlined}
\delta\hspace{-0.0in}{v}_\phi^{(0)}=\frac{1}{2\mathsf{F}}\frac{\partial}{\partial r}\Bigg[\frac{\delta\hspace{-0.01in}p^{(0)}}{\uprho}\Bigg],\label{eq:0vphi}
\end{multlined}
\end{equation}
\begin{equation}
\begin{multlined}
\delta\hspace{-0.0in}{v}_z^{(0)}=0,\label{eq:0vz}
\end{multlined}
\end{equation}
\begin{equation}
\begin{multlined}
\delta\uprho^{(0)}=-\frac{\partial [\delta\hspace{-0.01in}p^{(0)}]}{\partial z},\text{ and}\label{eq:0pe}
\end{multlined}
\end{equation}
\begin{equation}
\begin{multlined}
\boldsymbol{\nabla}\cdot\delta\hspace{-0.0in}\vec{v}^{(0)}=0.\label{eq:0pc}
\end{multlined}
\end{equation}
Note that the solutions on the order $\mathsf{E}^0$ are exactly the ones previously achieved in \citep{melatos2008,melatos2010,walin,epstein}. These solutions, given by \eqref{eq:0vr}-\eqref{eq:0pc}, correspond to the formation of a viscous boundary layer (also referred to as the {Rayleigh shear layer}) on the top and bottom faces of the cylinder on a time-scale $O(\mathsf{E}^{0}\Omega^{-1})$. Within this viscous boundary layer, the flow moves radially outward due to the gradient in the azimuthal velocity and the resulting imbalance between centrifugal and pressure gradient forces \citep{walin,epstein}.
\subsection{$\mathbf{O}({\mathbf{E^\frac{\textbf{1}}{\textbf{2}}}})$ solutions}
\label{section:sol1}
In solving for the $O(\mathsf{E}^\frac{1}{2})$ solutions, we assume that
\begin{equation}
\begin{multlined}
\frac{\delta\hspace{-0.01in}p^{(0)}}{p}\gg\frac{\delta\hspace{-0.01in}p^{(1)}}{p}\sim 0,\;\frac{\delta\uprho^{(0)}}{\uprho}\gg\frac{\delta\uprho^{(1)}}{\uprho}\sim 0\label{eq:1pr}
\end{multlined}
\end{equation}
The $O(\mathsf{E}^\frac{1}{2})$ terms in \eqref{eq:vr}-\eqref{eq:pc} yield
\begin{equation}
\begin{multlined}
\delta\hspace{-0.0in}{v}_r^{(1)}=\frac{1}{4\mathsf{F}}\frac{\partial\upchi}{\partial r},\label{eq:1vr}
\end{multlined}
\end{equation}
\begin{equation}
\begin{multlined}
\delta\hspace{-0.0in}{v}_\phi^{(1)}=\frac{1}{4\mathsf{F}r}\frac{\partial\upchi}{\partial \phi},\label{eq:1vphi}
\end{multlined}
\end{equation}
\begin{equation}
\begin{multlined}
\delta\hspace{-0.0in}{v}_z^{(1)}=\frac{\eta(z)}{\mathsf{F}\mathsf{N}^2(z)}\frac{\partial\upchi}{\partial z}+\Bigg[\frac{-\partial_z\eta}{\mathsf{F}\mathsf{N}^2(z)}-1\Bigg]\upchi,\text{ and}\label{eq:1vz}
\end{multlined}
\end{equation}
\begin{equation}
\begin{multlined}
\frac{\partial}{\partial t}\Bigg[\frac{\delta\uprho^{(0)}}{\uprho}\Bigg]+\boldsymbol{\nabla}\cdot\delta\hspace{-0.0in}\vec{v}^{(1)}=\mathsf{K_s}(z)\delta\hspace{-0.0in}{v}_z^{(1)},\label{eq:1pc}
\end{multlined}
\end{equation}
where, we have defined $\upchi$ as
\begin{equation}
\begin{multlined}
\upchi=-\frac{\partial}{\partial t}\Bigg[\frac{\delta\hspace{-0.01in}p^{(0)}}{\uprho}\Bigg]\sim O(\mathsf{E}^0).\label{eq:chi}
\end{multlined}
\end{equation}
These set of solutions represent the process of {Ekman pumping} -- the flow in the viscous boundary layer, given by \eqref{eq:0vr}-\eqref{eq:0pc}, sets a secondary motion in the interior\footnote{The important development is the excitation of flow in \textit{z}-direction in the boundary layer, given by \eqref{eq:1vz}.}, by which the fluid is pulled into the viscous boundary layer from the interior to replace the radial outward flow in it \citep{walin,epstein}. Note that the results on the order $O(\mathsf{E}^\frac{1}{2})$, given by \eqref{eq:1vr}-\eqref{eq:1pc}, are different from those in previous works with respect to the expression for $\delta\hspace{-0.0in}{v}_z^{(1)}$ \eqref{eq:1vz} and the continuity equation \eqref{eq:1pc}. This affects all future calculations and results.
\subsection{More on the scale-based solutions}
\label{section:understand}
We will skip the discussion of the $O(\mathsf{E}^1)$ solutions since they occur on much larger time-scales of $O(\mathsf{E}^{-1}\Omega^{-1})$. These $O(\mathsf{E}^{-1}\Omega^{-1})$ solutions correspond to the eventual 'spin-up' of the entire interior bulk matter when the interior bulk sets in complete co-rotation or steady differential co-rotation with the crust, as previously mentioned \citep{walin,epstein}. This sub-process on much larger time-scales is irrelevant to our discussion since it does not contribute to the gravitational wave emission. To recap the scale-based solutions, the sudden spin-up of the rotating cylinder leads to the formation of a viscous boundary layer at the top and bottom faces of the rotating cylinder. This viscous layer forms on a time-scale of $O(\mathsf{E}^{0}\Omega^{-1})$. The velocity field within this layer pushes the fluid radially outward across the layer, given by \eqref{eq:0vr}-\eqref{eq:0vz}. This $O(\mathsf{E}^0)$ flow excites {Ekman pumping} in the interior on a time-scale of $O(\mathsf{E}^{-\frac{1}{2}}\Omega^{-1})$, pushing the fluid radially inward and vertically into the boundary layer, given by \eqref{eq:1vr}-\eqref{eq:1vz}. Note that the vertical velocity of the $O(\mathsf{E}^{-\frac{1}{2}})$ flow, given by \eqref{eq:1vz}, is non-zero. This vertical velocity is constrained by the continuity law applied to the viscous layer \citep{epstein,melatos2008,walin,melatos2010}, such that
\begin{equation}
\begin{multlined}
\delta v_z|_{z=\pm 1}=\pm\frac{1}{2}\mathsf{E}^\frac{1}{2}[\boldsymbol{\nabla}\times(\delta\vec{v} - \vec{v}_ \mathsf{B})]_z|_{z=\pm 1}.\label{eq:boundary}
\end{multlined}
\end{equation}
where, $\vec{v}_ \mathsf{B}$ is the dimensionless velocity of the boundary layer in the frame rotating at $\Omega$. In this rotating frame, $\vec{v}_ \mathsf{B}=r\vec{e}_\phi$ \citep{melatos2010}. Note that we have assumed that the boundary layer is rigidly co-rotating with the cylinder with angular frequency $\Omega+\delta\Omega$ without any slippage. The above expression \eqref{eq:boundary} describes the continuity of the vertical flow across and inside the viscous boundary layer as a function of flow just outside the layer. This process occurs on a time-scale of $O( \mathsf{E}^\frac{1}{2})$, which is reflected in the magnitude term $ \mathsf{E}^\frac{1}{2}$ in \eqref{eq:boundary}. We also find that the process of {Ekman pumping} continues until the local velocity field $\delta\vec{v}$ becomes equal to the boundary velocity $\vec{v}_ \mathsf{B}$. This is followed by spin-up of the entire interior on much larger time-scales of $O( \mathsf{E}^{-1}\Omega^{-1})$. Furthermore, the magnitude term of $ \mathsf{E}^\frac{1}{2}$ can be understood in terms of scaling arguments. The viscous term in the dimensionless Navier-Stokes equation \eqref{eq:snavierstokes} is given by 
\begin{equation}
\begin{multlined}
\mathsf{F}\hspace{0.01in}\mathsf{E}\boldsymbol{\nabla}^2\sim\mathsf{F}\hspace{0.01in}\mathsf{E}\Bigg[\frac{1}{\delta\mathrm{L}}\Bigg]^2\sim O(\mathsf{F}),\label{eq:vb}
\end{multlined}
\end{equation}
where, $\delta\mathrm{L}$ is the scale of the thickness of the viscous boundary layer. Clearly, from the relation given above, $\delta\mathrm{L}=O(\mathsf{E}^\frac{1}{2})$; also see the detailed discussion by \citet{epstein} on this subject. The characteristic thickness of the boundary layer and the time-scale of {Ekman pumping} are both attributable to the magnitude term $\mathsf{E}^\frac{1}{2}$ in \eqref{eq:boundary}.
\subsection{The characteristic equation}
\label{section:equation}
Considering the $O(\mathsf{E}^0)$ and $O(\mathsf{E}^\frac{1}{2})$ solutions obtained in the previous section(s), we combine them to write the following differential equation with terms up to order $O(\mathsf{E}^\frac{1}{2})$ and $O(\mathsf{F}^{\,0})$:
\begin{equation}
\begin{multlined}
\frac{1}{r}\frac{\partial}{\partial r}\Bigg[r\frac{\partial\upchi}{\partial r}\Bigg]+ \frac{1}{r^2}\frac{\partial^2\upchi}{\partial\phi^2}-\Bigg[\frac{4\eta(z)\mathsf{K_s}(z)}{\mathsf{N}^2(z)}\Bigg]\frac{\partial\upchi}{\partial z}+\\\frac{4\eta(z)}{\mathsf{N}^2(z)}
\frac{\partial^2\upchi}{\partial z^2}=\Bigg[\frac{\partial_z\eta-\partial^2_z\eta}{\mathsf{N}^2(z)}\Bigg]\upchi.\!\!\!\!\!\!\!\label{eq:main}
\end{multlined}
\end{equation}
The above characteristic equation can be solved via the standard {\fontfamily{ppl}\selectfont \textit{method of separation of variables}} to yield
\begin{equation}
\begin{multlined}
\upchi(r,\phi,z,t)=\sum_{\upalpha=0}^{\infinity}\sum_{\gamma=1}^{\infinity}
\mathsf{J}_\upalpha(\uplambda_{\upalpha\gamma}r)\Bigg[\frac{\mathrm{A}_{\upalpha\gamma}(t)-i\mathrm{B}_{\upalpha\gamma}(t)}{2}\\ \times e^{i\upalpha\phi}+\frac{\mathrm{A}_{\upalpha\gamma}(t)+i\mathrm{B}_{\upalpha\gamma}(t)}{2}e^{-i\upalpha\phi}\Bigg]\hspace{0.01in}\mathrm{Z}_{\upalpha\gamma}(z),\;\;\;\;\;\label{eq:solution}
\end{multlined}
\end{equation}
where, $\uplambda_{\upalpha\gamma}$ is the $\gamma$th zero of the $\upalpha$th Bessel mode ($\mathsf{J}_\upalpha$), and $\mathrm{A}_{\upalpha\gamma}(t)$, $\mathrm{B}_{\upalpha\gamma}(t)$ are the associated {\fontfamily{ppl}\selectfont \textit{Bessel-Fourier coefficients}} which depend upon the assumed steady-state solution, which we will see shortly. The flow is constrained by a trivial boundary condition which requires no penetration through the side walls, i.e. $\delta v^{(0)}_r|_{r=1}=0$. This simply translates to $\partial_\phi\upchi|_{r=1}=0$ for $\forall$ $\phi$,  via \eqref{eq:0vr}. Moreover, $\mathrm{Z}_{\upalpha\gamma}(z)$ is the solution to the following differential equation: 
\begin{equation}
\begin{multlined}
\frac{4\eta(z)}{\mathsf{N}^2(z)}\frac{\partial^2\mathrm{Z}_{\upalpha\gamma}(z)}{\partial z^2}-\frac{4\eta(z)\mathsf{K_s}(z)}{\mathsf{N}^2(z)}\frac{\partial\mathrm{Z}_{\upalpha\gamma}(z)}{\partial z}-\\ \Bigg[\frac{\partial_z\eta-\partial^2_z\eta}{\mathsf{N}^2(z)}+\uplambda^2_{\upalpha\gamma}\Bigg]\mathrm{Z}_{\upalpha\gamma}(z)=0,\!\!\!\!\!\!\!\!\!\label{eq:Z}
\end{multlined}
\end{equation}
which depends on $\mathsf{N}^2$, which is turn depends exclusively on $\mathsf{K_s}$ and $\eta$. When $\mathsf{K_s}$ (or, $v_\mathrm{eq}$) and $\eta$ (or, $v_\mathrm{c}$) are constants, $\mathrm{Z}_{\upalpha\gamma}(z)$ takes the simple form given below, 
\begin{equation}
\begin{multlined}
\mathrm{Z}_{\upalpha\gamma}(z)=\frac{(\mathsf{FN}^2-\mathcal{B}_{-})e^{\mathcal{B}_{+}z}-(\mathsf{FN}^2-\mathcal{B}_{+})
e^{\mathcal{B}_{-}z}}{(\mathsf{FN}^2-\mathcal{B}_{-})e^{\mathcal{B}_{+}}-(\mathsf{FN}^2-\mathcal{B}_{+})
e^{\mathcal{B}_{-}}},\label{eq:Zc}
\end{multlined}
\end{equation}
where,
\begin{equation}
\begin{multlined}
\mathcal{B}_{\pm}=\frac{1}{2}\big[\mathsf{K_s}\pm\big(\mathsf{K}^2_{\mathsf{s}}+\eta^{-1}\mathsf{N}^2\uplambda^2_{\upalpha
\gamma}\big)^\frac{1}{2}\big].\label{eq:B}
\end{multlined}
\end{equation}
It must be noted that, following \citep{melatos2008}, we have temporarily and seemingly arbitrarily\footnote{The function $\mathrm{Z}_{\upalpha\gamma}(z)$ must be explicitly re-normalised to lie in the range $[0,1]$, since \eqref{eq:chi} dictates that $\upchi$ -- as a dimensionless variable -- must be at most of the order $O(\mathsf{E}^0)\sim 1$. This requires $\mathrm{Z}_{\upalpha\gamma}(z)$ to be of the same order in magnitude.} assumed $\mathrm{Z}_{\upalpha\gamma}(1)=1$. Moreover, we also assume $v_z|_{z=0}\sim v^{(1)}_z|_{z=0}=0$ to ensure symmetric flow across the $z=0$ plane\footnote{The boundary condition on axial flow, i.e. setting $v_z|_{z=0}\sim v^{(1)}_z|_{z=0}=0$ in \eqref{eq:1vz}, is equivalent to specifying $\mathrm{Z}_{\upalpha\gamma}(z)$ at $z=0$.}, given the relation prescribed in \eqref{eq:boundary}. This is precisely the result obtained by \citet{melatos2008} and \citet{melatos2010}.
\subsection{Temporal evolution}
\label{section:time}
The temporal evolution of {Ekman pumping} is governed by the boundary condition given in \eqref{eq:boundary} \citep{melatos2008,epstein}. Taking the first-order derivative of \eqref{eq:boundary} and using the results from the $O(\mathsf{E}^0)$ and $O(\mathsf{E}^\frac{1}{2})$ solutions, we find the exponentially decaying time-dependence\footnote{Refer to section {\ref{A.1}} in {{Appendix}} for details} of $\upchi$ as,
\begin{equation}
\begin{multlined}
\upchi(r,\phi,z,t)=\sum_{\upalpha=0}^{\infinity}\sum_{\gamma=1}^{\infinity}
\mathsf{J}_\upalpha(\uplambda_{\upalpha\gamma}r)\Bigg[\frac{\mathrm{A}_{\upalpha\gamma}-i\mathrm{B}_{\upalpha\gamma}}{2}\times\\ e^{i\upalpha\phi}+\frac{\mathrm{A}_{\upalpha\gamma}+i\mathrm{B}_{\upalpha\gamma}}{2}e^{-i\upalpha\phi}\Bigg]\hspace{0.01in}\mathrm{Z}_{\upalpha\gamma}(z)e^{-\omega_{\upalpha\gamma} t},\!\!\!\!\!\!\label{eq:full}
\end{multlined}
\end{equation}
where, momentarily assuming $\mathrm{Z}_{\upalpha\gamma}(1)$ to be an arbitrary value that we will define shortly, we get
\begin{equation}
\begin{multlined}
\omega_{\upalpha\gamma}=\frac{1}{4\mathsf{F}}\uplambda_{\upalpha\gamma}^2\mathrm{Z}_{\upalpha\gamma}(1)\Bigg[\frac{\eta(1)}{\mathsf{F}\mathsf{N}^2(1)}\frac{\partial\mathrm{Z}_{\upalpha\gamma}}{\partial z}\Bigg|_{z=1}+\\\Bigg\{\frac{-\partial_z\eta|_{z=1}}{\mathsf{F}\mathsf{N}^2(1)}-1\Bigg\}\mathrm{Z}_{\upalpha\gamma}(1)\Bigg]^{-1}.\!\!\!\!\!\!\label{eq:omega}
\end{multlined}
\end{equation}
Note that for the simple case of $\mathsf{K_s}(z),\eta\sim$ constant and $\mathrm{Z}_{\upalpha\gamma}(1)=1$, \eqref{eq:omega} reduces to
\begin{equation}
\begin{multlined}
\omega_{\upalpha\gamma}=\frac{\uplambda^2_{\upalpha\gamma}\Big[(\mathsf{FN}^2-\mathcal{B}_{-})e^{\mathcal{B}_{+}}-(\mathsf{FN}^2-\mathcal{B}_{+})
e^{\mathcal{B}_{-}}\Big]}{(4\mathsf{F}\hspace{0.01in}\mathsf{K}+\uplambda^2_{\upalpha\gamma})(e^{\mathcal{B}_{+}}-e^{\mathcal{B}_{-}})}.\label{eq:omegac}
\end{multlined}
\end{equation}
Further, given the explicit dependence of $\upchi$ on time, we integrate \eqref{eq:full} over $t\in[t,{\infinity})$ and get
\begin{equation}
\begin{multlined}
\frac{\delta\hspace{-0.01in}p^{(0)}(r,\phi,z,t)}{\uprho(z)}=\frac{\delta\hspace{-0.01in}p^{(0)}_{t\rightarrow{\hspace{0.01in}\infinity}}(r,\phi,z)}{\uprho(z)}+\sum_{\upalpha=0}^{\infinity}\sum_{\gamma=1}^{\infinity}
\omega_{\upalpha\gamma}^{-1}\times\\\mathsf{J}_\upalpha(\uplambda_{\upalpha\gamma}r)[\mathrm{A}_{\upalpha\gamma}\ccos(\upalpha\phi)+\mathrm{B}_{\upalpha\gamma}\csin(\upalpha\phi)]\hspace{0.01in}\times\\\mathrm{Z}_{\upalpha\gamma}(z)e^{-\omega_{\upalpha\gamma} t},\label{eq:final}
\end{multlined}
\end{equation}
where, the first term on the right-hand side is the constant of integration evaluated at $t\rightarrow\infinity$, i.e. the steady-state pressure profile of the spun-up cylinder. The relation given in \eqref{eq:final} encodes the variation of pressure perturbations up to the leading order in magnitude as a function of time.
\vspace{-0.1in}
\subsection{Initial and final conditions}
\label{section:initialandfinalconditions}
We are left with one intrinsic degree of freedom in our model in form of initial and final conditions in time, i.e. state of perturbations immediately following the glitch at $t=0$ and when {Ekman pumping} stops as $t\rightarrow\infinity$, respectively. In principle, we only require one boundary condition in time -- once $\mathrm{A}_{\upalpha\gamma}$, $\mathrm{B}_{\upalpha\gamma}$ are known -- since the state of modes at $t\rightarrow\infinity$ is coupled to their state at $t=0$ by the relation \eqref{eq:final}, and vice-versa. In this case, however, we require both the initial and final conditions in time to calculate $\mathrm{A}_{\upalpha\gamma}$, $\mathrm{B}_{\upalpha\gamma}$ since they are unknown. For example, two of the most general choices are: \textbf{{\fontfamily{ppl}\selectfont \textit{a}})} one can assume a scenario where the perturbations modes continuously grow from an axisymmetric state in the post-glitch phase at $t=0$ and reach a steady non-axisymmetric state as $t\rightarrow\infinity$, and remain in that state. This leads to emission of gravitational waves even in the steady state at $t\rightarrow\infinity$, and is somewhat unphysical. In fact, this equates to the scenario of 'semi-rigidity', where the top and bottom faces of the cylinder rotate differentially at $t\rightarrow\infinity$, potentially causing the crust to crack \citep{melatos2008}. On the contrary, \textbf{{\fontfamily{ppl}\selectfont \textit{b}})} an alternative scenario is when the perturbation modes are instantaneously excited at $t=0$ and eventually decay as $t\rightarrow\infinity$, which is more physical than the former choice. This choice disallows for any residual non-axisymmetry in the bulk, ensures zero residual steady-state emission, and also incorporates the feature of rigidity between the two faces of the cylinder \citep{melatos2010}. Both these possibilities are encoded our choice of assumed boundary conditions at $t=0$ and $t\rightarrow\infinity$. Hence, we assume the more physical set of initial and final conditions where the modes originate arbitrarily and instantaneously at $t=0$, and decay from some unknown initial value $\delta\mathsf{P}_0$ to a symmetric steady-state $\delta\mathsf{P}_{\infinity}$ as $t\rightarrow\infinity$ according to \eqref{eq:final}. Note that the steady state solution at  $t\rightarrow\infinity$ is an axisymmetric state with no angular or $z$-dependence but only radial dependence, given by $\delta\mathsf{P}_{\infinity}=r^2$ in dimensionless form \citep{melatos2010}. This axisymmetric state doesn't lead to any gravitational wave emission, as previously stated. 

Finally, in order to calculate $\mathrm{A}_{\upalpha\gamma}$, $\mathrm{B}_{\upalpha\gamma}$, we write
\begin{equation}
\begin{multlined}
{\!}\delta\mathsf{P}_0=\delta\mathsf{P}_{\infinity} + \sum_{\upalpha=0}^{\infinity}\sum_{\gamma=1}^{\infinity}
\omega_{\upalpha\gamma}^{-1}\mathsf{J}_\upalpha(\uplambda_{\upalpha\gamma}r)[\mathrm{A}_{\upalpha\gamma}\ccos(\upalpha\phi) +\mathrm{B}_{\upalpha\gamma}\\ \!\!\!\!\!\!\!\!\times\csin(\upalpha\phi)]\mathrm{Z}_{\upalpha\gamma}(z)=\sum_{\upalpha=0}^{\infinity}\mathrm{C}_\upalpha r^\upalpha(r^2-1)\ccos(\upalpha\phi)\,\mathrm{Z}_{\upalpha\gamma}(z),\;\;\;\;\;\;\;\;\;\;\label{eq:P0-P8}
\end{multlined}
\end{equation}
where, wherever suitable from this point onward, we will abbreviate for simplicity,
$$\delta\mathsf{P}_{t'}\equiv\displaystyle{\frac{\delta\hspace{-0.01in}p^{(0)}_{t\rightarrow{\hspace{0.01in}t'}}(r,\phi,z)}{\uprho(z)}}.$$
The assumed form of the initial arbitrary perturbations $\delta\mathsf{P}_0$ in \eqref{eq:P0-P8} is a sum of non-axisymmetric modes satisfying the boundary conditions \citep{melatos2010}. $\mathrm{C}_\upalpha$ are the relative weights of modes with respect to the loudest mode, excited at $t=0$, and they will be set equal to 1 in the calculations in section \ref{section:emission}. Note that any assumed form of $\delta\mathsf{P}_0$ must be constrained by the boundary conditions in space, i.e. no penetration allowed across the side walls, and be a solution to the Navier-Stokes equation by satisfying the relations in \eqref{eq:pnavierstokes}-\eqref{eq:penergy}. Our assumption of $\delta\mathsf{P}_0$ guarantees the decay of all modes at $t\rightarrow\infinity$, while it also ensures that the flow vanishes at the lateral surface at $r=\mathrm{L}$. We have assumed trivial $z$-dependence and $\phi$-dependence in \eqref{eq:P0-P8} for simplicity \citep{melatos2008,melatos2010}, without potentially corrupting the generality of the solutions. The associated {Bessel-Fourier coefficients} $\mathrm{A}_{\upalpha\gamma}$ and $\mathrm{B}_{\upalpha\gamma}$ can now be calculated\footnote{Refer to section {\ref{A.2}} in {{Appendix}} for details} as an implicit function of $z$ as follows,
\begin{equation}
\begin{multlined}
\mathrm{A}_{\upalpha\gamma}=\frac{2\omega_{\upalpha\gamma}}{\uppi\mathsf{J}^2_{\upalpha+1}(\uplambda_{\upalpha\gamma})}\int_0^{2\uppi}\mathrm{d}\phi\int_0^1\mathrm{d}z\int_{0}^{1}r\,\mathrm{d}r\,\times\\\mathsf{J}_
\upalpha(\uplambda_{\upalpha\gamma}r)\,\ccos(\upalpha\phi)\,[\delta\mathsf{P}_0-\delta\mathsf{P}_{\infinity}]\,\mathrm{Z}^{-1}_{\upalpha\gamma}(z)=\\\frac{2\mathrm{C}_\upalpha\,\omega_{\upalpha\gamma}}{\mathsf{J}^2_{\upalpha+1}(\uplambda_{\upalpha\gamma})}\int_{0}^{1}\mathrm{d}r\,
r^{\upalpha+1}(r^2-1)\mathsf{J}_\upalpha(\uplambda_{\upalpha\gamma}r),\!\!\!\!\!\!\!
\,\label{eq:fba}
\end{multlined}
\end{equation}
and,
\begin{equation}
\begin{multlined}
\mathrm{B}_{\upalpha\gamma}=\frac{2\omega_{\upalpha\gamma}}{\uppi\mathsf{J}^2_{\upalpha+1}(\uplambda_{\upalpha\gamma})}\int_0^{2\uppi}\mathrm{d}\phi\int_0^1\mathrm{d}z\int_{0}^{1}r\,\mathrm{d}r\times\\\mathsf{J}_
\upalpha(\uplambda_{\upalpha\gamma}r)\,\csin(\upalpha\phi)\,[\delta\mathsf{P}_0-\delta\mathsf{P}_{\infinity}]\,\mathrm{Z}^{-1}_{\upalpha\gamma}(z)=0.\!\!\!\!\!\label{eq:fbb}
\end{multlined}
\end{equation}
In principle, the {Bessel-Fourier coefficients} may not be constants. In fact, they could be functions of $\phi$ and $z$ depending on the chosen initial conditions in \eqref{eq:P0-P8}. However, since we chose trivial dependence on $\phi$ and $z$ in our assumed initial conditions in \eqref{eq:P0-P8}, $\mathrm{A}_{\upalpha\gamma}$ and $\mathrm{B}_{\upalpha\gamma}$ remain constant. The first few values of $\mathrm{A}_{\upalpha\gamma}$ are: $\mathrm{A}_{11}=-0.706\,\omega_{11}$, $\mathrm{A}_{21}=-0.521\,\omega_{21}$, $\mathrm{A}_{12}=0.154\,\omega_{12}$, and $\mathrm{A}_{22}=0.148\,\omega_{22}$.
\subsection{Final solutions}
\label{section:solutions}
We restore the dimensions and calculate the final {{velocity}}, {{density}}, and {{pressure}} fields in the inertial rest frame instead of the rotating frame. The density profile in the inertial frame is given by
\begin{equation}
\begin{multlined}
\uprho(r,\phi,z,t)=\uprho_\mathrm{0}\uprho_e(z/\mathrm{L})+\uprho_\mathrm{0}\frac{(\delta\Omega)\Omega\hspace{0.01in}\mathrm{L}}{g}\times\\\sum_{\upalpha=0}^{\infinity}\sum_{\gamma=1}^{\infinity}\omega_{\upalpha\gamma}^{-1}\mathrm{C}_\upalpha\mathrm{A}_{\upalpha\gamma}
\mathsf{J}_\upalpha\Bigg(\frac{\uplambda_{\upalpha\gamma}r}{\mathrm{L}}\Bigg)\ccos[\upalpha(\phi-\Omega t)]\times\\\partial_z[-\mathrm{L}\,\mathrm{Z}_{\upalpha\gamma}(z/\mathrm{L})\uprho_e(z/\mathrm{L})]e^{-\mathsf{E}^\frac{1}{2}\omega_{\upalpha\gamma}\Omega t},\label{eq:density}
\end{multlined}
\end{equation}
whereas, from \eqref{eq:0pe},
\begin{equation}
\begin{multlined}
\delta\uprho^{(0)}_{t\rightarrow{\hspace{0.01in}0}}(r,\phi,z)=-\partial_z[\uprho(z)\delta\mathsf{P}_0]=\\-\Bigg[\sum_{\upalpha=0}^{\infinity}\mathrm{C}_\upalpha\,r^\upalpha(r^2-1)\ccos(\upalpha\phi)\Bigg]\frac{\partial[\uprho(z)]}{\partial z}.\!\!\!\!\!\!\!\!\!\label{eq:density0}
\end{multlined}
\end{equation}
$\delta\hspace{-0.01in}v^{(0)}_r(r,\phi,z,{t\rightarrow{\hspace{0.01in}0}})$ and $\delta\hspace{-0.01in}v^{(0)}_\phi(r,\phi,z,{t\rightarrow{\hspace{0.01in}0}})$ can be similarly calculated from \eqref{eq:0vr} and \eqref{eq:0vphi} respectively. The pressure profile, and the velocity field up to order $O(\mathsf{E}^\frac{1}{2})$ are given by,
\begin{equation}
\begin{multlined}
p(r,\phi,z,t)=\uprho_\mathrm{0}g\mathrm{L}\,p_e(z/\mathrm{L})+\\\Bigg[\uprho_\mathrm{0}{(\delta\Omega)\Omega\hspace{0.01in}\mathrm{L}^2}\sum_{\upalpha=0}^{\infinity}\sum_{\gamma=1}^{\infinity}\omega_{\upalpha\gamma}^{-1}\mathrm{C}_\upalpha\mathrm{A}_{\upalpha\gamma}
\mathsf{J}_\upalpha\Bigg(\frac{\uplambda_{\upalpha\gamma}r}{\mathrm{L}}\Bigg)\mathrm{Z}_{\upalpha\gamma}(z/\mathrm{L})\times\\\ccos[\upalpha(\phi-\Omega t)]\,\uprho_e(z/\mathrm{L})e^{-\mathsf{E}^\frac{1}{2}\omega_{\upalpha\gamma}\Omega t}\Bigg],\;\label{eq:pressure}
\end{multlined}
\end{equation}
\begin{equation}
\begin{multlined}
\delta v_r\sim \delta v^{(0)}_r(r,\phi,z,t)=\frac{1}{2}(\delta\Omega)\hspace{0.01in}\mathrm{L}^2\sum_{\upalpha=0}^{\infinity}\sum_{\gamma=1}^{\infinity}\frac{\upalpha}{r}\omega_{\upalpha\gamma}^{-1}
\mathrm{C}_\upalpha\times\\\mathrm{A}_{\upalpha\gamma}
\mathsf{J}_\upalpha\Bigg(\frac{\uplambda_{\upalpha\gamma}r}{\mathrm{L}}\Bigg)\ccos[\upalpha(\phi-\Omega t)]\,\mathrm{Z}_{\upalpha\gamma}(z/\mathrm{L})e^{-\mathsf{E}^\frac{1}{2}\omega_{\upalpha\gamma}\Omega t},\;\;\;\label{eq:vrf}
\end{multlined}
\end{equation}
\begin{equation}
\begin{multlined}
\delta v_\phi\sim\delta v^{(0)}_\phi(r,\phi,z,t)=\Omega r + \frac{1}{2}(\delta\Omega)\hspace{0.01in}\mathrm{L}\sum_{\upalpha=0}^{\infinity}\sum_{\gamma=1}^{\infinity}\omega_{\upalpha\gamma}^{-1}
\times\\\mathrm{C}_\upalpha\mathrm{A}_{\upalpha\gamma}
\uplambda_{\upalpha\gamma}\,\partial_r\!\Bigg[\mathrm{L}\,\mathsf{J}_\upalpha\Bigg(\frac{\uplambda_{\upalpha\gamma}r}{\mathrm{L}}\Bigg)\Bigg]\ccos[\upalpha(\phi-\Omega t)]\,\times\\\mathrm{Z}_{\upalpha\gamma}(z/\mathrm{L})e^{-\mathsf{E}^\frac{1}{2}\omega_{\upalpha\gamma}\Omega t},\;\label{eq:vphif}
\end{multlined}
\end{equation}
and\footnote{We have left the expression in condensed form since the contribution is of the order  $O(\mathsf{E}^{\frac{1}{2}})$ only, which is lower than the magnitudes we want to explore.},
\begin{equation}
\begin{multlined}
\delta v_z\sim v^{(1)}_z(r,\phi,z,t)=\frac{1}{\mathsf{F}\mathsf{N}^2(z)}\frac{\partial\upchi}{\partial z}-\upchi= O(\mathsf{E}^{\frac{1}{2}})\indent\\\text{(in dimensionless units).}\label{eq:vzf}
\end{multlined}
\end{equation}
\section{Gravitational wave emission}
\label{section:emission}
In this section, we describe the gravitational wave emission from mass-quadrupole and current-quadrupole moments of the non-axisymmetric flow derived in section \ref{section:solution}.
\subsection{Gravitational wave emission via mass-quadrupole}
\label{section:emission mass}
The density, pressure and velocity fields calculated in the previous section lead to gravitational wave emission if the mass distribution and fluid flow are non-axisymmetric in nature. Gravitational wave emission is attributable to a non-axisymmetric distribution of mass that has a non-zero mass-quadrupole moment with at least second-order non-vanishing time-derivative. We derive the gravitational wave emission for the leading order quadrupole term ($\upalpha = 2$) straightaway\footnote{Refer to section {\ref{A.3}} in {{Appendix}} for details of the calculation.} for the $+$ and $\times$ polarizations for a {\fontfamily{ppl}\selectfont \textit{polar observer}} -- for an observer located at a distance $d_\mathsf{s}$ along the axis of rotation of the neutron star,
\begin{equation}
\begin{multlined}
h^{\mathrm{M}^\mathsf{P}}_{+}\!(t)=h^\mathrm{M}_\mathrm{o}\!\sum_{\gamma=1}^{\infinity}\kappa_{2\gamma}
\Bigg[-4\omega_{2\gamma}\mathsf{E}^{\frac{1}{2}}\,\csin(2\Omega t)+\\\hspace{1in}(4-\mathsf{E}\omega^2_{2\gamma})\ccos(2\Omega t)\Bigg]e^{-\mathsf{E}^\frac{1}{2}\omega_{2\gamma}\Omega t},\label{eq:plusp}
\end{multlined}
\end{equation}
\begin{equation}
\begin{multlined}
h^{\mathrm{M}^\mathsf{P}}_{\times}\!(t)=h^\mathrm{M}_\mathrm{o}\!\sum_{\gamma=1}^{\infinity}\kappa_{2\gamma}
\Bigg[-4\omega_{2\gamma}\mathsf{E}^{\frac{1}{2}}\,\ccos(2\Omega t)-\\\hspace{1in}(4-\mathsf{E}\omega^2_{2\gamma})\,\csin(2\Omega t)\Bigg]e^{-\mathsf{E}^\frac{1}{2}\omega_{2\gamma}\Omega t},\label{eq:crossp}
\end{multlined}
\end{equation}
where the full expression of $\kappa_{\upalpha\gamma}$ is too lengthy to quote here and is given in section {\ref{A.5}} in {{Appendix}} and $\mathrm{C}_\upalpha$ are set to 1. The characteristic dimensionless strain $h_\mathrm{o}$, and $t_{\upalpha\gamma}$ -- the relaxation time-scale for the $\{\upalpha,\gamma\}\mathrm{th}$ mode -- are given by,
\begin{equation}
\begin{multlined}
h^\mathrm{M}_\mathrm{o}=\uppi\uprho_\mathrm{o}\Omega^4\mathrm{L}^6\epsilon\frac{\mathrm{G}}{c^4d_{\mathsf{s}}g},\indent\label{eq:h0}
\end{multlined}
\end{equation}
\begin{equation}
\begin{multlined}
t_{\upalpha\gamma}=\mathsf{E}^{-\frac{1}{2}}\Omega^{-1}\omega^{-1}_{\upalpha\gamma}, \label{eq:time1}
\end{multlined}
\end{equation}
where, $\kappa_{2\gamma}$ and $h^\mathrm{M}_{\mathrm{o}}$ are both constant quantities. We transform the expressions \eqref{eq:plusp}-\eqref{eq:crossp} for time-series amplitudes to the more useful Fourier space for a polar observer as follows\footnote{$\delta_\mathrm{D}$ is the {\fontfamily{ppl}\selectfont \textit{Kronecker Delta function}} with units of Hz$^{-1}$.}: 
\begin{align*}
|h^{\mathrm{M}^\mathsf{P}}_{+}\!(\omega)|^2={h^\mathrm{M}_\mathrm{o}}^2\sum_{\gamma=1}^{\infinity}{|\kappa_{2\gamma}|^2}\Bigg\{\bigg[t^{-2}_{2\gamma}(4+t^{-2}_{2\gamma}\Omega^{-2})^2+\indent\\\omega^2
(4-t^{-2}_{2\gamma}\Omega^{-2})^2\bigg]\bigg[(4\Omega^2+t^{-2}_{2\gamma}
-\omega^2)^2+(2\omega t^{-1}_{2\gamma})^2\bigg]^{-1}\Bigg\},\tag{66}\label{eq:fplusp}
\end{align*}\setcounter{equation}{66}
\begin{equation}
\begin{multlined}
|h^{\mathrm{M}^\mathsf{P}}_{\times}\!(\omega)|^2={h^\mathrm{M}_\mathrm{o}}^2\sum_{\gamma=1}^{\infinity}{|\kappa_{2\gamma}|^2}\Bigg\{\bigg[4\Omega^2(4+t^{-2}_{2\gamma}\Omega^{-2})^2+\;\;\\16\omega^2 t^{-2}_{2\gamma}\Omega^{-2}\bigg]\bigg[(4\Omega^2+t^{-2}_{2\gamma}
-\omega^2)^2+(2\omega t^{-1}_{2\gamma})^2\bigg]^{-1}\Bigg\}.\label{eq:fcrossp}
\end{multlined}
\end{equation}
Clearly, $|h^{\mathrm{M}^\mathsf{P}}_{+}(\omega)|$ and $|h^{\mathrm{M}^\mathsf{P}}_{\times}(\omega)|$ exhibit resonance at $\omega_\mathrm{R}^2=4\Omega^2+t^{-2}_{2\gamma}$. A similar calculation can be made for an {\fontfamily{ppl}\selectfont \textit{equatorial observer}}, and the corresponding results are given by,
\begin{equation}
\begin{multlined}
\!h^{\mathrm{M}^\mathsf{E}}_{+}\!(t)=\frac{1}{2}h^\mathrm{M}_\mathrm{o}\!\sum_{\gamma=1}^{\infinity}\kappa_{2\gamma}
\Bigg[\!-4\omega_{2\gamma}\mathsf{E}^{\frac{1}{2}}\,\csin(2\Omega t)+\\\hspace{1in}(4-\mathsf{E}\omega^2_{2\gamma})\ccos(2\Omega t)\Bigg]e^{-\mathsf{E}^\frac{1}{2}\omega_{2\gamma}\Omega t},\indent\label{eq:pluse}
\end{multlined}
\end{equation}
\vspace{-0.25in}
\begin{equation}
\begin{multlined}
\!h^{\mathrm{M}^\mathsf{E}}_{\times}(t)=2h^\mathrm{M}_\mathrm{o}\sum_{\gamma=1}^{\infinity}\kappa_{1\gamma}
\Bigg[2\omega_{1\gamma}\mathsf{E}^{\frac{1}{2}}\,\ccos(\Omega t)+\\\hspace{1.09in}(1-\mathsf{E}\omega^2_{1\gamma})\,\csin(\Omega t)\Bigg]e^{-\mathsf{E}^\frac{1}{2}\omega_{1\gamma}\Omega t}.\indent\label{eq:crosse}
\end{multlined}
\end{equation}
It is important to note the change of oscillating frequency for the $\times$ polarization from $2\Omega$ in case of a {{polar observer}} to $\Omega$ in case of an {equatorial observer}. Further, additional $1\gamma$ modes are seen by an {equatorial observer}, besides the $2\gamma$ modes that appear in the emission spectrum. In Fourier space for an equatorial observer, we have
\begin{align*}
|h^{\mathrm{M}^\mathsf{E}}_{+}\!(\omega)|^2=\frac{1}{4}{h^\mathrm{M}_\mathrm{o}}^2\sum_{\gamma=1}^{\infinity}{|\kappa_{2\gamma}|^2}\Bigg\{\bigg[t^{-2}_{2\gamma}(4+t^{-2}_{2\gamma}\Omega^{-2})^2+\indent\\\omega^2
(4-t^{-2}_{2\gamma}\Omega^{-2})^2\bigg]\bigg[(4\Omega^2+t^{-2}_{2\gamma}
-\omega^2)^2+(2\omega t^{-1}_{2\gamma})^2\bigg]^{-1}\Bigg\},\;\;\tag{70}\label{eq:fpluse}
\end{align*}
\vspace{-0.25in}
\begin{align*}
|h^{\mathrm{M}^\mathsf{E}}_{\times}\!(\omega)|^2=4{h^\mathrm{M}_\mathrm{o}}^2\sum_{\gamma=1}^{\infinity}{|\kappa_{1\gamma}|^2}\Bigg\{\bigg[\Omega^2(1+t^{-2}_{1\gamma}\Omega^{-2})^2+\indent\\4\omega^2t^{-2}_{1\gamma}\Omega^{-2}\bigg]\bigg[(\Omega^2+t^{-2}_{1\gamma}
-\omega^2)^2+(2\omega t^{-1}_{1\gamma})^2\bigg]^{-1}\Bigg\}.\;\tag{71}\label{eq:fcrosse}
\end{align*}\setcounter{equation}{71}\\
\noindent In this case, $|h^{\mathrm{M}^\mathsf{E}}_{+}(\omega)|$ exhibits resonance at $\omega_\mathrm{R}^2=4\Omega^2+t^{-2}_{2\gamma}$, while $|h^{\mathrm{M}^\mathsf{E}}_{\times}(\omega)|$ exhibits resonance at  $\omega_\mathrm{R}^2=\Omega^2+t^{-2}_{1\gamma}$. It is worth noting that the factors $\kappa_{1\gamma}$,  $\kappa_{2\gamma}$ decrease in magnitude with increasing index $\gamma$, and we can truncate the above expressions at leading order $\gamma=1$. The maximum order-of-magnitude value of the amplitude of the emitted gravitational waves for both polarizations at a given frequency $\omega$ then depends strongly on the characteristic magnitude ${h^\mathrm{M}_\mathrm{o}}$ and its amplification by the frequency terms in the Fourier transforms. There also exists a weak dependency on the pre-factors $|\kappa_{1\gamma}|$ and  $|\kappa_{2\gamma}|$\footnote{The 'weak' dependency in this case refers to the fact that $|\kappa_{1\gamma}|$ and  $|\kappa_{2\gamma}|$ are not as sensitive to variations in $\mathsf{K_s}$ or $\mathsf{N}^2$, as we will see in later sections.}. 
\\ \\
In figure \ref{fig:freqmp} and figure \ref{fig:freqme}, we plot the frequency characteristics\footnote{In order to show the frequency characteristics, we abbreviate the remaining factors for simplicity such that ${\Bigg[\frac{\displaystyle {|h^{\mathcal{L}^\mathsf{Y}}_{\mp}\!(\omega)|}}{\displaystyle {\;h^\mathcal{L}_\mathrm{o}|\kappa_{\upalpha\gamma}|}}\Bigg]^2 \equiv \mathsf{G}^{\mathcal{L}^\mathsf{Y}}_{\mp | \upalpha\gamma}}$.} of the emitted signal amplitudes for  $|h^{\mathrm{M}^\mathsf{P}}_{+}(\omega)|$, $|h^{\mathrm{M}^\mathsf{P}}_{\times}(\omega)|$, $|h^{\mathrm{M}^\mathsf{E}}_{+}(\omega)|$ and $|h^{\mathrm{M}^\mathsf{E}}_{\times}(\omega)|$.
\end{multicols}
$\;$
\begin{multicols}{2}
\subsection{Gravitational wave emission via current-quadrupole}
\label{section:emission current}
Gravitational wave emission, as traditionally understood, from mass-quadrupole occurs when the associated oscillating mass-quadrupole moment excites gravitational waves. However, time-variation in the intrinsic mass-distribution (also known as the {\fontfamily{ppl}\selectfont \textit{mass-currents}}) of the bulk matter could also lead to gravitational wave radiation through `current-quadrupole' contribution \citep{kip,melatos2010}. This effect is a subset of the {\fontfamily{ppl}\selectfont \textit{gravitomagnetic effects}} -- the electromagnetic equivalent in gravitation. Similar to the case of electromagnetism, where electric charges and current multipoles emit electromagnetic radiation, time-varying mass-current multipoles also emit gravitational wave radiation, besides the well-known emission from mass-quadrupole moment. We straightaway produce the expressions for the $+$ and $\times$ polarization following \citet{kip}, \citet{current} and \citet{melatos2010} for a {polar observer} as follows\footnote{Refer to section {\ref{A.4}} in {{Appendix}} for more details of the calculation, and for expressions of the pre-factors $\mathsf{V}_{1\gamma}$ and $\mathsf{V}_{2\gamma}$.},
\begin{equation}
\begin{multlined}
h^{\mathrm{C}^\mathsf{P}}_{+}(t)=h^\mathrm{C}_\mathrm{o}\sum_{\gamma=1}^{\infinity}\mathsf{V}_{2\gamma}\Big[-4t^{-1}_{2\gamma}\Omega^{-1}\ccos(2\Omega t)  -\\(4-t^{-2}_{2\gamma}\Omega^{-2})\csin(2\Omega t)\Big]e^{-t^{-1}_{2\gamma} t},\label{eq:plusc}
\end{multlined}
\end{equation}
\begin{equation}
\begin{multlined}
h^{\mathrm{C}^\mathsf{P}}_{\times}(t)=h^\mathrm{C}_\mathrm{o}\sum_{\gamma=1}^{\infinity}\mathsf{V}_{2\gamma}\Big[-4t^{-1}_{2\gamma}\Omega^{-1}\csin(2\Omega t) +\\(4-t^{-2}_{2\gamma}\Omega^{-2})\ccos(2\Omega t)\Big]e^{-t^{-1}_{2\gamma} t},\label{eq:crossc}
\end{multlined}
\end{equation}
and, for an {equatorial observer} by,
\begin{equation}
\begin{multlined}
h^{\mathrm{C}^\mathsf{E}}_{+}(t)=2h^\mathrm{C}_\mathrm{o}\sum_{\gamma=1}^{\infinity}\mathsf{V}_{1\gamma}\Big[2t^{-1}_{1\gamma}\Omega^{-1}\ccos(\Omega t) +\;\;\;\;\;\\  (1-t^{-2}_{1\gamma}\Omega^{-2})\csin(\Omega t)\Big]e^{-t^{-1}_{1\gamma} t},\label{eq:plusc1}
\end{multlined}
\end{equation}
\vspace{-0.15in}
\begin{equation}
\begin{multlined}
h^{\mathrm{C}^\mathsf{E}}_{\times}(t)=\frac{1}{2}h^\mathrm{C}_\mathrm{o}\sum_{\gamma=1}^{\infinity}\mathsf{V}_{2\gamma}\Big[-4t^{-1}_{2\gamma}\Omega^{-1}\csin(2\Omega t) +\\(4-t^{-2}_{2\gamma}\Omega^{-2})\ccos(2\Omega t)\Big]e^{-t^{-1}_{2\gamma} t},\label{eq:crossc1}
\end{multlined}
\end{equation}
where,
\begin{equation}
\begin{multlined}
h^\mathrm{C}_\mathrm{o}=2\uppi\uprho_\mathrm{o}\Omega^3\mathrm{L}^6\,\epsilon\frac{\mathrm{G}}{3c^5d_\mathsf{s}}.\label{eq:h0c}
\end{multlined}
\end{equation}
Note that we have restricted ourselves to the leading-order quadrupole term $l=2$ of the mass-current multipole expansion. Once more, we write the above expressions for {{polar}} and {{equatorial}} observers in the Fourier space. In case of a {polar observer}, this reduces to
\end{multicols}
\begin{multicols}{2}
\begin{figure}[H]
\centering
\includegraphics[width=78mm]{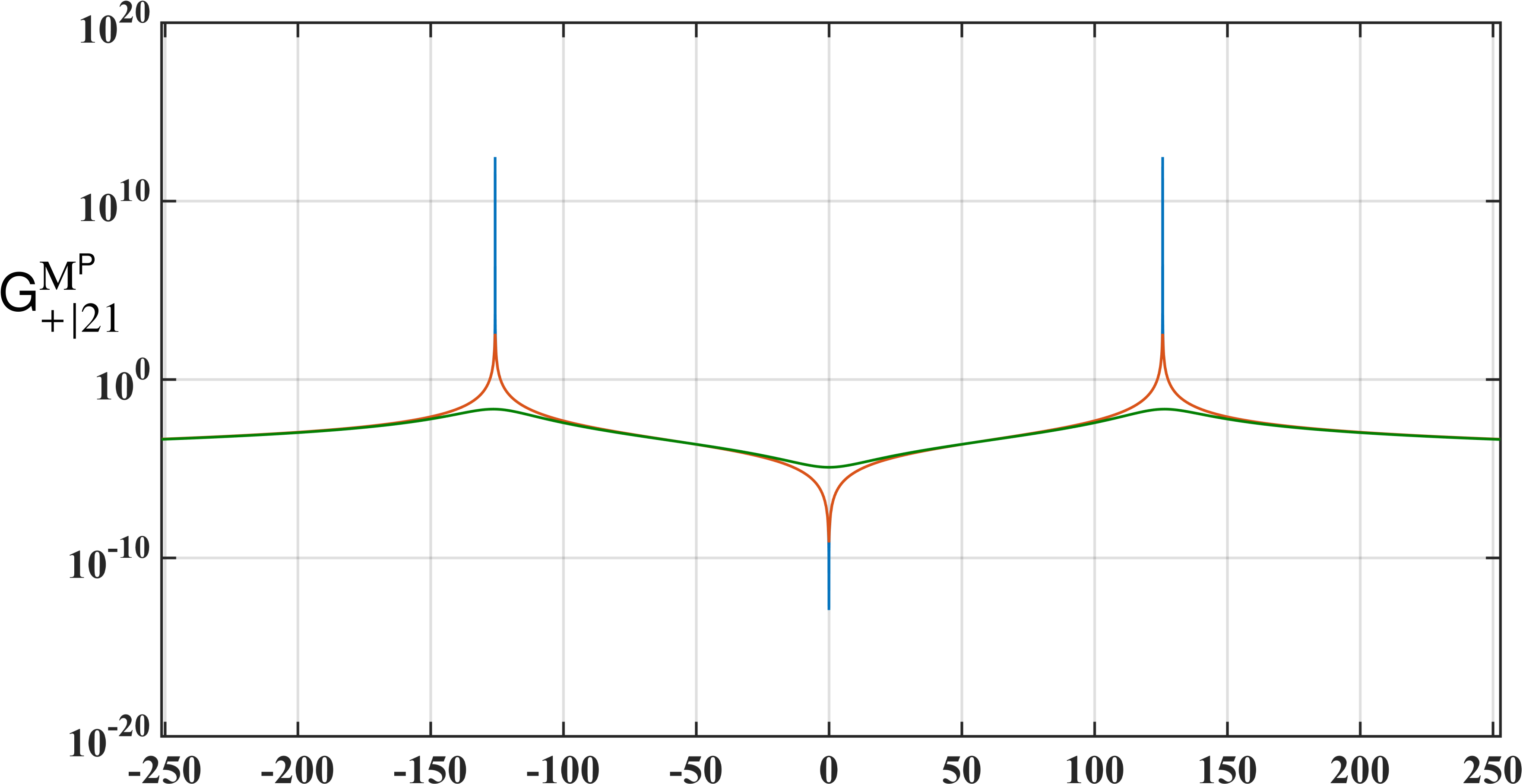} 
\includegraphics[width=78mm]{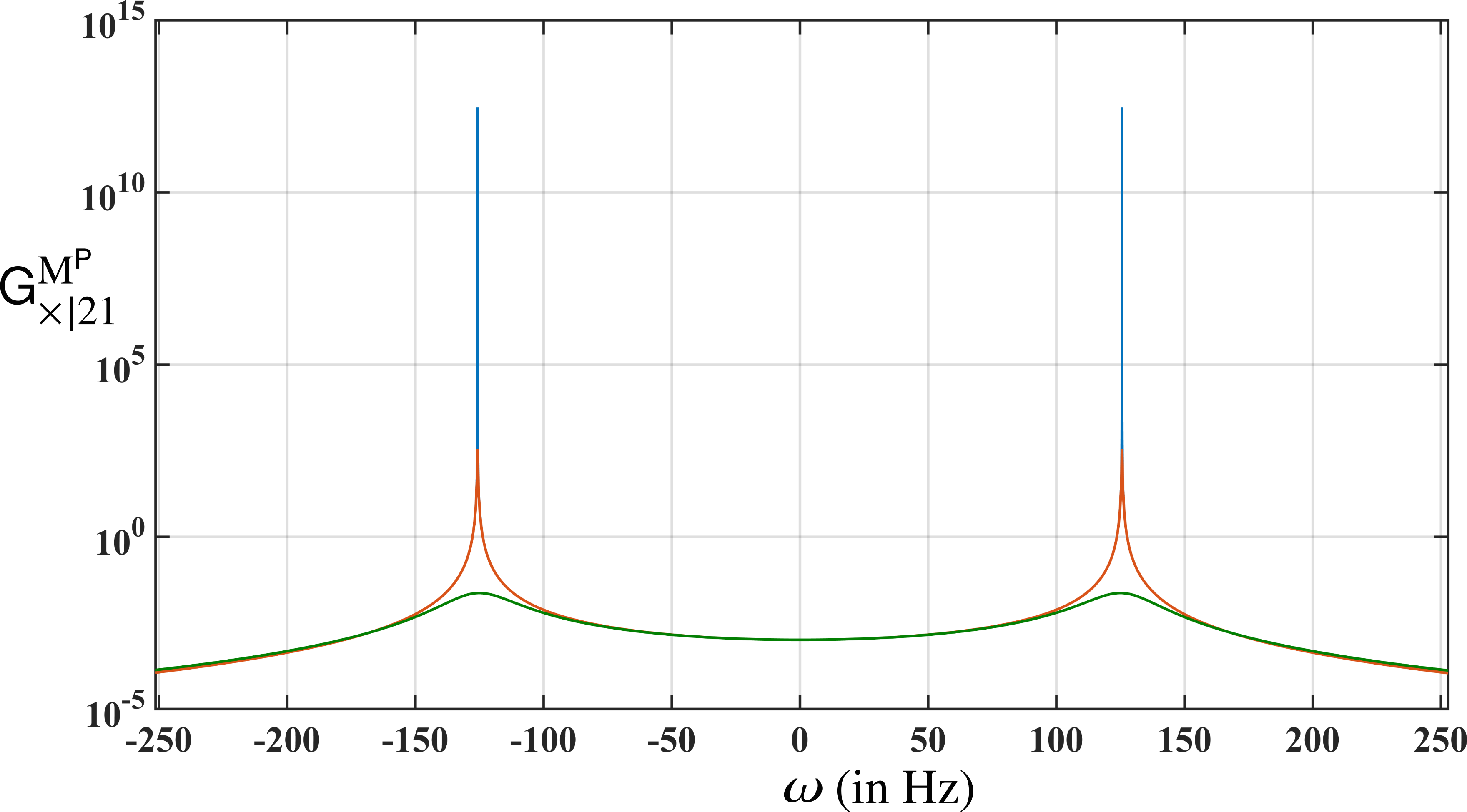} 
\vspace{-0.2in}
\caption{{\small Frequency characteristics of emitted signals for {polar observer}:} {{\small 3 different set of data are plotted for for $\Omega/2\uppi=10\,$Hz and $\partial_z\eta\sim 10^{-11}$. The respective color-coded time-scales are {9.8} (blue), {1.1$\times 10^{-4}$} (red) and {8.5$\times 10^{-7}$} (green) days. The corresponding resonant frequencies are {$\pm125.66371\,$}, {$\pm125.66375\,$} and {$\pm126.39611\,$}Hz}. Note that the values of time-scales are calculated for specifically chosen physical parameters of the system -- $v_\mathrm{c}$, $\partial_z\eta$, $v_\mathrm{eq}$, $\mathsf{K}$, and $\mathsf{F}$, in order to cover a large range of time-scales. A similar result for an {equatorial observer} is shown in figure \ref{fig:freqme}.}}
\label{fig:freqmp}
\end{figure}
\begin{figure}[H]
\centering
\includegraphics[width=78mm]{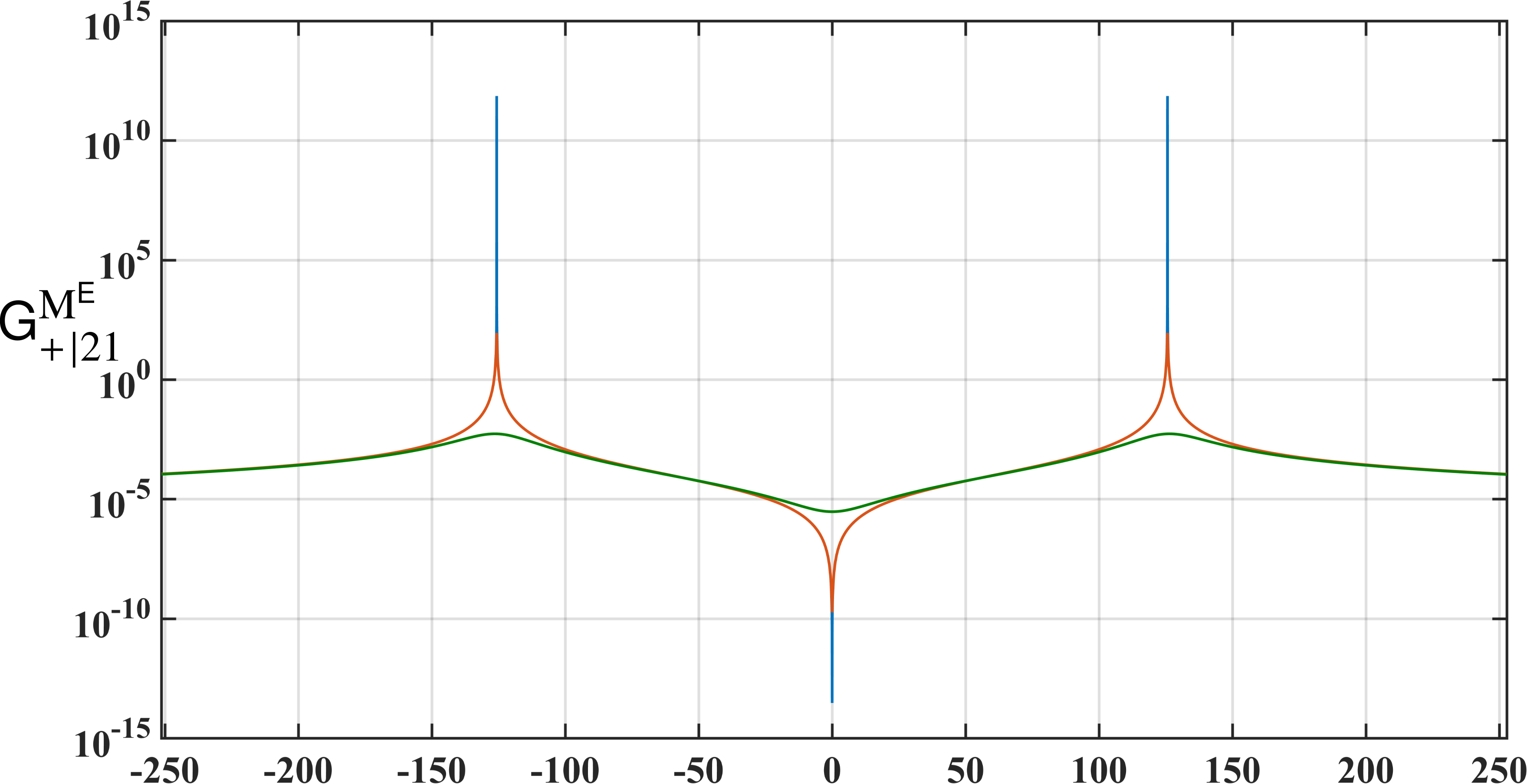} 
\includegraphics[width=78mm]{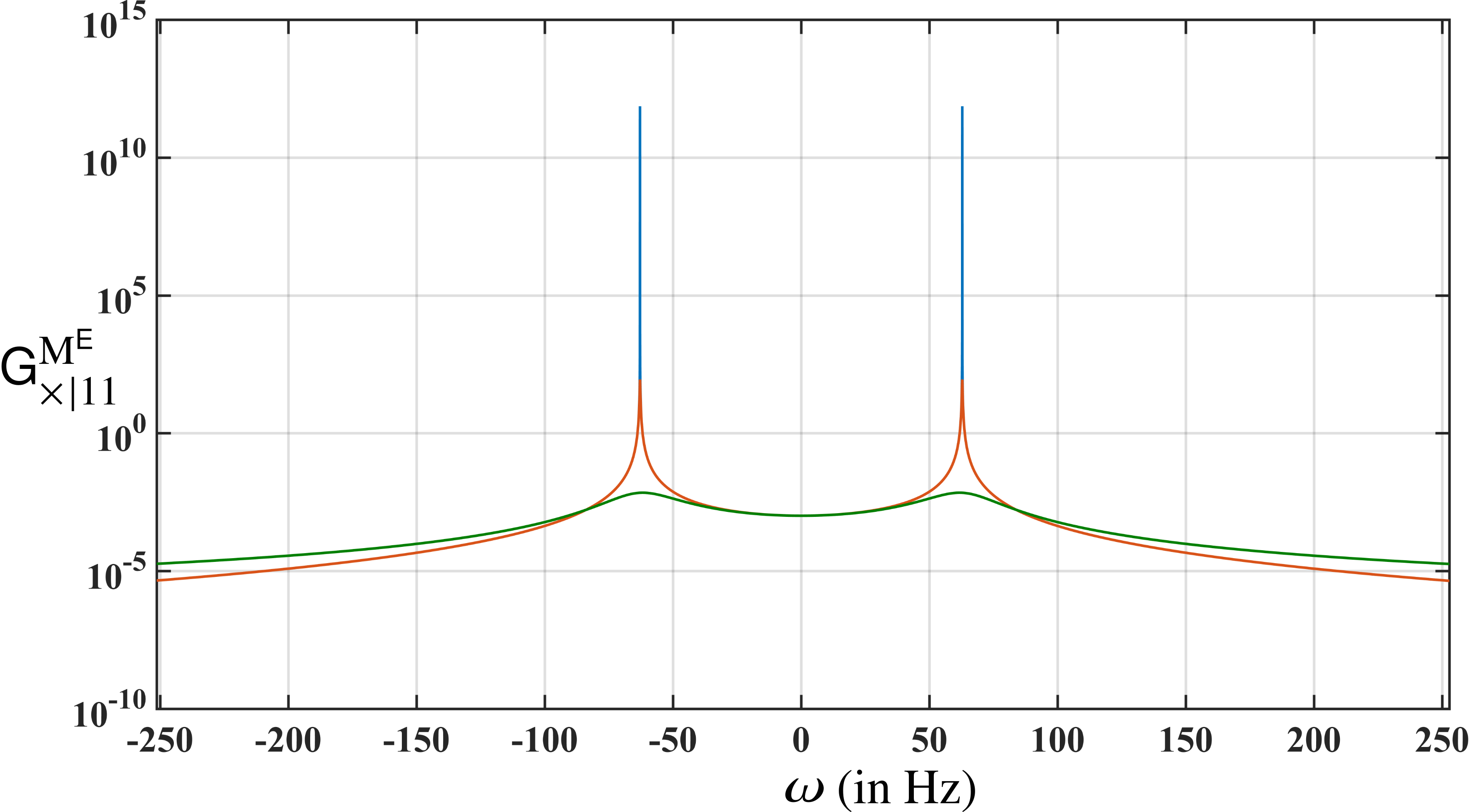} 
\vspace{-0.2in}
\caption{{\small Frequency characteristics of emitted signals for {equatorial observer}:} {{\small 3 different set of data are plotted for for $\Omega/2\uppi=10\,$Hz and $\partial_z\eta\sim 10^{-11}$. The respective color-coded time-scales are {9.8} (blue), {1.1$\times 10^{-4}$} (red) and {8.5$\times 10^{-7}$} (green) days. Moreover, the corresponding resonant frequencies for $\times$ polarization are {$\pm62.83186\,$}, {$\pm62.83188\,$}, {$\pm63.19805\,$}Hz}, whereas the resonant frequencies for the $+$ polarization remain exactly the same as they were for the case of a {polar observer}. Other physical parameters are chosen to be the same as in figure \ref{fig:freqmp}.}}
\label{fig:freqme}
\end{figure}
\end{multicols}
\vspace{-0.4in}
\begin{multicols}{2}
\begin{align*}
|h^{\mathrm{C}^\mathsf{P}}_{+}\!(\omega)|^2={h^\mathrm{C}_\mathrm{o}}^2\sum_{\gamma=1}^{\infinity}|\mathsf{V}_{2\gamma}|^2 \Bigg\{\bigg[4\Omega^2(4+t^{-2}_{2\gamma}\Omega^{-2})^2+\indent\\16\omega^2
t^{-2}_{2\gamma}\Omega^{-2}\bigg]\bigg[(4\Omega^2+t^{-2}_{2\gamma}
-\omega^2)^2+(2\omega t^{-1}_{2\gamma})^2\bigg]^{-1}\Bigg\},\tag{75}\label{eq:pluscp}
\end{align*}\setcounter{equation}{75}\vspace{-0.20in}
\begin{align*}
|h^{\mathrm{C}^\mathsf{P}}_{\times}\!(\omega)|^2={h^\mathrm{C}_\mathrm{o}}^2\sum_{\gamma=1}^{\infinity}|\mathsf{V}_{2\gamma}|^2\Bigg\{\bigg[t^{-2}_{2\gamma}(4+t^{-2}_{2\gamma}\Omega^{-2})^2+\;\indent\\\omega^2
(4-t^{-2}_{2\gamma}\Omega^{-2})^2\bigg]\bigg[(4\Omega^2+t^{-2}_{2\gamma}
-\omega^2)^2+(2\omega t^{-1}_{2\gamma})^2\bigg]^{-1}\Bigg\}.\tag{76}\label{eq:crosscp}
\end{align*}\setcounter{equation}{76}\\ 
\noindent It should be noted that the resonant frequencies for current-quadrupole contribution from $|h^{\mathrm{M}^\mathsf{P}}_{+}(\omega)|$ and $|h^{\mathrm{M}^\mathsf{P}}_{\times}(\omega)|$ are the same as they were for mass-quadrupole contribution. Further, for the case of {\fontfamily{ppl}\selectfont \textit{equatorial observers}},
\begin{align*}
|h^{\mathrm{C}^\mathsf{E}}_{+}\!(\omega)|^2=4{h^\mathrm{C}_\mathrm{o}}^2\sum_{\gamma=1}^{\infinity}|\mathsf{V}_{1\gamma}|^2\Bigg\{\bigg[\Omega^2(1+t^{-2}_{1\gamma}\Omega^{-2})^2+
\indent\\4\omega^2t^{-2}_{1\gamma}\Omega^{-2}\bigg]\bigg[(\Omega^2+t^{-2}_{1\gamma}
-\omega^2)^2+(2\omega t^{-1}_{1\gamma})^2\bigg]^{-1}\Bigg\},\tag{77}\label{eq:plusce}
\end{align*}\setcounter{equation}{77}
\begin{align*}
\,|h^{\mathrm{C}^\mathsf{E}}_{\times}\!(\omega)|^2=\frac{1}{4}{h^\mathrm{C}_\mathrm{o}}^2\sum_{\gamma=1}^{\infinity}|\mathsf{V}_{2\gamma}|^2\Bigg\{\bigg[t^{-2}_{2\gamma}(4+t^{-2}_{2\gamma}\Omega^{-2})^2+\indent\\\omega^2
(4-t^{-2}_{2\gamma}\Omega^{-2})^2 \bigg]\bigg[(4\Omega^2+t^{-2}_{2\gamma}
-\omega^2)^2+(2\omega t^{-1}_{2\gamma})^2\bigg]^{-1}\Bigg\}.\;\tag{78}\label{eq:crossce}
\end{align*}\setcounter{equation}{78}\\
\noindent We see that the emitted signals from the mass-quadrupole and the current-quadrupole are similar in nature in terms of the resonant frequencies and the general behavior of the frequency responses\footnote{The characteristic amplitudes for mass and current quadrupole are related by $\displaystyle {\frac{|h^\mathrm{C}_\mathrm{o}|}{|h^\mathrm{M}_\mathrm{o}|} = \frac{2g}{3\Omega c}}$.}. However, there is a notable switch in the $+$ and $\times$ polarizations. Additionally, $\mathsf{V}_{1\gamma}$ and $\mathsf{V}_{2\gamma}$ pre-factors now appear instead of $\kappa_{1\gamma}$ and $\kappa_{2\gamma}$, besides the different characteristic amplitudes. Lastly, the frequency characteristics for  $|h^{\mathrm{C}^\mathsf{P}}_{+}(\omega)|$, $|h^{\mathrm{C}^\mathsf{P}}_{\times}(\omega)|$, $|h^{\mathrm{C}^\mathsf{E}}_{\times}(\omega)|$ and $|h^{\mathrm{C}^\mathsf{E}}_{+}(\omega)|$ follow the same shapes as shown previously in figure \ref{fig:freqmp} and figure \ref{fig:freqme}.
\subsection{A verdict on parameter space}
\label{section:vtime}
It is clear from the general expressions of $\kappa_{\upalpha\gamma}$ (in section {\ref{A.5}} in {{Appendix}}) and $\mathsf{V}_{\upalpha\gamma}$ (in section {\ref{A.4}} in {{Appendix}}) that their calculations are cumbersome to perform unless we could make some simplifying assumptions. Ideally, one would like to explore the range of parameter space where the first derivative of $\eta$ -- i.e. $\partial_z\eta$ in \eqref{eq:omegac}, and $\partial_z\mathsf{K_s}$ are constants, and follow
\begin{equation}
\begin{multlined}
|\partial_z\eta| \ll \mathsf{F}\mathsf{N}^2 \ll \eta < 1\indent\text{for}\;\;\;\forall\;z \in (0, 1],\\
|\partial_z\mathsf{K_s}| \ll |\mathsf{K_s}|\;\;\;\text{for}\;\;\;\forall\;z \in (0, 1]. \indent\indent\label{eq:partialeta}
\end{multlined}
\end{equation}
Such a choice of a regime is physically reasonable and it makes the calculations analytically feasible, without compromising the generality of the model. These assumptions allow us to reduce the parameter space and explore the model in its simplest form. Meanwhile, since we do not have any prior functional forms of $v_\mathrm{c}$ and $v_\mathrm{eq}$ with respect to $z$-coordinate, we assume a simple scenario where $v_\mathrm{c}$ is linear in $z$ and takes the form\footnote{Note that any functional form of $v_\mathrm{c}(z)$ can be reduced to this expression at the leading order as long as $|\partial_z\hspace{-0.0175in}v_\mathrm{c}|\ll v_\mathrm{c}^\mathrm{o}$. This is equivalent to a ``stiff'' polytropic equation of state with the polytropic exponent $\gamma\rightarrow 1$.},
\begin{equation}
\begin{multlined}
v_\mathrm{c}(z) = v_\mathrm{c}^\mathrm{o} + z\times\partial_z\hspace{-0.0175in}v_\mathrm{c},\label{eq:vcnew}
\end{multlined}
\end{equation}
while, at the same time, $\mathsf{N}^2$ is taken to be a constant. These assumptions leave $v_\mathrm{eq}$ implicitly varying in $z$ according to \eqref{eq:bvf}. It must be noted that this doesn't imply constancy of $\eta$. In fact, it is simply that $\partial_z\eta \sim 2v_\mathrm{c}(z)\partial_z\hspace{-0.0175in}v_\mathrm{c}$, and $\partial^2_z\eta \sim 2(\partial_z\hspace{-0.0175in}v_\mathrm{c})^2$. Lastly, we are left with $\mathsf{N}^2$, $v_\mathrm{c}^\mathrm{o}$ and $\partial_z\hspace{-0.0175in}v_\mathrm{c}$ as free parameters in our model. $\mathsf{K_s}$ (or, $v_\mathrm{eq}$) in this case becomes a dependent parameter varying in $z$ according to \eqref{eq:bvf}, as previously stated. Thus, we restrict ourselves to the domain where
\begin{equation}
\begin{multlined}
|\partial_z\hspace{-0.0175in}v_\mathrm{c}| \ll v_\mathrm{c}^\mathrm{o} < 1\implies|\partial_z\mathsf{K_s}| \ll |\mathsf{K_s}|, \partial_z\eta \sim 2v^\mathrm{o}_\mathrm{c}\partial_z\hspace{-0.0175in}v_\mathrm{c},\label{eq:partialeta2}
\end{multlined}
\end{equation}
Under such assumptions, the calculations for the factors $\kappa_{\upalpha\gamma}$ and $\mathsf{V}_{\upalpha\gamma}$ become analytic and relatively simpler\footnote{See section {\ref{A.5}} in {{Appendix}} for details, and for full expressions of $\kappa_{\upalpha\gamma}$ and $\mathsf{V}_{\upalpha\gamma}$.}. The simplification occurs because $\partial_z\eta$ is now invariant in $z$ according to \eqref{eq:partialeta2}. To further validate our choice, we find that numerical errors dominate significantly when calculating $\kappa_{\upalpha\gamma}$ and $\mathsf{V}_{\upalpha\gamma}$ numerically, especially toward lower ranges of $v_\mathrm{c}^\mathrm{o}$. These numerical errors are catalysed by large corresponding magnitudes of $\mathsf{K_s}$ when $v_\mathrm{c}^\mathrm{o}$ becomes very small. This effect is shown in detail in section {\ref{A.6}} in {{Appendix}} where we have compared numerical and analytic results for $\kappa_{\upalpha\gamma}$ and $\mathsf{V}_{\upalpha\gamma}$, assuming \eqref{eq:partialeta2} to be true. In nutshell, the analytic approximation in \eqref{eq:partialeta2} enables us to selectively explore the more crucial aspects of the improved model, such as $\partial_z\hspace{-0.0175in}v_\mathrm{c}$, while ignoring the less crucial degrees of freedom of the system, such as spatial variations in $\mathsf{N}^2$. The complete reduced expressions for $\kappa_{\upalpha\gamma}$ and $\mathsf{V}_{\upalpha\gamma}$ are given in section {\ref{A.5}} in {{Appendix}}.

It must be noted that such an assumption of constancy of $\partial_z\eta$ isn't applied while calculating $\omega_{\upalpha\gamma}$ and the corresponding time-scales $t_{\upalpha\gamma}$, via \eqref{eq:omega}. However, the time-scales $t_{\upalpha\gamma}$ are not prone to the errors from numerical computations, as opposed to $\kappa_{\upalpha\gamma}$ and $\mathsf{V}_{\upalpha\gamma}$. It remains straightforward to compute them numerically and accurately. Nonetheless, the approximated expression for the time-scales is given in section {\ref{A.5}} in {{Appendix}} [see \eqref{eq:approxt}].
\section{Time-scales of emitted signals and corresponding amplitudes}
\label{section:times}
In this section, we explore the decay time-scales of the emitted signals. We see from the expressions in \eqref{eq:plusp}-\eqref{eq:crosse} for the mass-quadrupole contribution, and \eqref{eq:plusc}-\eqref{eq:crossc1} for the current-quadrupole contribution, that the decay time-scale $t_{\upalpha\gamma}$ for a given \{$\upalpha$, $\gamma$\} mode -- as defined previously in \eqref{eq:h0} -- is given by,
 \begin{equation}
\begin{multlined}
t_{\upalpha\gamma}=\mathsf{E}^{-\frac{1}{2}}\Omega^{-1}\omega^{-1}_{\upalpha\gamma}.\label{eq:timescale}
\end{multlined}
\end{equation}
The emitted gravitational wave signal amplitude at a given frequency $\omega$ depends intrinsically on the time-scale; this is shown in the expressions in \eqref{eq:plusp}-\eqref{eq:crossce}. Following the discussion in the previous section, we have 3 independent parameters to vary: $v^\mathrm{o}\mathrm{c}$, $\partial_z\hspace{-0.0175in}v_\mathrm{c}$ and $\mathsf{N^2}$, under the analytic approximations introduced by \eqref{eq:partialeta2}. In figure \ref{hplot}, we plot the characteristics for the involved time-scales $t_{11}$ and $t_{21}$, and the corresponding gravitational wave amplitudes for mass-quadrupole and current-quadrupole contributions at resonant frequencies i.e. $\omega = \omega_\mathrm{R}$ (denoted by subscript \textbf{R}). Note that the resonant frequencies $\omega_\mathrm{R}$ are also a function of $t_{\upalpha\gamma}$, as shown in section \ref{section:emission mass}. This corresponds to the effect where $|h^{\mathrm{M}^\mathsf{E}}_{\times}|$, $|h^{\mathrm{C}^\mathsf{E}}_{+}|$ emit at different resonant frequencies and different time-scales than $|h^{\mathrm{M}^\mathsf{P}}_{\times}|$, $|h^{\mathrm{M}^\mathsf{P}}_{+}|$, $|h^{\mathrm{M}^\mathsf{E}}_{+}|$, $|h^{\mathrm{C}^\mathsf{P}}_{\times}|$, $|h^{\mathrm{C}^\mathsf{P}}_{+}|$ and $|h^{\mathrm{C}^\mathsf{E}}_{\times}|$, as shown in figure \ref{hplot}. We also find that only a very small fraction of mechanical energy [$O(10^{-9}-10^{-7})$] from the glitch is converted into gravitational wave emission\footnote{The energetics of the emitted amplitudes is discussed in detail in section {\ref{energetics}}.}.

Note that in the case of $\partial_z\hspace{-0.0175in}v_\mathrm{c} < 0$ in figure \ref{hplot} (rightmost panels), the apparent outlier in the plots for $\partial_z\hspace{-0.0175in}v_\mathrm{c} = -10^{-4}c\mathrm{L}^{-1}$ is an artifact of low resolution in parameter space. In figure \ref{fig:last}, we show the characteristics in the vicinity of the outlier for clarity.\vspace{-0.1in}
\subsection{Growing modes}
\label{growingmodes}
\begin{figure}[H]
\includegraphics[width=69.75mm, height= 94.2mm, angle=90]{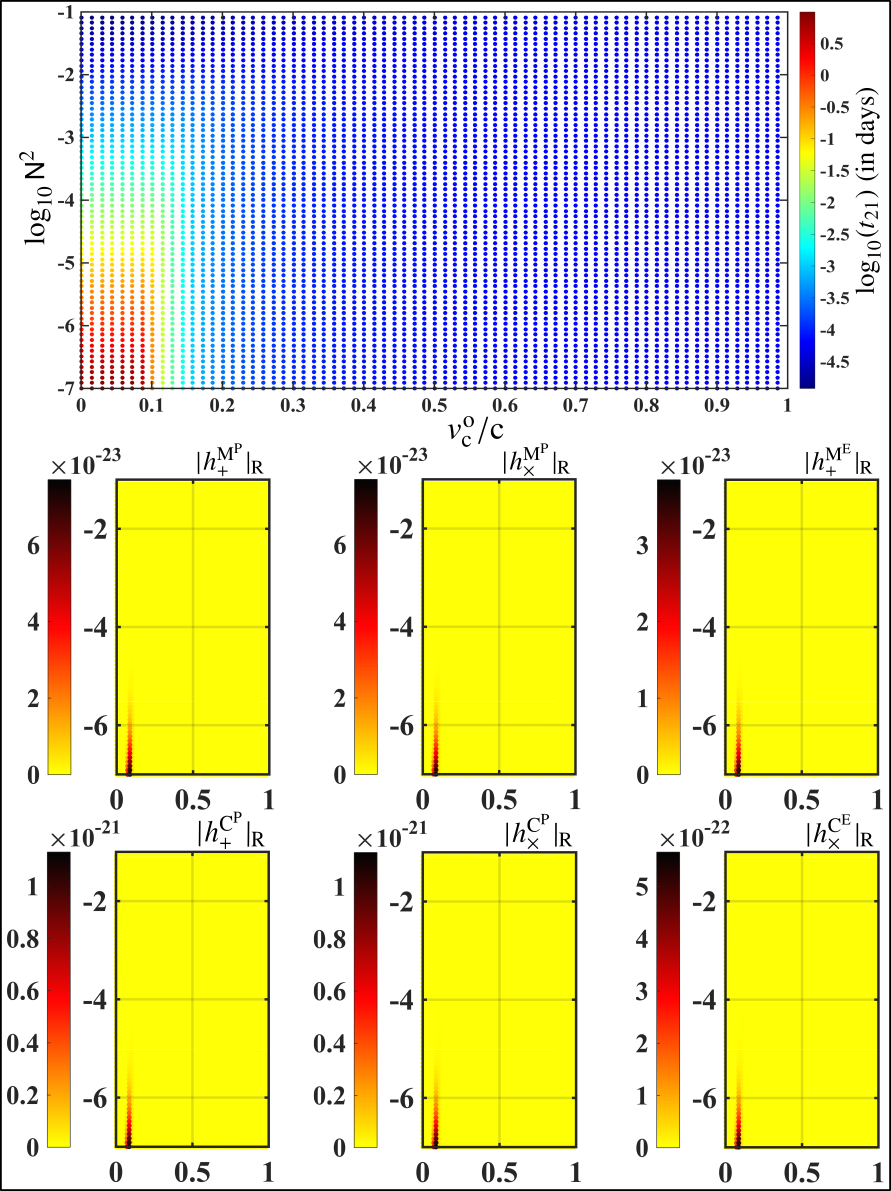} 
\includegraphics[width=69.75mm, height= 63.7mm, angle=90]{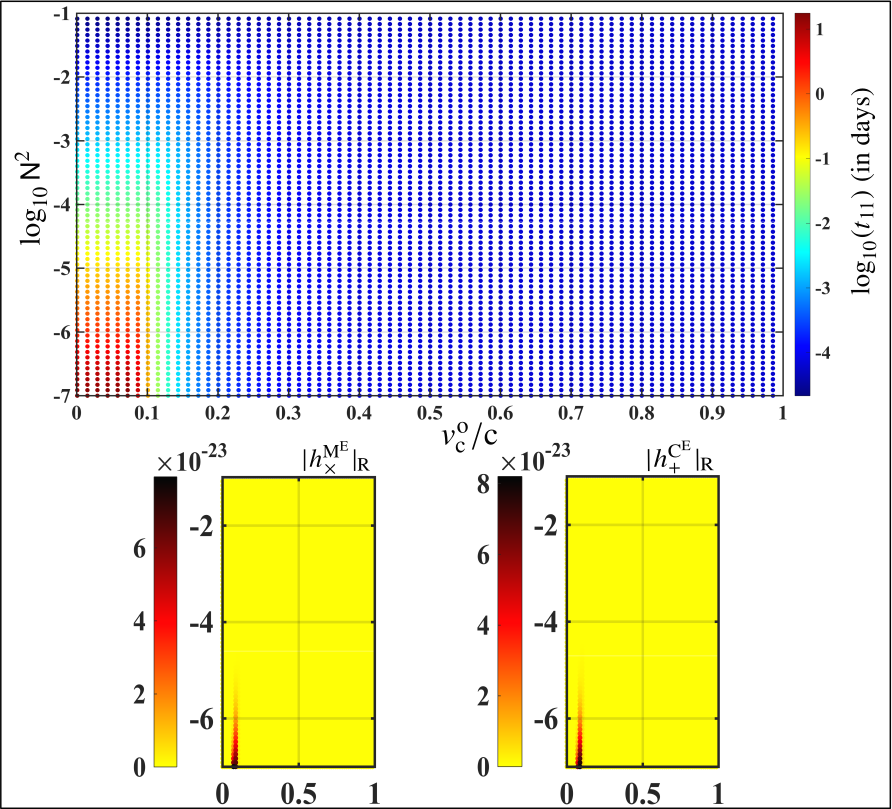} 
\caption{\small{Emitted gravitational wave strain ({\fontfamily{ppl}\selectfont \textit{turn page sideways}}): The top panel represents the time-scales for the $t_{21}$ mode and the corresponding gravitational wave strain amplitudes for 3 sets of values for\hspace{0.0175in} $\partial_z\hspace{-0.0175in}v_\mathrm{c}$ at $0$, $10^{-3}c\mathrm{L}^{-1}$, $-10^{-4}c\mathrm{L}^{-1}$ (left to right) respectively. The bottom panel shows the $t_{11}$ modes, \hspace{0.015in}and corresponding gravitational wave strain amplitudes. \hspace{0.015in}The parameters are set to: $f=100\,$Hz, $\mathsf{E}=10^{-7}$, $\epsilon = 10^{-4}$, $d_\mathsf{s}=1.0\,$kpc, $\mathrm{L}=10^{4}$m, $g=10^{12}\mathrm{m/sec^2}$, $\uprho_\mathrm{o}\hspace{-0.015in}=10^{17}\mathrm{kg/m^3}$. All positive time-scales as well as the corresponding emitted amplitudes are marked by $\bullet$, while the negative time-scales and corresponding amplitudes are marked by $\boldsymbol{+}$. Negative time-scales correspond to the scenario of {growing modes}{\textcolor{blue}{\protect\footnotemark[20]}}.}}
\label{hplot}
\end{figure}

\begin{figure}[H]
\centering
\includegraphics[width=150mm]{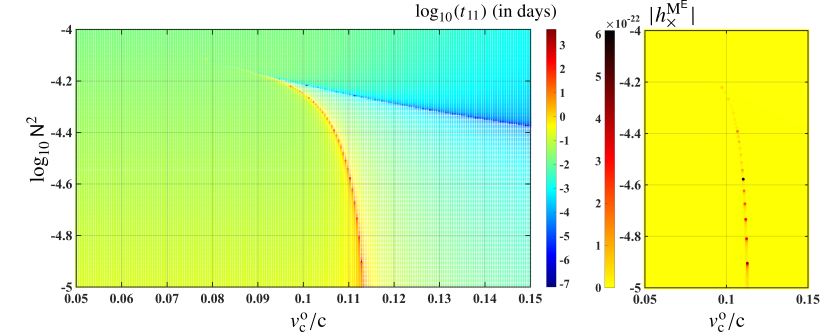} 
\caption{{\small Outlier zoomed in:} {{\small The characteristics of the parameter space in the vicinity of the apparent outlier in figure \ref{hplot} for the case of $\partial_z\hspace{-0.0175in}v_\mathrm{c} = -10^{-4}c\mathrm{L}^{-1}$ (rightmost panels in figure \ref{hplot}) are shown in higher resolution. All positive time-scales as well as corresponding emitted amplitudes are marked by $\bullet$, while the negative time-scales and corresponding strain amplitudes are marked by $\boldsymbol{+}$.}}} 
\label{fig:last}
\end{figure}

In figure \ref{hplot}, for $\partial_z\hspace{-0.0175in}v_\mathrm{c} < 0$ (rightmost panels), we see that it is possible for the system to exhibit {\fontfamily{ppl}\selectfont \textit{growing modes}}. The {growing modes} refer to the cases where perturbations become unstable and grow monotonically, denoted by $\boldsymbol{+}$ marker in figure \ref{hplot} and figure \ref{fig:last}. {growing modes} are characterised by negative time-scales, i.e. $t_{\upalpha\gamma} < 0$. In standard Oceanography and Fluid Mechanics literature, {growing modes} are associated with convection and overturning\footnote{See {\fontfamily{ppl}\selectfont \textit{An Introduction to Dynamic Meteorology}} by {J.R. Holton}, and {\fontfamily{ppl}\selectfont \textit{Waves in Fluids}} by {J. Lighthill}.}. They represent a system that gains energy from
\end{multicols}
\pagebreak
\begin{figure}[H]
\includegraphics[width=69.75mm, height= 94.2mm, angle=90]{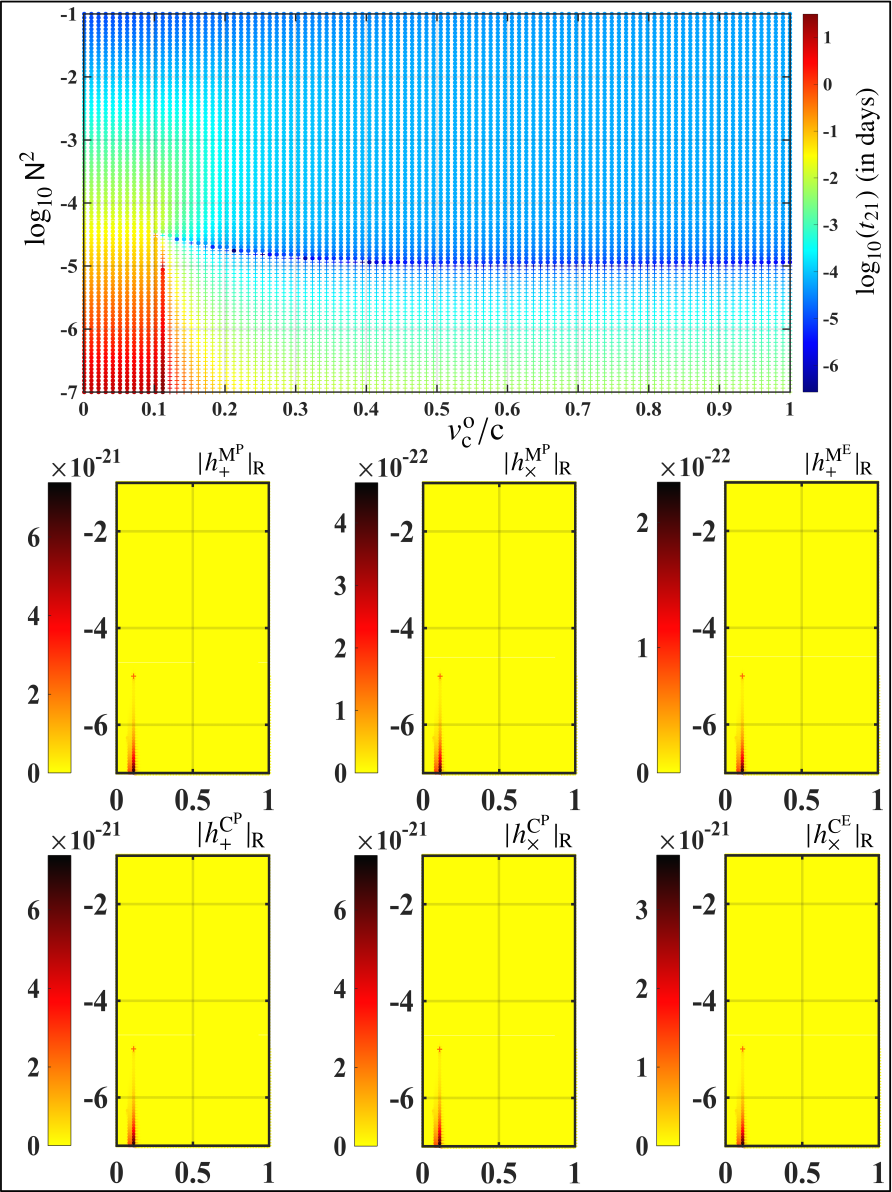} 
\includegraphics[width=69.75mm, height= 63.7mm, angle=90]{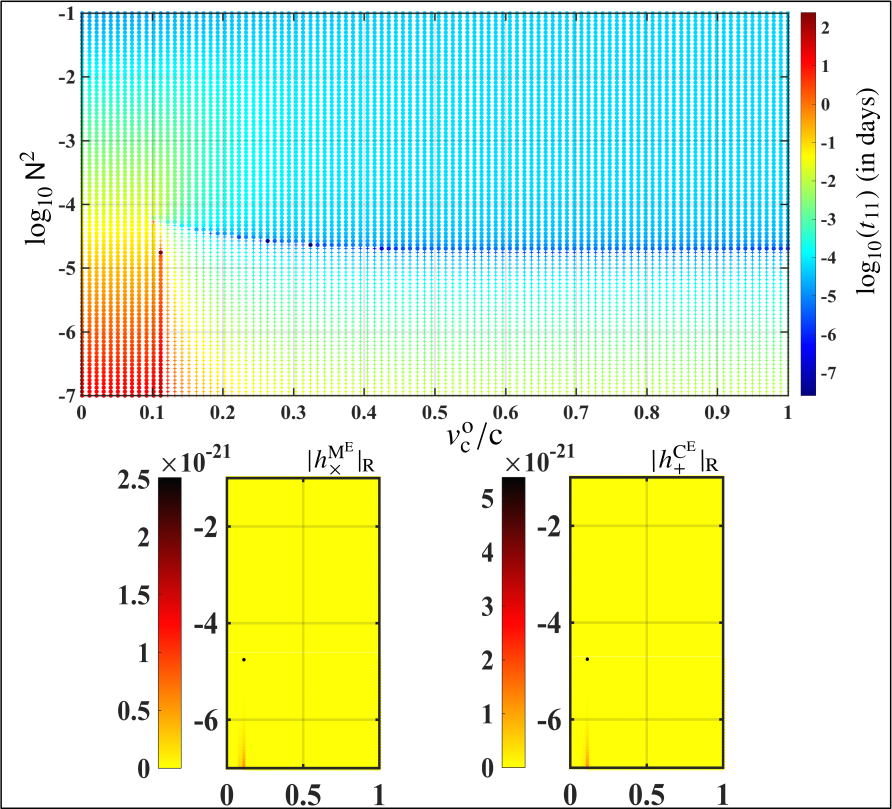} 
\end{figure}
\vspace{-0.3in} 
\begin{figure}[H]
\includegraphics[width=69.75mm, height= 94.2mm, angle=90]{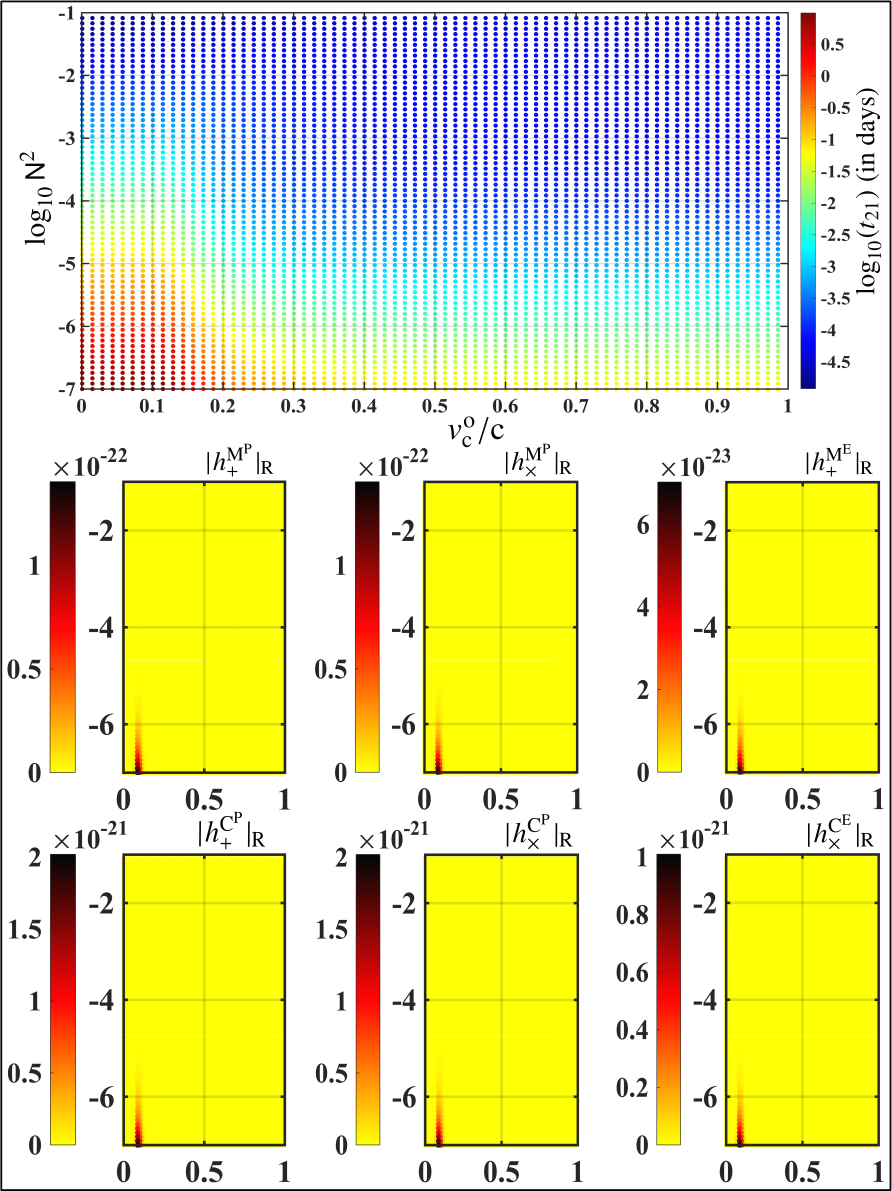} 
\includegraphics[width=69.75mm, height= 63.7mm, angle=90]{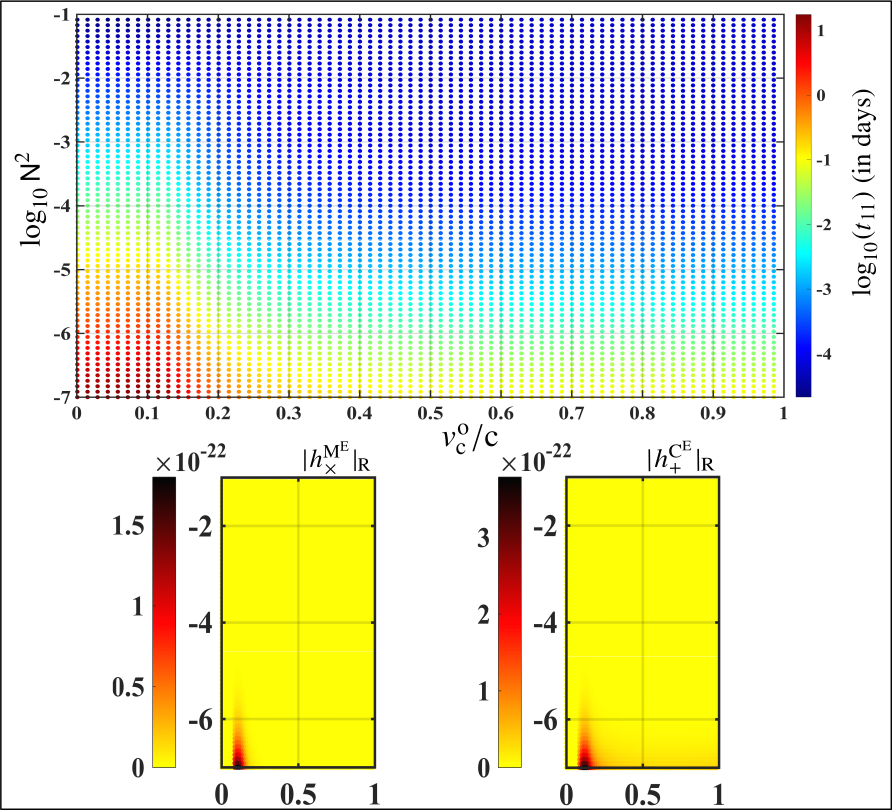} 
\end{figure}
\vspace{-0.3in} 
\pagebreak
\begin{multicols}{2}
\noindent its surroundings and is not in an adiabatic state. However, this is not true for our system since it is an isolated neutron star. For this reason, these {growing modes} are unphysical\footnote{A brief explanation and interpretation of the existence of {growing modes} is discussed in section {\ref{A.8}} in {{Appendix}} and section \ref{conclusions} respectively.} and the corresponding regions in the parameter space are gravitationally inaccessible.\vspace{-0.05in}
\section{Discussion}
\label{conclusions}
\begin{figure}[H]
\centering
\includegraphics[width=150mm]{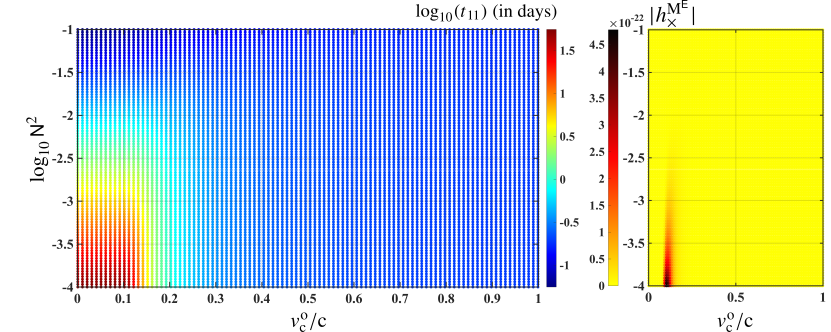} 
\caption{{\small Sensitivity to $\mathsf{E}$:} {{\small The characteristics of emitted signals for $\mathsf{E} = 10^{-14}$ and $\partial_z\hspace{-0.0175in}v_\mathrm{c} = 0$ are shown, setting $f=100\,$Hz, $\epsilon = 10^{-4}$, $d_\mathsf{s}=1\,$kpc, $\mathrm{L}=10^{4}$m, $g=10^{12}\mathrm{m/sec^2}$, $\uprho_\mathrm{o}=10^{17}\mathrm{kg/m^3}$. Note the emission of loud amplitudes within the range $\mathsf{N}^2 \in (10^{-4}, 10^{-2})$; the equivalent classical {{Brunt-V{\"a}is{\"a}l{\"a} frequency}} $\mathsf{N}^2_\mathrm{c}$ for this range lies within permitted physical expectations.}}} 
\label{fig:E}
\end{figure}

To conclude our study, one can broadly make the following reiterations and conclusions. We have extended the previous works by \citet{melatos2008} and \citet{melatos2010}, by incorporating a more general {equation of state} (characterized by $v_\mathrm{c}$) and {stratification length} (characterized by $\mathsf{K_s}$) in sections \ref{section:equationmain} and \ref{section:solution}. We derived the expected time-scales of emission of gravitational wave signals and the corresponding strain amplitudes from mass-quadrupole and current-quadrupole formalisms in section \ref{section:emission}. In order to better visualize the results, we explored the properties of emission in $\mathsf{N}^2$ and $v_\mathrm{c}^\mathrm{o}$ parameter space by making some simplifying approximations given by \eqref{eq:partialeta2} in section \ref{section:vtime}. The results are shown in figure \ref{hplot}, where we find that it is possible for such a hydrodynamic system to emit gravitational waves at a ground-based detector with a strain amplitude greater than $O(10^{-25})$ for a source at a distance of roughly 1 kpc. The corresponding time-scales for the loudest signals are as long as $O(300)$ days, also shown in figure \ref{hplot}. The results in figure \ref{hplot} are explored for favorable values of physical parameters, such as at glitch magnitude $\epsilon = O(10^{-4})$, $d_\mathsf{s} = 1\,$kpc and $f = 100\,$Hz. The analysis yields a strain amplitude as high as $O(10^{-21})$ toward lower magnitudes of $\mathsf{N}^2$, i.e. $\mathsf{N}^2 \leq O(10^{-5})$, and $v_\mathrm{c}^\mathrm{o}$ approximately equaling $0.09c$ -- $0.11c$, for the majority of individual amplitudes\footnote{See section {\ref{A.8}} in {{Appendix}} for more details.}. Besides, in broader range of values of $\mathsf{N}^2$ and $v_\mathrm{c}^\mathrm{o}$ different from the aforementioned ranges, we expect emission of the order of $O(10^1 - 10^{1.5})$ days in duration with amplitudes in the range of $O(10^{-23.5} - 10^{-26.5})$. It must be noted that the current-quadrupole contribution tends to be larger than the corresponding mass-quadrupole contribution to the emitted signal, as shown in section \ref{section:emission} and figure \ref{hplot}. This is largely because of the characteristic amplitude $h_\mathrm{o}^\mathrm{C}$ being larger than $h_\mathrm{o}^\mathrm{M}$ by a factor\footnote{This factor yields a value of the order $O(10^{1} - 10^{2})$ for $\Omega=O(10^2$Hz), assuming $g = O(10^{12}\text{m/sec}^2)$.} of $\displaystyle \frac{2g}{3\Omega c}$. Furthermore, very low values of $\mathsf{N}^2$ (as low as $10^{-6} - 10^{-7}$) are debatable since no physical phenomenon account for such magnitudes of $\mathsf{N}^2$. Note that the `classical' {{Brunt-V{\"a}is{\"a}l{\"a} frequency}} $\mathsf{N}_\mathrm{c}^2$ is expected to lie loosely in the range of (0.01, 1) \citep{melatos2008}. The equivalent magnitude of the lower bound on redefined {{Brunt-V{\"a}is{\"a}l{\"a} frequency}} $\mathsf{N}^2$ is then given by:  $\mathsf{N}^2 \sim \eta_\mathrm{o}\mathsf{N}_\mathrm{c}^2 = 10^{-4}$, for $v_\mathrm{c}^\mathrm{o} = 0.1c$. Thus, very low values of $\mathsf{N}^2$ lie outside the current estimates on equivalent values of $\mathsf{N}_\mathrm{c}^2$. In fact, very loud signals of amplitude $O(10^{-25})$ and higher lie near the lower bound of current estimates on  $\mathsf{N}_\mathrm{c}^2$, roughly in the range $10^{-4} - 10^{-7}$ for $\mathsf{N}^2$. However, the value of {Ekman number} $\mathsf{E}$ is could lie anywhere in the range of $10^{-17} - 10^{-7}$  \citep{melatos2008,melatos2010,largeE1,largeE2,E1e,E2e,E3e,E4e,E1t}, whereas we have based our analysis on the assumption of $\mathsf{E}=10^{-7}$. The time-scales and the corresponding gravitational wave amplitudes depend on $\mathsf{E}$ such that, $\mathsf{E}\downarrow\,\implies t_{\upalpha\gamma}\!\uparrow\,\implies h_\mathrm{R}\!\uparrow$. Thus, for lower values of $\mathsf{E}$, stronger emissions could occur even at higher values of $\mathsf{N}^2$. This effect is shown in figure \ref{fig:E} where we have regenerated parts of figure \ref{hplot} for $\mathsf{E} = 10^{-14}$. Note that since {Ekman number} is directly proportional to the sheer viscosity of the bulk matter \citep{viscosity,largeE1} and inversely proportional to the square of its temperature \citep{E1t}, we expect higher values of E $(10^{-7})$ for colder neutron stars $(\mathrm{T}\sim10^6\,\mathrm{K})$ \citep{E1t}, and vice versa. Thus, in principle, hotter neutron stars should be better candidates for transient gravitational waves than colder neutron stars. However, this is not entirely true since it is expected that hotter and younger neutron stars undergo post-glitch relaxation via crust-core dynamics aided by magnetic field rather than bulk hydrodynamics \citep{crustcore}\citep{crustcore2}. 
\end{multicols}
\begin{multicols}{2}
\subsubsection{Detectability of emitted signals}
\label{detectability}
\begin{figure}[H]
\centering
\includegraphics[width=150mm]{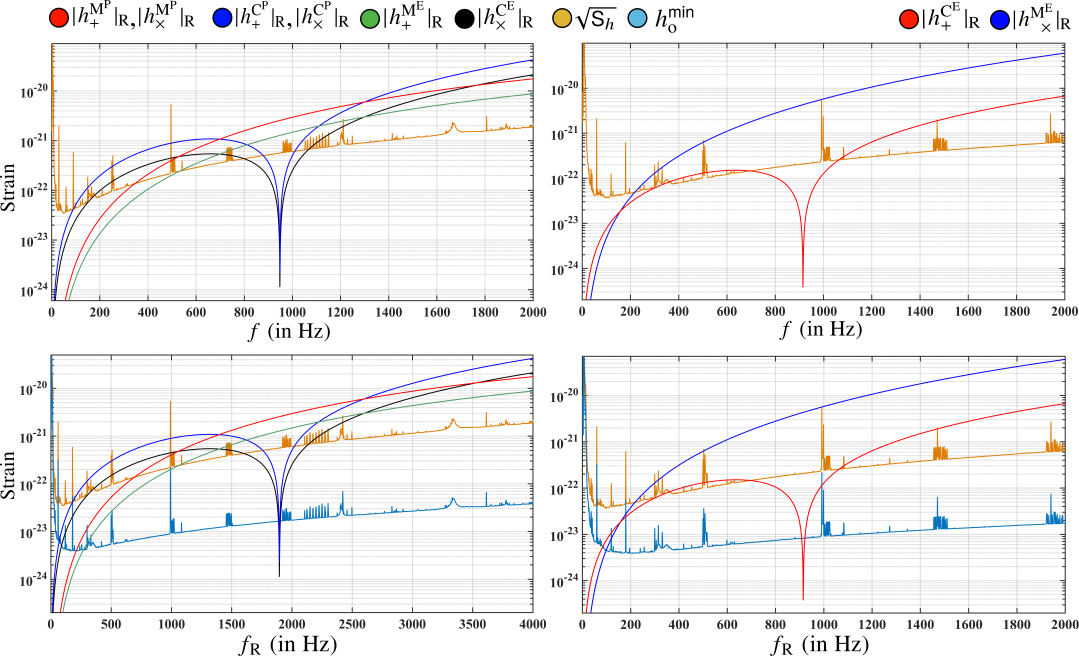} 
\caption[]{{\small Sensitivity vs. $f$:} {{\small The characteristics of emitted signals\textcolor{blue}{{\protect\footnotemark[20]}} $\;$and $h_\mathrm{o}^\mathsf{min}$ are shown as a function of neutron star's rotational frequency $f$ as well as the emitted resonant frequency $f_\mathrm{R}$, given by $f_\mathrm{R}=\omega_\mathrm{R}/2\uppi$. We have set $v_\mathrm{c}^\mathrm{o} = 0.1c$, $\mathsf{N}^2 = 10^{-4}$,  $\mathsf{E}=10^{-10}$, $\epsilon = 10^{-4}$, $d_\mathsf{s}=1\,$kpc, $\mathrm{L}=10^{4}$m, $g=10^{12}\mathrm{m/sec^2}$, $\uprho_\mathrm{o}=10^{17}\mathrm{kg/m^3}$, $\partial_z\hspace{-0.0175in}v_\mathrm{c}=0$. Note that the emitted amplitudes are largely insensitive to $\partial_z\hspace{-0.0175in}v_\mathrm{c}$ for the chosen points in \{$v_\mathrm{c}^\mathrm{o}$, $\mathsf{N}^2$\} parameter space. The {\fontfamily{ppl}\selectfont \textit{multi-detector amplitude spectral density}} $\sqrt{\mathsf{S}_h(\omega)}$ is calculated by taking the harmonic mean of the individual amplitude spectral densities of H1 (aLIGO Hanford) and L1 detectors (aLIGO Livingston) measured during initial days of the O1 run i.e. Sept 12 -- Oct 20, 2015.}}}
\label{fig:Sh}
\end{figure}

We can also derive characteristics of emitted signals as a function of rotational frequency $f$ of the neutron star. It has been shown that the minimum strain amplitude $h_\mathrm{o}^\mathsf{min}$ of a continuous gravitational wave detectable by a network of 2 detectors searched over a large parameter space with a coherent search duration of $\mathrm{T}_\mathsf{obs}$ hours during which the signal is present is given by \citep{Sh, DavidTransients}
\vspace{-0.1in}
\begin{equation}
\begin{multlined}
h_\mathrm{o}^\mathsf{min}(\omega)=\mathrm{K}_t\Bigg[\frac{\mathsf{S}_h(\omega)}{\mathrm{T_\mathsf{obs}}(\omega)}\Bigg]^{\frac{1}{2}},\label{eq:Sh}
\end{multlined}
\end{equation}
where, $\sqrt{\mathsf{S}_h(\omega)}$ is the {\fontfamily{ppl}\selectfont \textit{multi-detector amplitude spectral density}} for a network of 2 detectors (H1, L1), and $\mathrm{K}_t$ is roughly equal to 30. Given this relation, we can compare the strength of emitted gravitational wave signals with the strain detectable by aLIGO. We again restrict ourselves to emission at resonant frequencies only, i.e. $\omega = \omega_\mathrm{R}$. Note that we can express $h_\mathrm{o}^\mathsf{min}$ as a function of $f$ instead of $\omega$ since $ \omega_\mathrm{R}$ is an implicit function of $f$. This allows us to rewrite $h_\mathrm{o}^\mathsf{min}$ as,
\begin{equation}
\begin{multlined}
h_\mathrm{o}^\mathsf{min}(f)\sim30.0\;\Bigg[\frac{\mathsf{S}_h(f)}{t_{\upalpha\gamma}(f)}\Bigg]^{\frac{1}{2}}.\label{eq:Sh1}
\end{multlined}
\end{equation}
where, $t_{\upalpha\gamma}$ is expressed in hours\footnote{Refer to section {\ref{A.7}} in {{Appendix}} for discussion on properties of $t_{\upalpha\gamma}$ as a function of $f$.}. In figure \ref{fig:Sh}, we plot $h_\mathrm{o}^\mathsf{min}(f)$ and compare it with the emitted gravitational wave amplitudes\footnote{Note that the resonant frequencies of emitted modes for mass-quadrupole and current-quadrupole contributions, and for a given orientation of the observer (polar, equatorial, or otherwise), depend on the polarizations ($+$ and $\times$), which in turn depend on the featuring time-scales $t_{\upalpha\gamma}$, as seen in figure \ref{hplot} and section \ref{section:emission}. The overall signal is a superposition of all such individual emissions shown in figure \ref{fig:Sh}, possibly at multiple resonant frequencies for a single source with a given orientation. In this regard, \eqref{eq:Sh1} assumes that these individual emissions are resolvable in frequency; this usually holds true when the featuring time-scales $t_{\upalpha\gamma}$ are not very small (see \ref{section:emission mass}, \ref{section:emission current}).} as a function of $f$. We have set the parameters $\mathsf{E}$, $v_\mathrm{c}^\mathrm{o}$ and $\mathsf{N}^2$ at nominal values of $10^{-10}$, $0.1c$ and $10^{-4}$ respectively. We find that for the selected region in parameter space in \ref{fig:Sh}, it is possible to detect the gravitational wave emission with current aLIGO sensitivity, especially in the mid to high frequency range.

One must carefully note that we have assumed an invariant $\mathsf{N}^2$ in space and time in order to simplify our results for easier graphical visualization and understanding. In principle, one could vary all featuring parameters, i.e. $\mathsf{N}^2$ or $\mathsf{K_s}$, $v_\mathrm{c}$, $v_\mathrm{eq}$, in all possible ways. This is because all analytically derived results in sections \ref{section:equationmain}-\ref{section:emission current} are general in nature and assume none of the approximations described in section \ref{section:vtime}. However, such a thorough and complete analysis will require extensive numerical computations and better priors on the parameter space. More importantly, the main aim of this study was to estimate the strength of the emitted gravitational wave signals and their time-scales as a function of spatial variation in the adiabatic sound speed $v_\mathrm{c}$ and stratification length $z_\mathsf{s}$. This is shown in detail in figure \ref{hplot} and figure \ref{fig:last}. We find that signal characteristics are more sensitive to small spatial variations in $v_\mathrm{c}$ and $\mathsf{K_s}$ in some regions of the parameter space than others. In fact, for these regions in the parameter space, the maximum duration of the emission increases by a factor of $300$ when $\partial_z\hspace{-0.0175in}v_\mathrm{c} = -10^{-4}c\mathrm{L}^{-1}$, as compared to when $\partial_z\hspace{-0.0175in}v_\mathrm{c} = 0$. The corresponding amplitudes also increase by a similar factor, as seen in figure \ref{hplot},\ref{fig:last}. In parts of the parameter space characterized by {growing modes}, no gravitational emission is possible due to hydrodynamic instability.
\end{multicols}
\begin{multicols}{2}
\subsubsection{\textbf{Energetics of the system}}
\label{energetics}
\begin{figure}[H]
\includegraphics[width=77mm]{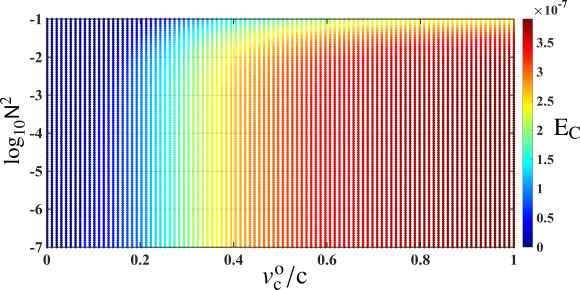}  
\caption{{\small Energetics:} {{\small We have set $\mathsf{E}=10^{-7}$, $\epsilon = 10^{-4}$, $d_\mathsf{s}=1\,$kpc, $\mathrm{L}=10^{4}$m, $g=10^{12}\mathrm{m/sec^2}$, $\uprho_\mathrm{o}=10^{17}\mathrm{kg/m^3}$, and $\partial_z\hspace{-0.0175in}v_\mathrm{c}=0$. We find that similar results, i.e. $\mathsf{E}_\mathsf{C}=O(10^{-7})$, are achieved when we set $0<|\partial_z\hspace{-0.0175in}v_\mathrm{c}|\ll 1$; this is due to the fact that $\kappa_{\upalpha\gamma}$ and $\mathrm{V}_{\upalpha\gamma}$ show very weak dependence on $\partial_z\hspace{-0.0175in}v_\mathrm{c}$ when $|\partial_z\hspace{-0.0175in}v_\mathrm{c}|\ll 1$.}}} 
\label{fig:energy}
\end{figure}

It is an interesting exercise to estimate the fraction of mechanical energy (from the glitch) that gets converted into gravitation wave emission. For instance, the total gravitational wave energy emitted by a waveform $h(t)\propto e^{i\Omega_\mathrm{w} t}e^{-\gamma_\mathrm{w} t}$ is given by \citep{kip,PrixTransients}\\
\begin{equation}
\begin{multlined}
\mathsf{E}_\mathsf{GW} = \frac{c^3}{8\mathrm{G}}[\Omega^2_\mathrm{w}+ \gamma^2_\mathrm{w}] d^2_\mathsf{s}\int^{\infinity}_0 |h(\omega)|^2\mathrm{d}\omega,\label{eq:energywave}
\end{multlined}
\end{equation}
where, we have used Parseval's theorem such that $$\int^{{\infinity}}_0 |h(t)|^2\mathrm{d}t = \frac{1}{2\uppi}\int^{\infinity}_0 |h(\omega)|^2\mathrm{d}\omega.$$
We can easily calculate $\mathsf{E}_\mathsf{GW}$ by integrating (numerically or analytically) the total emitted waveform\footnote{Note the total emission is a sum of the mass-quadrupole and current-quadrupole emission.} over time, or by integrating its Fourier transform in frequency space. Note that the expression \eqref{eq:energywave} assumes an isotropic distribution of signal as a function of the observation angle $i$. In our case, the emission is not isotropically distributed as a function of $i$. In fact, the amplitude for a given polarization varies as a linear combination of sines and cosines of $i$, as briefly discussed in section {\ref{A.4}} in {{Appendix}} \citep{melatos2010}. In order to simplify this to an order-of-magnitude estimate, the total emission can be constrained by an isotropic limit, such that\footnote{This approximation assumes that the amplitude measured by a polar observer is isotropically distributed as a function $i$. This is a reasonable assumption for an order-of-magnitude estimate of emitted energy considering that the observed amplitudes for polar and equatorial observers are of the same order of magnitude, as seen in figure \ref{fig:freqmp},\ref{fig:freqme} and figure \ref{fig:Sh}.}
\begin{equation}
\begin{multlined}
|h(\omega)|^2 \sim 2 \sum_{\mathcal{P}=+,\times}\Big[\sum_{\mathcal{L}=\mathrm{M},\mathrm{C}}|h^{\mathcal{L}^\mathsf{P}}_{\mathcal{P}}(\omega)|\Big]^2.\label{eq:energyapprox}
\end{multlined}
\end{equation}
Combining \eqref{eq:energywave} and \eqref{eq:energyapprox}, we get
{\vspace{-0.025in}}
\begin{equation}
\begin{multlined}
\mathsf{E}_\mathsf{GW} \sim \frac{c^3}{4\mathrm{G}}[\Omega^2_\mathrm{w}+ \gamma^2_\mathrm{w}] d^2_\mathsf{s}\int^{\infinity}_0\hspace{-0.1in} \sum_{\mathcal{P}=+,\times}\hspace{-0.05in}\Big[\sum_{\mathcal{L}=\mathrm{M},\mathrm{C}}\hspace{-0.05in}|h^{\mathcal{L}^\mathsf{P}}_{\mathcal{P}}(\omega)|\Big]^2\mathrm{d}\omega.\label{eq:energywaveapprox}
\end{multlined}
\end{equation} 
On the other hand, the total mechanical energy $\mathsf{E}_\mathsf{glitch}$ imparted by the glitch is written as\footnote{This approximation assumes that only the crust of the neutron star gains angular momentum from the glitch while the bulk fluid is decoupled from the crust at the time of the glitch. Moreover, we also assume that the crust is very thin compared to the radius of the cylinder and it contains only a fraction of the mass [$O(10^{-2})$] of the entire neutron star.}
\begin{equation}
\begin{multlined}
\mathsf{E}_\mathsf{glitch}\sim \Gamma\mathrm{M_{total}}\mathrm{L}^2\Omega_\mathrm{r}\Delta\Omega_\mathrm{r} = 2\uppi\epsilon\Gamma\uprho_\mathrm{o}\mathrm{L}^5\Omega^2_\mathrm{r},\label{eq:glitchenergy}
\end{multlined}
\end{equation}
where, $\Gamma$ is the fraction of total neutron star mass ($\mathrm{M_{total}}\sim2\uppi\uprho_\mathrm{o}\mathrm{L}^3$) contained within the crust; this is assumed to be a small fiducial value of $10^{-2}$. Then, the fraction of mechanical energy $\mathsf{E}_\mathsf{C}$ ($=\mathsf{E}_\mathsf{GW}/\mathsf{E}_\mathsf{glitch}$) converted into gravitational waves is given by
\begin{equation}
\begin{multlined}
\mathsf{E}_\mathsf{C} \sim \frac{c^3}{8\uppi\mathrm{G}} \frac{[\Omega^2_\mathrm{w}+ \gamma^2_\mathrm{w}] d^2_\mathsf{s}}{\epsilon\Gamma\uprho_\mathrm{o}\mathrm{L}^5\Omega^2_\mathrm{r}}\int^{\infinity}_0 \hspace{-0.1in}\sum_{\mathcal{P}=+,\times}\hspace{-0.05in}\Big[\sum_{\mathcal{L}=\mathrm{M},\mathrm{C}}\hspace{-0.05in}|h^{\mathcal{L}^\mathrm{P}}_{\mathcal{P}}(\omega)|\Big]^2\mathrm{d}\omega.\label{eq:fracenergy}
\end{multlined}
\end{equation} 
We find that the ratio $\mathsf{E}_\mathsf{C}$ yields values of the order of $O(10^{-7})$, assuming $\Omega_\mathrm{w} \sim 2\Omega_\mathrm{r}$. This suggests that a large fraction of the energy from the glitch is converted into the kinetic and potential energy of bulk fluid. We also note that the value of $\mathsf{E}_\mathsf{C}$ in the \{$\mathrm{N}^2$, $v^\mathrm{o}_\mathrm{c}$, $\partial_z\hspace{-0.0175in}v_\mathrm{c}$\} parameter space depends only on the pre-factors $\kappa_{\upalpha\gamma}$ and $\mathrm{V}_{\upalpha\gamma}$\footnote{Note that this dependence is generally biased toward $\mathrm{V}_{\upalpha\gamma}$ since the current-quadrupole emission is significantly louder than the mass-quadrupole emission.}. In figure \ref{fig:energy}, we show an example of the characteristics of $\mathsf{E}_\mathsf{C}$.
\subsubsection{\textbf{Choice of equation of state}}
\label{eos}
In \eqref{eq:equationofstate}, we assumed a simple form of the equation of state where the adiabatic speed of sound $v_\mathrm{c}$ is slowly and linearly varying with \textit{z}. It may be useful to compare this choice with a general polytropic equation of state with polytropic index $\mathrm{n}$, polytropic exponent $\gamma=(\mathrm{n}+1)/\mathrm{n}$, and polytropic constant  $\mathrm{K}_\mathrm{p}$, i.e. $p=\mathrm{K}_\mathrm{p}\uprho^{\gamma}$. The adiabatic speed of sound $v_\mathrm{p}$ for the polytropic equation is given by: $v^2_\mathrm{p}=\mathrm{K}_\mathrm{p}\uprho^{1/\mathrm{n}}$. Clearly, our model of equation of state resembles the polytropic model for $\mathrm{n}\rightarrow{\infinity}$. We also know that neutron star interiors are generally well-modeled for values of the polytropic index $\mathrm{n}\in(0.5,1.5)$. Thus, we should inquire whether our assumption of $\mathrm{n}\rightarrow{\infinity}$ is reasonable. In order to answer this question, consider that the gravitational wave emission is almost entirely dominated by the processes occurring at the viscous boundary layer, such as the exchange of fluid across this layer, as discussed in great detail in sections \ref{A.8}, \ref{A.1}, \ref{section:time}. These processes determine the time-scales of relaxation, which in turn determine the peak gravitational wave amplitudes (at resonant frequencies). We also note that the pre-factors $\kappa_{\upalpha\gamma}$ and $\mathrm{V}_{\upalpha\gamma}$ are largely insensitive to these processes, as discussed in section \ref{energetics}. Thus, our choice of equation of state particularly encodes physical processes at the viscous boundary layer. In the vicinity of this viscous boundary layer, i.e. $z\sim1$, any polytropic equation of state can be reduced to linear order in $z$. In this limit, the true form of the equation of state becomes irrelevant. For example, for a $\mathrm{n}=1$ polytrope, 
\begin{equation}
\begin{multlined}
v_\mathrm{p}|_{z\sim1}=\sqrt{\mathrm{K}_\mathrm{p}\uprho|_{z\sim1}} \sim v^{\mathrm{o}}_\mathrm{c} + \partial_z\hspace{-0.0175in}v_\mathrm{c}.\label{eq:polytrope}
\end{multlined}
\end{equation}
In particular, for typical crust density of $\uprho|_{z\sim1}\sim 10^{9}\,\textrm{kg}$ $\textrm{m}^{-3}$, $v^{\mathrm{o}}_\mathrm{c} \sim c$ and $\partial_z\hspace{-0.0175in}v_\mathrm{c} \sim 0$, we find $\mathrm{K}_\mathrm{p} \sim 10^7 \,\textrm{kg}^{-1}\textrm{m}^5\textrm{s}^{-2}$.\\

However, there are certain aspects that we have overlooked, such as the effects of the magnetic field and the superfluid nature of the core. The strong magnetic field in neutron stars affect the crust-core interactions and coupling dynamics of the superfluid \citep{crustcore}\citep{crustcore2}, possibly shortening the duration of the emission, especially in young and hot neutron stars such as the Vela pulsar \citep{vela}. Recent works by \citet{cab} have explored the effect of magnetic field on post-glitch relaxation phase but in a slightly different context. Note that we have assumed that non-axisymmetric modes are equally likely to be excited by the glitch as the axisymmetric ones ($\mathrm{C}_\upalpha=1$). If this assumption does not hold, the gravitational amplitudes should be re-scaled by the same factor. In conclusion, we believe this to be a viable model to predict the expected order of magnitude of the amplitude and duration of the emitted gravitational wave signals from glitching neutron stars that involve relaxation via {Ekman pumping}. It lays down a basic approach to predict the approximate internal state of the neutron star and first-order variations in it, if any such transient signal is detected by gravitational wave detectors from the post-glitch relaxation phase.
\section{Acknowledgments}
\label{acknowledgments}
I would like to extend my hearty thanks to Maria Alessandra Papa at the Max-Planck-Institut f{\"u}r Gravitationphysik for her support throughout, and to Andrew Melatos for his input through the course of this work. This document has LIGO DCC number {\small \qag{P1600091}}.
\begin{center}
$\ast$~$\ast$~$\ast$
\end{center}
\columnbreak
\begin{center}
$\,$
\end{center}
\end{multicols}
\begin{center}
$\,$
\end{center}
\pagebreak
\begin{multicols}{2}
\appendix
\section{Appendix}
\label{Appendix}
\subsection{Time-evolution}
\label{A.1}
\counterwithin{figure}{section}
In this section, we derive the time-evolution of the $\upchi(r,\phi,z,t)$ function. We begin by isolating the time-dependence in $\upchi(r,\phi,z,t)$ by separating the variables as follows,
\begin{equation}
\begin{multlined}
\upchi(r,\phi,z,t)\equiv\bar{\upchi}(r,\phi,z)\mathsf{T}(t).\tag{A.1.1}\label{A.1.1}
\end{multlined}
\end{equation}
In parallel, we take the time-derivative of \eqref{eq:boundary},
\begin{equation}
\begin{multlined}
\partial_t({\delta v_z})|_{z=\pm 1}=\mp\frac{1}{2}\mathsf{E}^\frac{1}{2}\partial_t(\boldsymbol{\nabla}\times\delta\vec{v})_z|_{z=\pm 1}=\\\pm\frac{1}{2}\mathsf{E}^\frac{1}{2}\Bigg[\frac{1}{r}\frac{\partial}{\partial r}(r\partial_t[\delta v_{\phi}])-\frac{1}{r}\frac{\partial}{\partial \phi}(\partial_t[\delta v_r])\Bigg]\Bigg|_{z=\pm 1}.\tag{A.1.2}\label{A.1.2}
\end{multlined}
\end{equation}
Ignoring all the $O(\mathsf{E}^1)$ or higher order terms on the right-hand side in \eqref{A.1.2} and using results from sections \ref{section:sol0}, \ref{section:sol1} and \eqref{eq:main}, we get
\begin{equation}
\begin{multlined}
\partial_t[{\delta v_z^{(1)}}]|_{z=\pm 1}=\mp\frac{1}{4\mathsf{F}}\Bigg[\frac{1}{r^2}\frac{\partial^2\upchi}{\partial \phi^2}+\frac{1}{r}\frac{\partial}{\partial r}(r\frac{\partial\upchi}{\partial r})\Bigg]\Bigg|_{z=\pm 1}\\=\pm\frac{1}{4\mathsf{F}}\uplambda_{\upalpha\gamma}^2\upchi|_{z=\pm 1}.\;\;\;\;\;\;\tag{A.1.3}\label{A.1.3}
\end{multlined}
\end{equation}
Moreover, we use \eqref{eq:1vz} to further simplify \eqref{A.1.3} as
\begin{equation}
\begin{multlined}
\Bigg[\frac{\eta(z)}{\mathsf{F}\mathsf{N}^2(z)}\frac{\partial\dt{\upchi}}{\partial z}+\Bigg\{\frac{-\partial_z\eta}{\mathsf{F}\mathsf{N}^2(z)}-1\Bigg\}\dt{\upchi}\Bigg]\Bigg|_{z=\pm 1}=\\\mp\frac{1}{4\mathsf{F}}\uplambda_{\upalpha\gamma}^2\upchi|_{z=\pm 1}.\tag{A.1.4}\label{A.1.4}
\end{multlined}
\end{equation}
Now, introducing the separation of variables from \eqref{A.1.1}, we re-write the above equation \eqref{A.1.4} as
\begin{equation}
\begin{multlined}
\Bigg[\frac{\eta(z)}{\mathsf{F}\mathsf{N}^2(z)}\frac{\partial{\bar{\upchi}}}{\partial z}+\Bigg\{\frac{-\partial_z\eta}{\mathsf{F}\mathsf{N}^2(z)}-1\Bigg\}\bar{\upchi}\Bigg]\Bigg|_{z=\pm 1}\dt{\mathsf{T}}(t)\\=\mp\frac{1}{4\mathsf{F}}\uplambda_{\upalpha\gamma}^2\bar{\upchi}|_{z=\pm 1}\mathsf{T}(t).\tag{A.1.5}\label{A.1.5}
\end{multlined}
\end{equation}
We further reduce the previous expression \eqref{A.1.5} by separating the variables into $(r,\phi)$ and $z$ to
\begin{equation}
\begin{multlined}
\Bigg[\frac{\eta(\pm1)}{\mathsf{F}\mathsf{N}^2(\pm1)}\frac{\partial\mathrm{Z}_{\upalpha\gamma}}{\partial z}\Bigg|_{z=\pm 1}+\Bigg\{\frac{-\partial_z\eta|_{z=\pm1}}{\mathsf{F}\mathsf{N}^2(\pm 1)}-1\Bigg\}\times\\\mathrm{Z}_{\upalpha\gamma}(\pm 1)\Bigg]\dt{\mathsf{T}}(t)=\mp\frac{1}{4\mathsf{F}}\uplambda_{\upalpha\gamma}^2\mathrm{Z}_{\upalpha\gamma}(\pm 1)\mathsf{T}(t).\tag{A.1.6}\label{A.1.6}
\end{multlined}
\end{equation}
The equation \eqref{A.1.6} can now be solved to yield $\mathrm{T}(t)$ as,
\begin{equation}
\begin{multlined}
\mathsf{T}(t)\propto e^{-\omega_{\upalpha\gamma}t},\tag{A.1.7}\label{A.1.7}
\end{multlined}
\end{equation}
such that $\omega_{\upalpha\gamma}$ is given by
\begin{equation}
\begin{multlined}
\omega_{\upalpha\gamma}=\frac{1}{4\mathsf{F}}\uplambda_{\upalpha\gamma}^2\mathrm{Z}_{\upalpha\gamma}(1)\Bigg[\frac{\eta(1)}{\mathsf{F}\mathsf{N}^2(1)}\frac{\partial\mathrm{Z}_{\upalpha\gamma}}{\partial z}\Bigg|_{z=1}+\\\Bigg\{\frac{-\partial_z\eta|_{z=1}}{\mathsf{F}\mathsf{N}^2(1)}-1\Bigg\}\mathrm{Z}_{\upalpha\gamma}(1)\Bigg]^{-1}.\tag{A.1.8}\label{A.1.8}
\end{multlined}
\end{equation}\columnbreak 
Note that $\mathrm{Z}_{\upalpha\gamma}(z)$ is symmetric about $z=0$ plane and we have evaluated the expression at $z=1$.
\subsection{Bessel-Fourier Coefficients}
\label{A.2}
In this section, we will calculate the {Bessel-Fourier coefficients} introduced in \eqref{eq:final}. We use the {{orthogonality property}} of the Bessel functions, which states that Bessel functions are orthogonal with respect to the inner product as follows\footnote{$\delta_{\gamma\upalpha}$ is the {{Dirac-delta function}}.},
\begin{equation}
\begin{multlined}
\langle\mathsf{J}_\upalpha(\uplambda_{\upalpha\gamma}r),\mathsf{J}_\upalpha(\uplambda_{\upalpha\upalpha}r)\rangle=\int_0^1r\,\mathsf{J}_\upalpha(\uplambda_{\upalpha\gamma}r)\mathsf{J}_\upalpha(\uplambda_{\upalpha\upalpha}r)\mathrm{d}r=\\\frac{1}{2}\delta_{\gamma\upalpha}[\mathsf{J}_{\upalpha+1}(\uplambda_{\upalpha\gamma}r)]^2.\;\;\;\;\tag{A.2.1}\label{A.2.1}
\end{multlined}
\end{equation}
For a Fourier-Bessel series of the form $f(r)=\sum_{\upalpha=1}^{\infinity}\mathrm{C}_\upalpha\mathsf{J}_\upalpha(\uplambda_{\upalpha\upalpha}r)$, the coefficients $\mathrm{C}_\upalpha$ can be calculated by taking projection of the function $f(r)$ over the corresponding Bessel functions as,
\begin{equation}
\mathrm{C}_\upalpha=\frac{\langle f(r),\mathsf{J}_\upalpha(\uplambda_{\upalpha\upalpha}r)\rangle}{\langle\mathsf{J}_\upalpha(\uplambda_{\upalpha\gamma}r),\mathsf{J}_\upalpha(\uplambda_{\upalpha\gamma}r)\rangle}.\tag{A.2.2}\label{A.2.2}
\end{equation}
Using the above relation in combination with \eqref{eq:P0-P8}, we can substitute for $f(r)$,
\begin{equation}
\begin{multlined}
f(r)=\delta\mathsf{P}_0-\delta\mathsf{P}_{\infinity}=\sum_{\upalpha=0}^{\infinity}\sum_{\gamma=1}^{\infinity}
\omega_{\upalpha\gamma}^{-1}\mathsf{J}_\upalpha(\uplambda_{\upalpha\gamma}r)\times\\ [\mathrm{A}_{\upalpha\gamma}\ccos(\upalpha\phi)+\mathrm{B}_{\upalpha\gamma}\csin(\upalpha\phi)]\hspace{0.01in}\mathrm{Z}_{\upalpha\gamma}(z)=\\\Bigg[\sum_{\upalpha=0}^{\infinity}\mathrm{C}_\upalpha r^\upalpha(r^2-1)\,\ccos(\upalpha\phi)\mathrm{Z}_{\upalpha\gamma}(z)\Bigg] - r^2,\tag{A.2.3}\label{A.2.3}
\end{multlined}
\end{equation}
which, when applied to \eqref{A.2.2}, gives
\begin{equation}
\begin{multlined}
\omega^{-1}_{\upalpha\gamma}\mathrm{A}_{\upalpha\gamma}\ccos(\upalpha\phi)\mathrm{Z}_{\upalpha\gamma}(z)=\frac{2}{\mathsf{J}^2_{\upalpha+1}(\uplambda_{\upalpha\gamma})}\int_{0}^{1}r\times\\\mathsf{J}_
\upalpha(\uplambda_{\upalpha\gamma}r)[\delta\mathsf{P}_0-\delta\mathsf{P}_{\infinity}]\,\mathrm{d}r.\tag{A.2.4}\label{A.2.4}
\end{multlined}
\end{equation}
We multiply both sides with $\ccos(\upalpha\phi)$ and integrate the resulting expression in $\phi$ and $z$ variables assuming that $\mathrm{A}_{\upalpha\gamma}$ is an absolute constant, and arrive at the following result:
\begin{equation}
\begin{multlined}
\mathrm{A}_{\upalpha\gamma}=\frac{2\omega_{\upalpha\gamma}}{\uppi\mathsf{J}^2_{\upalpha+1}(\uplambda_{\upalpha\gamma})}\int_0^{2\uppi}\mathrm{d}\phi\int_0^1\mathrm{d}z\int_{0}^{1}r\,\mathrm{d}r\times\\\mathsf{J}_
\upalpha(\uplambda_{\upalpha\gamma}r)\,\ccos(\upalpha\phi)\,[\delta\mathsf{P}_0-\delta\mathsf{P}_{\infinity}]\,\mathrm{Z}^{-1}_{\upalpha\gamma}(z).\;\;\;\;\;\tag{A.2.5}\label{eq:A.2.5}
\end{multlined}
\end{equation}
Similarly, for $\mathrm{B}_{\upalpha\gamma}$,
\begin{equation}
\begin{multlined}
\mathrm{B}_{\upalpha\gamma}=\frac{2\omega_{\upalpha\gamma}}{\uppi\mathsf{J}^2_{\upalpha+1}(\uplambda_{\upalpha\gamma})}\int_0^{2\uppi}\mathrm{d}\phi\int_0^1\mathrm{d}z\int_{0}^{1}r\,\mathrm{d}r\times\\\mathsf{J}_
\upalpha(\uplambda_{\upalpha\gamma}r)\,\csin(\upalpha\phi)\,[\delta\mathsf{P}_0-\delta\mathsf{P}_{\infinity}]\,\mathrm{Z}^{-1}_{\upalpha\gamma}(z).\;\;\;\;\;\tag{A.2.6}\label{eq:A.2.6}
\end{multlined}
\end{equation}
\subsection{Quadrupole moment formalism}
\label{A.3}
In this section, we will underline the formalism for calculating the expressions \eqref{eq:plusp}-\eqref{eq:fcrosse} for gravitational wave emission. In the reference frame of a {polar observer} at a distance $d$, the components of the gravitational wave strain in Einstein's quadrupole moment formalism in the transverse traceless gauge (abbreviated as `TT') are given by
\begin{equation}
\begin{multlined}
h_{+}(t)=h^{\mathsf{TT}}_{xx}(t)=-h^{\mathsf{TT}}_{yy}(t)=\frac{\mathrm{G}}{c^4d}[\,\ddt{\mathrm{I}}_{xx}(t)-\ddt{\mathrm{I}}_{yy}(t)],\tag{A.3.1}\label{A.3.1}
\end{multlined}
\end{equation}
\begin{equation}
\begin{multlined}
h_{\times}(t)=h^{\mathsf{TT}}_{xy}(t)=\frac{\mathrm{2G}}{c^4d}\ddt{\mathrm{I}}_{xy}(t),\tag{A.3.2}\label{A.3.2}
\end{multlined}
\end{equation}
where, ${\mathrm{I}}_{ik}(t)$ is the reduced quadrupole moment of inertia, and it is given in terms of stress-energy tensor component $\mathsf{T}^{00}$ by,
\begin{equation}
\begin{multlined}
{\mathrm{I}}_{ik}(t)=\frac{1}{c^2}\int\mathrm{d}^3\vec{x}\Bigg[x_i x_k-\delta_{ik}\frac{|\vec{x}|^2}{3}\Bigg]\mathsf{T}^{00}(\vec{x},t).\tag{A.3.3}\label{A.3.3}
\end{multlined}
\end{equation}
Combining \eqref{A.3.1}, \eqref{A.3.2} and \eqref{A.3.3}, we get
\begin{equation}
\begin{multlined}
h^\mathsf{P}_{+}(t)=\frac{\mathrm{G}}{c^6d}\int\mathrm{d}^3\vec{x}\,[x^2-y^2]\ddt{\mathsf{T}}^{\,00}(\vec{x},t)=\\ \frac{\mathrm{G}}{c^6d}\int\mathrm{d}^3\vec{r}\,r^2\ccos(2\phi)\ddt{\mathsf{T}}^{\,00}_\mathsf{NA}(\vec{r},t),\!\!\!\!\!\! \tag{A.3.4}\label{A.3.4}
\end{multlined}
\end{equation}
\begin{equation}
\begin{multlined}
h^\mathsf{P}_{\times}(t)=\frac{\mathrm{2G}}{c^6d}\int\mathrm{d}^3\vec{x}\,[xy]\ddt{\mathsf{T}}^{\,00}(\vec{x},t)=\\\frac{\mathrm{G}}{c^6d}\int\mathrm{d}^3\vec{r}\,r^2\csin(2\phi)\ddt{\mathsf{T}}^{\,00}_\mathsf{NA}(\vec{x},t),\!\!\!\!\!\!\!\!\!\!\!\!\!\!\!\!\!\!\tag{A.3.5}\label{A.3.5}
\end{multlined}
\end{equation}
where, the sub-script $\mathsf{NA}$ refers to non-axisymmetric terms. Moreover, in case of a perfect fluid,  we neglect the viscous terms while evaluating $\mathsf{T}^{\upmu\upnu}$ since they are of the order $O(\mathsf{E})$, and the stress-energy tensor component ${\mathsf{T}}^{\,00}$ is then given by
\begin{equation}
\begin{multlined}
{\mathsf{T}}^{\,00}=\Bigg[\uprho+\frac{p}{c^2}\Bigg]u^0u^0+pg^{00},\tag{A.3.6}\label{A.3.6}
\end{multlined}
\end{equation}
where, the 0-component $u^0$ of the 4-velocity $\vec{u}$ is given by
\begin{equation}
\begin{multlined}
u^0=\frac{c}{\sqrt{1-\frac{\displaystyle\vec{v}\cdot\vec{v}}{\displaystyle c^2}}}.\tag{A.3.7}\label{A.3.7}
\end{multlined}
\end{equation}
We break the expression \eqref{A.3.6} into separate terms describing the  constitutive equilibrium and perturbative terms, i.e. $\uprho\rightarrow\uprho_e+\epsilon\delta\uprho$, $p\rightarrow p_e+\epsilon\delta\hspace{-0.01in}p$ and $\vec{v}\rightarrow\vec{v}_r+\delta\vec{v}$, as described in section \ref{section:spinup}. Here, $\vec{v}_r$ is simply the velocity of a fluid element given in cylindrical coordinates by $\vec{v}_r=(0,\Omega r, 0)$, assuming co-rotation with the neutron star crust. Note that equilibrium state is axisymmetric in nature and doesn't contribute to the signal emission. The contributing non-axisymmetric terms in $\mathsf{T}^{00}$ are then given by\footnote{Here, we have assumed $g^{00}=-1$ and $|v^2|\ll c^2$.}
\begin{equation}
\begin{multlined}
{\mathsf{T}}^{\,00}_\mathsf{NA}=\epsilon\delta\uprho c^2+(\uprho_e c^2+p_e)\Bigg[2\frac{\delta\vec{v}\cdot\vec{v}_r}{c^2} +\frac{\delta\vec{v}\cdot\delta\vec{v}}{c^2}\Bigg]+\\ \epsilon(\delta\uprho c^2+\delta\hspace{-0.01in}p)\Bigg[2\frac{\delta\vec{v}\cdot\vec{v}_r}{c^2}+ \frac{\vec{v}_r\cdot\vec{v}_r}{c^2}+\frac{\delta\vec{v}\cdot\delta\vec{v}}{c^2}\Bigg]\sim\!\!\!\!\!\!\!\!\!\!\!\!\!\!\!\!\!\\\epsilon\delta\uprho c^2+(\uprho_e c^2+p_e)\Bigg[2\frac{\delta\vec{v}\cdot\vec{v}_r}{c^2}+\frac{\delta\vec{v}\cdot\delta\vec{v}}{c^2}\Bigg]+\indent\\(\epsilon\delta\uprho c^2+\delta\hspace{-0.01in}p)\Bigg[ \frac{\vec{v}_r\cdot\vec{v}_r}{c^2}\Bigg].\tag{A.3.8}\label{A.3.8}
\end{multlined}
\end{equation}
Note that there exists no explicit factor of $\epsilon$ when it comes to $\delta\vec{v}$, as discussed previously in section \ref{section:spinup}. The factor of $\epsilon$ in the order of magnitude of $\delta\vec{v}$ is implicitly contained within $\delta\vec{v}$. Further, combining the expressions \eqref{A.3.4}, \eqref{A.3.5} and \eqref{A.3.8}, we calculate the gravitational wave emission up to the order $O(\epsilon^1)$ given by \eqref{eq:plusp}-\eqref{eq:fcrosse}.
\vspace{-0.1in} 
\subsection{Current-quadrupole moment}
\label{A.4}
In this section, we briefly describe the method to derive strain amplitude for the current-quadrupole contribution quoted in \eqref{eq:plusc}-\eqref{eq:h0c}. We follow \citep{kip,current,melatos2010}, and make appropriate modifications corresponding to our assumption of spatially varying stratification length and adiabatic sound speed. The general expression for the $+$ and $\times$ polarizations contributed by the current-quadrupole moment (labeled by the super-script $\mathrm{C}$) for a general observer at distance $d$ is given by \citep{melatos2010,current,kip},
\begin{equation}
\begin{multlined}
h^\mathrm{C}_{+}(t)=\frac{\mathrm{G}}{2c^5d}\Bigg[\frac{5}{2\uppi}\Bigg]^{\frac{1}{2}}\big[\mathrm{Im}\{\ddt{\mathrm{C}}^{21}(t)\}\csin(i)+\\\mathrm{Im}\{\ddt{\mathrm{C}}^{22}(t)\}\ccos(i)\big],\tag{A.4.1}\label{A.4.1}
\end{multlined}
\end{equation} 
\vspace{-0.1in} 
\begin{equation}
\begin{multlined}
\;h^\mathrm{C}_{\times}(t)=\frac{\mathrm{G}}{4c^5d}\Bigg[\frac{5}{2\uppi}\Bigg]^{\frac{1}{2}}\big[\mathrm{Re}\{\ddt{\mathrm{C}}^{21}(t)\}\csin(2i)+\\\hspace{-0.3in}\mathrm{Re}\{\ddt{\mathrm{C}}^{22}(t)\}\;[1+\ccos^2(i)]\big],\tag{A.4.2}\label{A.4.2}
\end{multlined}
\end{equation} 
where, $\mathrm{C}^{l\upnu}(t)$ are the $(l,\upnu)$-multipoles of the mass-current distribution. Note that we have only considered the leading order quadrupole moment $(l=2)$, which is the lowest multipole moment that contributes to the gravitational wave emission via its non-vanishing second-order time-derivative $\ddt{\mathrm{C}}^{2\upnu}(t)$. The presence of additional $c^{5}$ factor, as opposed to $c^{4}$ in case of the mass-quadropole moment, suggests that the current-quadrupole contribution is much smaller than the mass-quadrupole moment. This is true for systems with low density. However, for high-density systems such as a neutron star, current-quadrupole emission may be larger than mass-quadrupole contribution, as described in section \ref{section:emission current}. We have also ignored the $\upnu=0$ mode which contributes at the order of $O(\mathsf{E}^1)$ while retaining the more significant $\upnu=1,2$ modes. Moreover, $i$ denotes the angle between neutron star's rotation axis and the observer's line of sight, such that $i=0$ for a polar observer, and $i=90^\circ$ for an equatorial observer. The $\mathrm{C}^{2\upnu}(t)$ terms are explicitly given by \citep{melatos2010}\footnote{In case of current-quadrupole contribution, it is possible to have continuous emission of gravitational waves at large time-scales, $t \gg t_{2\upnu}$, as shown by \citet{melatos2010}. This continuous residual emission is not artificial (cf. \citet{melatos2008}). In calculating the expression for $\mathrm{C}^{2\upnu}(t)$, we have ignored terms responsible for this residual continuous contribution since we concern ourselves solely with transient emission.}
\begin{equation}
\begin{multlined}
\mathrm{C}^{2\upnu}(t)=\frac{(-1)^{\upnu+1}8\uppi(10\uppi)^\frac{1}{2}}{15\upnu\,\uprho^{-1}_{\mathrm{o}}\mathrm{L}^{-6}(\delta\Omega)^{-1}}\sum_{\gamma=1}^{\infinity}\mathsf{V}_{\upnu\gamma}e^{-(\mathsf{E}^\frac{1}{2}\omega_{\upnu\gamma}+i\upnu)\Omega t},\tag{A.4.3}\label{A.4.3}
\end{multlined}
\end{equation}
where,
\begin{equation}
\begin{multlined}
\mathsf{V}_{\upnu\gamma}={2}\mathrm{A}_{\upnu\gamma}\omega^{-1}_{\upnu\gamma}\int_0^\mathrm{1}\mathrm{d}r\int_0^1\mathrm{d}z\,r^{\upnu+1}z^{2-\upnu}\times\\\hspace{0.75in}\hat{\mathbf{U}}\Bigg[
\mathsf{J}_\upnu({\uplambda_{\upnu\gamma}r})\mathrm{Z}_{\upnu\gamma}(z)\uprho_e(z)\Bigg].\tag{A.4.4}\label{A.4.4}
\end{multlined}
\end{equation}
Moreover, the operator $\hat{\mathbf{U}}$ is written as
\begin{equation}
\begin{multlined}
\hat{\mathbf{U}}=\Bigg[z\frac{\partial^2}{\partial r^2}+\frac{z}{r}\frac{\partial}{\partial r}-z\frac{\upnu^2}{r^2}-r\frac{\partial^2}{\partial r\partial z}\Bigg] +\\\hspace{1in} 2\mathsf{F}\Bigg[r^2\frac{\partial^2}{\partial z^2}-rz\frac{\partial^2}{\partial r\partial z}-2z\frac{\partial}{\partial z}\Bigg]\tag{A.4.5}.\label{A.4.5}
\end{multlined}
\end{equation}
Finally, the expressions for $+$ and $\times$ polarizations can now be reduced using the above relations to the expressions quoted in \eqref{eq:plusc}-\eqref{eq:h0c}.
\subsection{\boldmath{$\kappa_{\upnu\gamma}\;\text{and}\;\mathsf{V}_{\upnu\gamma}$}}
\label{A.5}
In this section, we quote the full expression of $\kappa_{\upnu\gamma}$\footnote{The pre-factor of 2 in $\kappa_{\upnu\gamma}$ comes from extending the symmetric integral to $z\in[-1,1]$.}.
\end{multicols}
\begin{equation}
\begin{multlined}
\kappa_{\upnu\gamma}={2}\omega^{-1}_{\upnu\gamma}{\mathrm{A}_{\upnu\gamma}}
\Bigg[\int_0^{1}\mathrm{d}r\,r^3\mathsf{J}_\upnu({\uplambda_{\upnu\gamma}r})\int_0^1\mathrm{d}z\,\partial_z[-\mathrm{Z}_{\upnu\gamma}(z)\uprho_e(z)]+\mathsf{K}\int_0^1\mathrm{d}r\,r^4
\partial_r[\mathsf{J}_\upnu({\uplambda_{\upnu\gamma}r})]\times\\\int_0^\mathrm{L}\mathrm{d}z\,\Bigg[1+\frac{\mathsf{K}}{\mathsf{K_s}(z)}\Bigg]\mathrm{Z}_{\upnu\gamma}(z) \uprho_e(z)+\frac{\Omega^2\mathrm{L}^2}{c^2}\int_0^1\mathrm{d}r\,r^5
\mathsf{J}_\upnu({\uplambda_{\upnu\gamma}r})\times\\\int_0^1\mathrm{d}z\,\Big[\partial_z[-\mathrm{Z}_{\upnu\gamma}(z)\uprho_e(z)]+\mathsf{K}\,\mathrm{Z}_{\upnu\gamma}(z)\uprho_e(z)\Big]\Bigg]\tag{A.5.1}\label{eq:kappa}
\end{multlined}
\end{equation}
\begin{multicols}{2}
\noindent Moreover, following the assumptions described in \eqref{eq:partialeta2} in section \ref{section:vtime}, the above expression for $\kappa_{\upnu\gamma}$ can be further reduced to a simpler and easier form. The simplifying assumptions lead to the case where all coefficients in \eqref{eq:Z} become effectively invariant with respect to the \textit{z}-coordinate. This leaves the solution for $\mathrm{Z}_{\upnu\gamma}(z)$ straightforward to achieve. Moreover, the integrals in the exponents involving $\mathsf{K_s}$ in \eqref{eq:kappa} are dissolved, and the resulting exponential terms can be folded into $\mathrm{Z}_{\upnu\gamma}(z)$ to yield
\end{multicols}
\begin{equation}
\begin{multlined}
\kappa_{\upnu\gamma}={2}\omega^{-1}_{\upnu\gamma}{\mathrm{A}_{\upnu\gamma}}
\Bigg[\mathcal{L}_1\int_0^{1}\mathrm{d}r\,r^3\mathsf{J}_\upnu({\uplambda_{\upnu\gamma}r})+\mathsf{K}\Bigg[1+\frac{\mathsf{K}}{\mathsf{K_s}}\Bigg]\mathcal{L}_2\int_0^1\mathrm{d}r\,r^4
\partial_r[\mathsf{J}_\upnu({\uplambda_{\upnu\gamma}r})]+\\\frac{\Omega^2\mathrm{L}^2}{c^2}\Bigg[\mathcal{L}_1 + \mathsf{K}
\mathcal{L}_2\Bigg]\int_0^1\mathrm{d}r\,r^5
\mathsf{J}_\upnu({\uplambda_{\upnu\gamma}r})\Bigg],\tag{A.5.2}\label{eq:kappar}
\end{multlined}
\end{equation}
\vspace{0.01in}
\begin{multicols}{2}
\noindent where, $\mathcal{L}_1$ and $\mathcal{L}_2$ are given in terms of $\mathcal{K}_{\pm}$ by
\begin{equation}
\begin{multlined}
\mathcal{L}_{1}=\frac{(\mathsf{FN}^2-\mathcal{K}_{-})[1-e^{-\mathcal{K}_{-}}]
-(\mathsf{FN}^2-\mathcal{K}_{+})[1-e^{-\mathcal{K}_{+}}]}{(\mathsf{FN}^2-\mathcal{K}_{-})e^{\mathcal{K}_{+}}-(\mathsf{FN}^2-\mathcal{K}_{+})
e^{\mathcal{K}_{-}}},\tag{A.5.3}\label{eq:L1}
\end{multlined}
\end{equation}
\begin{equation}
\begin{multlined}
\;\\
\mathcal{L}_{2}=\frac{(\mathsf{FN}^2-\mathcal{K}_{-})\frac{\displaystyle 1 \displaystyle -\displaystyle e^{-\mathcal{K}_{-}}}{\mathcal{K}_{-}}
-(\mathsf{FN}^2-\mathcal{K}_{+})\frac{ \displaystyle 1 \displaystyle - \displaystyle e^{-\mathcal{K}_{+}}}{\mathcal{K}_{+}}}{(\mathsf{FN}^2-\mathcal{K}_{-})e^{\mathcal{K}_{+}}-(\mathsf{FN}^2-\mathcal{K}_{+})
e^{\mathcal{K}_{-}}}.\tag{A.5.4}\label{eq:L2}
\end{multlined}
\end{equation}
Further, $\mathcal{K}_{\pm}$ in \eqref{eq:L1} and \eqref{eq:L2} is given by
\begin{equation}
\begin{multlined}
\mathcal{K}_{\pm}=\frac{1}{2}\big[\mathsf{K_s}\pm\big(\mathsf{K}^2_{\mathsf{s}}+\eta_{\mathrm{o}}^{-1}[\mathsf{N}^2\uplambda^2_{\upnu
\gamma}+\partial_z\eta-\partial^2_z\eta]\big)^\frac{1}{2}\big],\tag{A.5.5}\label{eq:D}
\end{multlined}
\end{equation}
where, $\eta_{\mathrm{o}} = (v^\mathrm{o}_\mathrm{c}/c)^2$, $\partial_z\eta \sim 2v^\mathrm{o}_\mathrm{c}\partial_z\hspace{-0.0175in}v_\mathrm{c}$ and $\partial^2_z\eta \sim 2(\partial_z\hspace{-0.0175in}v_\mathrm{c})^2$. Similarly, we calculate the reduced expression for $\mathsf{V}_{\upnu\gamma}$ in terms of $\mathcal{L}^{(\mathrm{g})}_{3}$, $\mathcal{L}^{(\mathrm{g})}_{4}$ and $\mathcal{L}^{(\mathrm{g})}_{5}$. We define $\mathcal{L}^{(\mathrm{g})}_{3}$, $\mathcal{L}^{(\mathrm{g})}_{4}$ and $\mathcal{L}^{(\mathrm{g})}_{5}$ as follows:
\begin{equation}
\begin{multlined}
\mathcal{L}^{(\mathrm{g})}_{3}=\frac{(\mathsf{FN}^2-\mathcal{K}_{-})\mathcal{H}_{\mathrm{g}}(\mathcal{K}_{-})
-(\mathsf{FN}^2-\mathcal{K}_{+})\mathcal{H}_{\mathrm{g}}(\mathcal{K}_{-})}{(\mathsf{FN}^2-\mathcal{K}_{-})e^{\mathcal{K}_{+}}-(\mathsf{FN}^2-\mathcal{K}_{+})
e^{\mathcal{K}_{-}}},\tag{A.5.6}\label{eq:L3}
\end{multlined}
\end{equation}
\begin{equation}
\begin{multlined}
\mathcal{L}^{(\mathrm{g})}_{4}=\frac{(\mathsf{FN}^2-\mathcal{K}_{-}){\displaystyle\frac{\mathcal{H}_{\mathrm{g}}(\mathcal{K}_{-})}{\mathcal{K}^{-1}_{-}}}
-(\mathsf{FN}^2-\mathcal{K}_{+}){\displaystyle\frac{\mathcal{H}_{\mathrm{g}}(\mathcal{K}_{+})}{\mathcal{K}^{-1}_{+}}}}{(\mathsf{FN}^2-\mathcal{K}_{-})e^{\mathcal{K}_{+}}-(\mathsf{FN}^2-\mathcal{K}_{+})
e^{\mathcal{K}_{-}}},\tag{A.5.7}\label{eq:L4}
\end{multlined}
\end{equation}
\begin{equation}
\begin{multlined}
\mathcal{L}^{(\mathrm{g})}_{5}=\frac{(\mathsf{FN}^2-\mathcal{K}_{-}){\displaystyle\frac{\mathcal{H}_{\mathrm{g}}(\mathcal{K}_{-})}{\mathcal{K}^{-2}_{-}}}
-(\mathsf{FN}^2-\mathcal{K}_{+}){\displaystyle\frac{\mathcal{H}_{\mathrm{g}}(\mathcal{K}_{+})}{\mathcal{K}^{-2}_{+}}}}{(\mathsf{FN}^2-\mathcal{K}_{-})e^{\mathcal{K}_{+}}-(\mathsf{FN}^2-\mathcal{K}_{+})
e^{\mathcal{K}_{-}}},\tag{A.5.8}\label{eq:L5}
\end{multlined}
\end{equation}\newpage
{\noindent}where, $\mathcal{H}_\mathrm{g}(\mathcal{K}_{-})$ is defined by the integral given below\footnote{Note that the occurrences of $(\mathrm{g})$ in expressions of $\mathcal{L}^{(\mathrm{g})}_{3}$,  $\mathcal{L}^{(\mathrm{g})}_{3}$ and $\mathcal{L}^{(\mathrm{g})}_{3}$ are intended as {super-scripts} and {not exponents}.}
\begin{equation}
\begin{multlined}
\mathcal{H}_\mathrm{g}(\mathcal{K}_{\pm}) = \int_0^1\mathrm{d}z\,z^{\mathrm{g}}e^{-\mathcal{K}_{\pm}z}.\tag{A.5.9}\label{eq:gamma}
\end{multlined}
\end{equation}
The resulting complete expression for $\mathsf{V}_{\upnu\gamma}$ is then expanded and written in terms of $\mathcal{L}^{(\mathrm{g})}_{3}$, $\mathcal{L}^{(\mathrm{g})}_{4}$ and $\mathcal{L}^{(\mathrm{g})}_{5}$ as follows:
\end{multicols}
\begin{equation}
\begin{multlined}
\mathsf{V}_{\upnu\gamma} ={2}\mathrm{A}_{\upnu\gamma}\omega^{-1}_{\upnu\gamma}\Bigg[\mathcal{L}^{(3-\upnu)}_{3}\int_0^\mathrm{1}\mathrm{d}r\,r^{\upnu-1}
\big[r^2\partial^2_r[\mathsf{J}_\upnu({\uplambda_{\upnu\gamma}r})]+r\partial_r[\mathsf{J}_\upnu({\uplambda_{\upnu\gamma}r})]
-\upnu^2\mathsf{J}_\upnu({\uplambda_{\upnu\gamma}r})\big] + \mathcal{L}^{(2-\upnu)}_{4}\times\\\int_0^\mathrm{1}\mathrm{d}r\,r^{\upnu+2}
\partial_r[\mathsf{J}_\upnu({\uplambda_{\upnu\gamma}r})]+2\mathsf{F}\Big[\mathcal{L}^{(2-\upnu)}_{5}\int_0^\mathrm{1}\mathrm{d}r\,r^{\upnu+3}
\mathsf{J}_\upnu({\uplambda_{\upnu\gamma}r}) + \mathcal{L}^{(3-\upnu)}_{4}\int_0^\mathrm{1}\mathrm{d}r\,r^{\upnu+1}
\big[r\partial_r[\mathsf{J}_\upnu({\uplambda_{\upnu\gamma}r})]+2\mathsf{J}_\upnu({\uplambda_{\upnu\gamma}r})\big]\Big]\Bigg].\tag{A.5.10}\label{eq:V}
\end{multlined}
\end{equation}
\begin{multicols}{2}
\noindent Moreover, the approximated expression of $t_{\upnu\gamma}$ can also be calculated following \eqref{eq:partialeta2}, and is given by
\end{multicols}
\begin{equation}
\begin{multlined}
t_{\upnu\gamma}=\frac{4\mathsf{E}^{-\frac{1}{2}}\Omega^{-1}\mathsf{F}^2\mathsf{N}^2\Big[(\mathsf{FN}^2-\mathcal{K}_{-})e^{\mathcal{K}_{+}}-(\mathsf{FN}^2-\mathcal{K}_{+})
e^{\mathcal{K}_{-}}\Big]}{\uplambda^2_{{\upnu\gamma}}\Big[(\eta_\mathrm{1}\mathcal{K}_{+} - \partial_z\eta|_{z=1} - \mathsf{FN}^2)(\mathsf{FN}^2-\mathcal{K}_{-})e^{\mathcal{K}_{+}}-(\eta_\mathrm{1}\mathcal{K}_{-} - \partial_z\eta|_{z=1} - \mathsf{FN}^2)(\mathsf{FN}^2-\mathcal{K}_{+})
e^{\mathcal{K}_{-}}\Big]},\tag{A.5.11}\label{eq:approxt}
\end{multlined}
\end{equation}
\begin{multicols}{2}
\noindent where, $$\eta_\mathrm{1} \sim \eta_\mathrm{o} + \partial_z\eta + {\displaystyle\frac{1}{2}}\partial_z^2\eta;\,\,\partial_z\eta|_{z=1} \sim \partial_z\eta + \partial_z^2\eta;$$ given, $$\eta_{\mathrm{o}} = (v^\mathrm{o}_\mathrm{c}/c)^2,\,\,\partial_z\eta \sim 2v^\mathrm{o}_\mathrm{c}\partial_z\hspace{-0.0175in}v_\mathrm{c}\;\mathrm{and} \;\partial^2_z\eta \sim 2(\partial_z\hspace{-0.0175in}v_\mathrm{c})^2.$$
\end{multicols}
\vspace{-0.1in}
\subsection{Error characterization}
\label{A.6}
\begin{figure}[H]
\centering
\includegraphics[width=140mm]{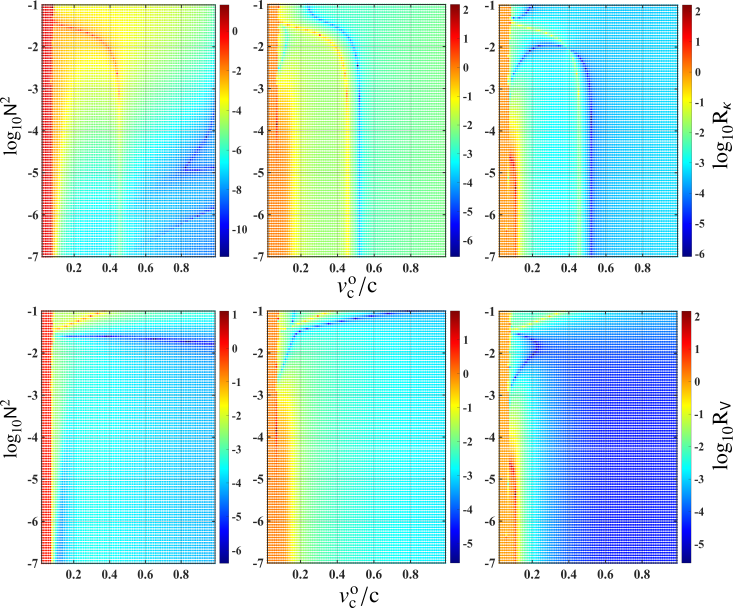} 
\caption{{\small Error characteristics:} {{\small 3 different sets of data are plotted for the $t_{21}$ mode at a rotational frequency of $100\;$Hz, and for a set of values of $\partial_z\hspace{-0.0175in}v_\mathrm{c}$ of $0$, $10^{-3}c\mathrm{L}^{-1}$ and $-10^{-4}c\mathrm{L}^{-1}$ from left to right. Other relevant physical parameters are chosen from astrophysical priors, as in all previous figures. The top panel represents $\mathrm{R_\kappa}$ while the bottom panel represents $\mathrm{R}_\mathsf{V}$. Note that larger values of $\mathrm{R_\kappa}$ and $\mathrm{R}_\mathsf{V}$ signify large mismatch between approximated analytic and numerical results.}}} 
\label{fig:A1}
\end{figure}
\begin{multicols}{2}
\noindent In this section, we show the comparison between analytically approximated and numerically computed results for $\kappa_{\upnu\gamma}$ and $\mathsf{V}_{\upnu\gamma}$. We have explored the results for the $t_{21}$ mode since this is sufficient for our purposes. We define the differences between analytically approximated ($\kappa^\mathrm{t}_{\upnu\gamma}$, $\mathsf{V}^\mathrm{t}_{\upnu\gamma}$) and numerical ($\kappa^\mathrm{n}_{\upnu\gamma}$, $\mathsf{V}^\mathrm{n}_{\upnu\gamma}$) results as follows,
\begin{equation}
\begin{multlined}
\mathrm{R_\kappa} = \Bigg|\frac{\kappa^\mathrm{t}_{\upnu\gamma} - \kappa^\mathrm{n}_{\upnu\gamma}}{\kappa^\mathrm{t}_{\upnu\gamma} + \kappa^\mathrm{n}_{\upnu\gamma}}\Bigg|,\tag{A.6.1}\label{eq:resK}
\end{multlined}
\end{equation}
\begin{equation}
\begin{multlined}
\mathrm{R}_\mathsf{V} = \Bigg|\frac{\mathsf{V}^\mathrm{t}_{\upnu\gamma} - \mathsf{V}^\mathrm{n}_{\upnu\gamma}}{\mathsf{V}^\mathrm{t}_{\upnu\gamma} + \mathsf{V}^\mathrm{n}_{\upnu\gamma}}\Bigg|.\tag{A.6.2}\label{eq:resV}
\end{multlined}
\end{equation}
In figure \ref{fig:A1}, we plot the characteristics of $\mathrm{R_\kappa}$ and $\mathrm{R}_\mathsf{V}$ for three cases, i.e when $\partial_z\hspace{-0.0175in}v_\mathrm{c}\in \{0, 10^{-3}c\mathrm{L}^{-1}, -10^{-4}c\mathrm{L}^{-1}\}$. The leftmost panels show a baseline mismatch between approximated analytic values and numerically calculated values of $\kappa_{\upnu\gamma}$ and $\mathsf{V}_{\upnu\gamma}$. Note that since $\partial_z\hspace{-0.0175in}v_\mathrm{c} = 0$ for these two panels, the numerical and approximated analytic results should not have a high mismatch. However, the results deviate from accuracy for certain regions in parameter space, especially for lower values of $v^\mathrm{o}_\mathrm{c}$. The center and rightmost panels show similar characteristics. Note that the mismatch in $\kappa_{\upnu\gamma}$ and $\mathsf{V}_{\upnu\gamma}$ follows somewhat similar characteristics to the time-scales plotted in figure \ref{hplot}. The underlying reason is fairly straightforward: larger time-scales occur when $\mathsf{K_s}$ becomes large in magnitude, and this large magnitude of $\mathsf{K_s}$ tends to throw off the numerical results from accuracy while the approximated analytic results continue to follow an accurate description. Note that the factor $\uprho_e(z)$ in the expressions of $\kappa_{\upnu\gamma}$ and $\mathsf{V}_{\upnu\gamma}$ tends to fall very rapidly with $z$ from a large value $\uprho_\mathrm{o}$ at $z = 0$ for large magnitudes of $\mathsf{K_s}$\footnote{While this effects the numerical results of $\kappa_{\upnu\gamma}$ and $\mathsf{V}_{\upnu\gamma}$, no such effect is present in the expression for time-scale $t_{\upnu\gamma}$.}. We also find that the numerical values of $\mathrm{Z}_{\upnu\gamma}(z)$ tend to wander inaccurately into negative domain from tolerance-induced numerical errors nearing $z = 0$. This small discrepancy between the values calculated by numerical methods and approximate analytic expressions is amplified by the larger value of $\uprho_e(z)$ nearing $z = 0$, especially when $\mathsf{K_s}$ is large, leading to a large mismatch. This affect also contributes to figure \ref{fig:A1} for the case of $\partial_z\hspace{-0.0175in}v_\mathrm{c} = 0$, i.e. leftmost panels.
\subsection{\boldmath{$t_{\upnu\gamma}\;\text{vs}\;f$}}
\label{A.7}
\begin{figure}[H]
\centering
\includegraphics[width=78mm]{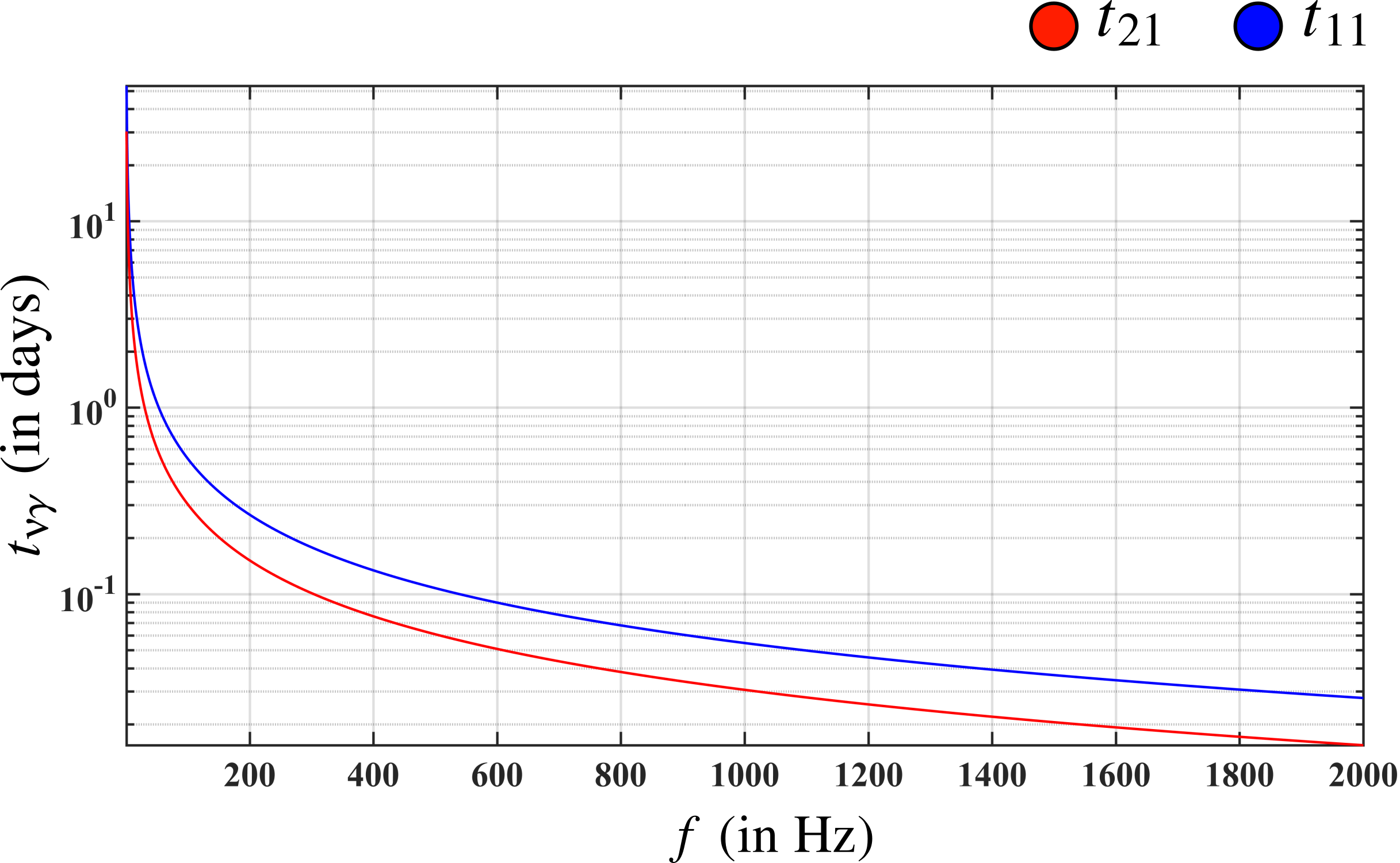} 
\caption{{\small $t_{\upnu\gamma}$ characteristics:} {{\small $t_{21}$ and $t_{11}$ are plotted as a function of neutron star's rotational frequency $f$. We have set $v_\mathrm{c}^\mathrm{o} = 0.1c$, $\mathsf{N}^2 = 10^{-4}$,  $\mathsf{E}=10^{-10}$, $\epsilon = 10^{-4}$, $d_\mathsf{s}=1\,$kpc, $\mathrm{L}=10^{4}$m, $g=10^{12}\mathrm{m/sec^2}$, $\uprho_\mathrm{o}=10^{17}\mathrm{kg/m^3}$, $\partial_z\hspace{-0.0175in}v_\mathrm{c}=0$.}}} 
\label{fig:tE}
\end{figure}

\noindent In this section, we elaborate on the characteristics of emitted signals as a function of neutron star's rotational frequency $f$. In figure \ref{fig:tE}, we plot the time-scales for \{2,1\} and \{1,1\} modes, i.e. $t_{21}$ and $t_{11}$, as a function of $f$. These time-scales have been calculated and implicitly included in the results via \eqref{eq:Sh1} in figure \ref{fig:Sh}. We can conclude from figure \ref{fig:tE} that these time-scales may span orders of magnitudes. For \eqref{eq:Sh1} to be a valid measure of minimum detectable strain for such signals, the observation time for the coherent search must be larger than these time-scales, i.e. $\mathrm{T}_\mathsf{obs} \ge t_{\upnu\gamma}$. 
\vspace{-0.07in}
\vspace{-0.1in}
\begin{figure}[H]
\centering
\includegraphics[width=78mm]{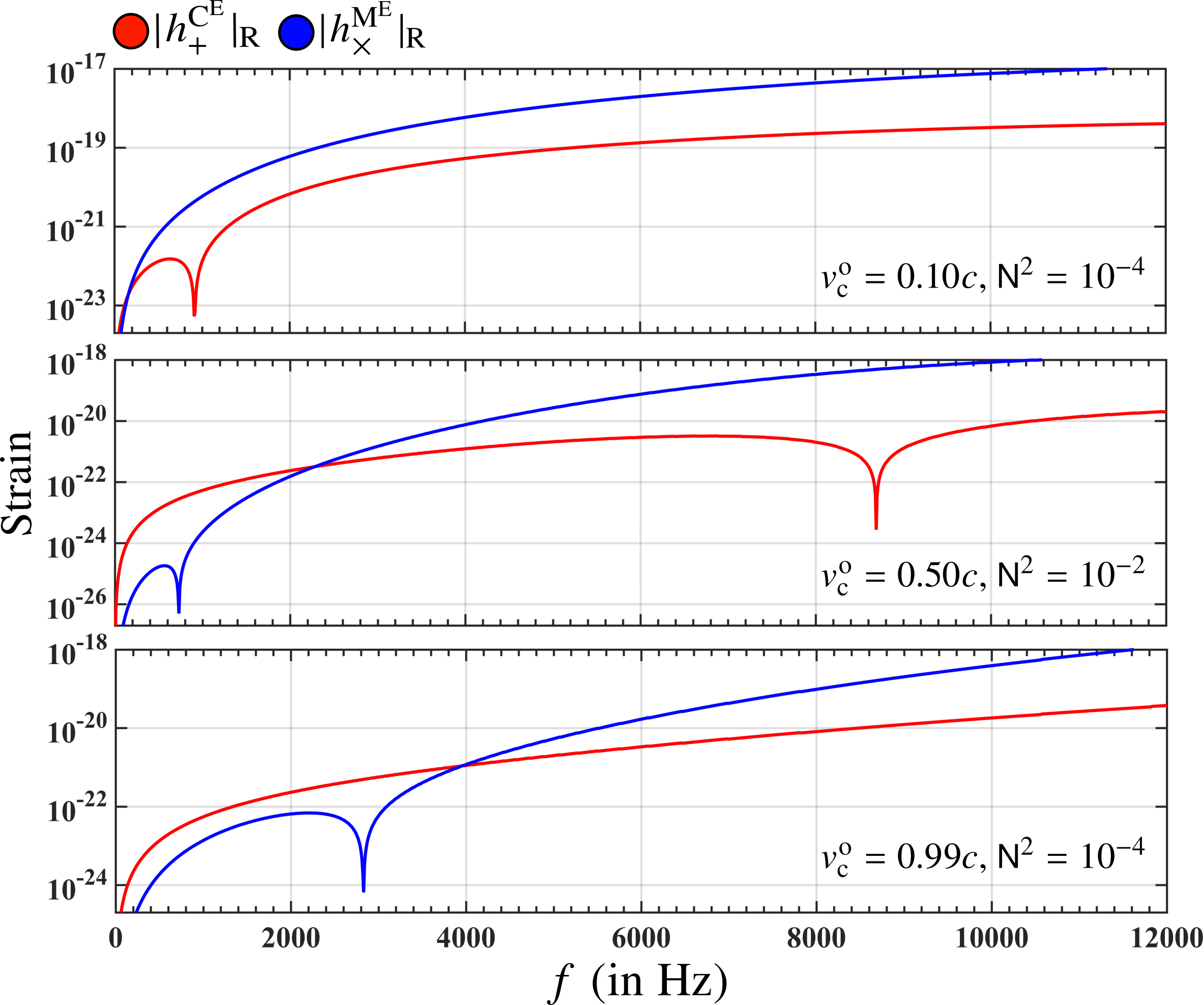} 
\caption{{\small Emission characteristics:} {{\small We have set $\mathsf{E}=10^{-10}$, $\epsilon = 10^{-4}$, $d_\mathsf{s}=1\,$kpc, $\mathrm{L}=10^{4}$m, $g=10^{12}\mathrm{m/sec^2}$, $\uprho_\mathrm{o}=10^{17}\mathrm{kg/m^3}$, $\partial_z\hspace{-0.0175in}v_\mathrm{c}=0$, and 3 different sets of values of $v_\mathrm{c}^\mathrm{o}$ and $\mathsf{N}^2$ are explored. Note the appearance of dips in one or both the contributions in frequency space i.e. mass-quadrupole and current-quadrupole emissions, due to variation in $v_\mathrm{c}^\mathrm{o}$ and $\mathsf{N}^2$.}}} 
\label{fig:KV}
\end{figure}
Moreover, there is a noticeable dip in emission from current-quadrupole contribution in the mid-frequency range in figure \ref{fig:Sh}. This dip is caused by $\mathsf{V}_{\upnu\gamma}$ becoming negative with increasing frequency. The sharp dip occurs due to the inclusion of $\mathsf{V}_{\upnu\gamma}$ in \eqref{eq:plusc}-\eqref{eq:crossc1} via its absolute magnitude. This affect is not limited to $\mathsf{V}_{\upnu\gamma}$ and current-quadrupole contribution only. In fact, the presence of this dip in current-quadrupole or mass-quadrupole emission depends upon the values of $v_\mathrm{c}^\mathrm{o}$ and $\mathsf{N}^2$. Like  $\mathsf{V}_{\upnu\gamma}$,  $\kappa_{\upnu\gamma}$ may also show similar effect for alternative values of $v_\mathrm{c}^\mathrm{o}$ and $\mathsf{N}^2$. Furthermore, the location of this dip in frequency space is found to vary with $v_\mathrm{c}^\mathrm{o}$ as well as $\mathsf{N}^2$. We also find that the location of this dip is less sensitive to variation in $\mathsf{N}^2$ as compared to $v_\mathrm{c}^\mathrm{o}$. In figure \ref{fig:KV} above, we plot a part of figure \ref{fig:Sh} (top-right panel) to demonstrate the aforementioned effect.
\subsection{Dependence of emission properties on \boldmath{$\partial_z\hspace{-0.0175in}v_\mathrm{c}$}}
\label{A.8}
The significant and critical affect of $\partial_z\hspace{-0.0175in}v_\mathrm{c}$ on the properties of the gravitational wave emission discussed in section \ref{section:times} and section \ref{conclusions} is best estimated by looking at \eqref{eq:boundary} and the discussion in section \ref{A.1}. For instance, \eqref{A.1.2} equates the rate of flow into the viscous boundary layer at the top and bottom faces of the cylinder with the rate of flow out of this boundary layer back into the bulk; this is a direct consequence of the conservation of mass across the viscous boundary layer \citep{epstein,melatos2008,walin,melatos2010}. This rate of exchange of fluid determines the dissipation time-scale of a certain perturbed \{$\upalpha,\gamma$\} mode -- faster exchange of fluid leads to faster dissipation of the perturbation. The value of $\partial_z\hspace{-0.0175in}v_\mathrm{c}$ bears a direct consequence on this process at the boundary layer. For example, from \eqref{A.1.8} and \eqref{eq:Z}, we see that $\partial_z\hspace{-0.0175in}v_\mathrm{c}$ contributes via the ``slope term'' ${\partial\mathrm{Z}_{\upalpha\gamma}}/{\partial z}$, and $\partial_z\eta$ term in the denominator in \eqref{A.1.8}. In fact, the ${\partial\mathrm{Z}_{\upalpha\gamma}}/{\partial z}$ term is the dominant determinant in deciding the speed of exchange since $|\partial_z\eta| \ll 1$. When the slope term is large and positive, the exchange of fluid is slow, as clearly seen in \eqref{A.1.8}. This is simply because the fluid flowing out of the boundary layer and back into the bulk has to work against high pressure gradient at $z=\pm 1$, which is set by the positive value of the slope term. Note that positive slope of $\mathrm{Z}_{\upalpha\gamma}$ implies decreasing pressure in $z$-direction, as seen in \eqref{eq:chi}. Similarly, when the slope term is positive but small, the exchange of fluid is faster since the pressure gradient decreases in value. Note that when the slope term becomes negative, we may see {growing modes} although this is neither a sufficient nor a necessary condition; the {growing modes} could also occur when $\mathrm{N}^2 < 0$ despite the slope being positive. This effect is seen in figure \ref{hplot}, where the value $\partial_z\hspace{-0.0175in}v_\mathrm{c}=-10^{-4}c\mathrm{L}^{-1}$ increases the characteristic time-scales as well as the corresponding gravitational wave amplitudes in some regions of the parameter space. It is important to remember here that this increase in gravitational wave amplitude occurs at the resonance frequency only, and the amplitudes decay in the side-bands. Hence, while increasing time-scales increase the gravitational wave amplitude at the resonance frequency, they also decrease the effective bandwidth of the signal frequency. \\
\end{multicols}
\begin{multicols}{2}
\bibliographystyle{plainnat}
\bibliography{reference}
\end{multicols}
\end{document}